\documentclass[11pt,english]{article}
\usepackage[T1]{fontenc}
\usepackage[latin9]{inputenc}
\usepackage{verbatim}
\usepackage{mathrsfs}
\usepackage{amsmath}
\usepackage{amsthm}
\usepackage{amssymb}
\usepackage{graphicx}
\usepackage{setspace}
\usepackage[authoryear]{natbib}
\usepackage{xargs}[2008/03/08]
\usepackage{tikz}
\usepackage[show]{simple_notes}
\usepackage{float}
\usepackage{threeparttable}
\usepackage{pdfpages}

\newcommand{\blind}{1}

\makeatletter
\theoremstyle{plain}
\newtheorem{thm}{\protect\theoremname}
  \theoremstyle{plain}
  \newtheorem{assumption}{\protect\assumptionname}
  \theoremstyle{plain}

  \theoremstyle{remark}
  \newtheorem*{rem*}{Remark}
  
  \numberwithin{thm}{section}

\@ifundefined{date}{}{\date{}}
\usepackage{ifpdf} 
\ifpdf 

 \IfFileExists{lmodern.sty}{\usepackage{lmodern}}{}

\fi 

\usepackage{multirow}
\usepackage{mathrsfs}
\usepackage{enumitem}

\usepackage{ifthen}
\usepackage{mnsymbol}
\usepackage{xcolor}
\usepackage{natbib} 
\def\LyX{\texorpdfstring{%
  L\kern-.1667em\lower.25em\hbox{Y}\kern-.125emX\@}
  {LyX}}

\newcommand{\mylabel}[2]{#2\def\@currentlabel{#2}\label{#1}}
\newcommand{\vertiii}[1]{{\left\vert\kern-0.25ex\left\vert\kern-0.25ex\left\vert #1 
    \right\vert\kern-0.25ex\right\vert\kern-0.25ex\right\vert}}

\usepackage[hidelinks]{hyperref}
\hypersetup{
    colorlinks,
    linkcolor={red!50!black},
    citecolor={blue!50!black},
    urlcolor={blue!80!black}
}

\usepackage{babel}
  \providecommand{\assumptionname}{Assumption}
  \providecommand{\propositionname}{Proposition}
\providecommand{\theoremname}{Theorem}

\global\long\def\expect{\mathbb{E}}
\global\long\def\prob{\mathrm{Pr}}

\global\long\def\real{\mathbb{R}}

\global\long\def\manifold{\mathcal{M}}

\global\long\def\hilbert{\mathcal{H}}

\global\long\def\asymplt{\lesssim}

\global\long\def\asympgt{\gtrsim}
\global\long\def\asympeq{\asymp}

\newcommandx\tangentspace[2][usedefault, addprefix=\global, 1=\manifold]{T_{#2}#1}

\global\long\def\vec#1{\mathbf{#1}}
\newcommandx\lpnorm[3][usedefault, addprefix=\global, 1=r, 2=]{\|#3\|_{\mathcal{L}^{#1}}^{#2}}
\newcommandx\lp[1][usedefault, addprefix=\global, 1=p]{\mathcal{L}^{#1}}
\global\long\def\transpose{\mathrm{\top}}

\newcommandx\vfnorm[3][usedefault, addprefix=\global, 1=\mu, 2=]{\|#3\|_{#1}^{#2}}
\newcommandx\vfinnerprod[2][usedefault, addprefix=\global, 1=\mu]{\llangle#2\rrangle_{#1}}

\global\long\def\tdomain{\mathcal{T}}

\newcommandx\opnorm[3][usedefault, addprefix=\global, 1=\mu, 2=]{\vertiii{#3}_{#1}^{#2}}
\newcommandx\fronorm[2][usedefault, addprefix=\global, 1=]{|#2|_{F}^{#1}}

\global\long\def\sci{C}

\def\dk{d_{\mathrm{K}}}
\def\law{\mathcal{L}}
\def\sci{\mathrm{SCR}}

\def\size{\textsc{size}}
\def\ts{\textstyle}
\def\nmin{n}
\def\R{\mathbb{R}}
\def\P{\mathbb{P}}
\def\ngroup{K}
\def\notdependon{not depending on $n$}

\definecolor{myblue}{rgb}{0.0, 0.2, 0.6}
\def\lin#1{\textcolor{black}{#1}}

\usepackage{xr}
\begin{document}

%

\def\spacingset#1{\renewcommand{\baselinestretch}%
	{#1}\small\normalsize} \spacingset{1}


\if1\blind
{
	\title{\bf High-dimensional MANOVA via Bootstrapping and its Application to Functional and Sparse Count Data}
		\author{Zhenhua Lin\thanks{
				Research partially supported by NUS start-up grant R-155-001-217-133.}\\
			Department of Statistics and Applied Probability, National University of Singapore\\
			\,\\
			Miles E.~Lopes\thanks{Research supported in part by NSF grant DMS-1915786.} \\
			Department of Statistics, University of California, Davis\\
			\, \\
			Hans-Georg M\"uller\thanks{Research supported in part by NSF grant DMS-2014626.} \\
			Department of Statistics, University of California, Davis}
	\maketitle
} \fi

\if0\blind
{
	\bigskip
	\bigskip
	\bigskip
	\begin{center}
		{\LARGE\bf  High-dimensional MANOVA via Bootstrapping and its Application to Functional and Sparse Count Data}
	\end{center}
	\medskip
} \fi

\begin{abstract}
	We propose a new approach to the problem of high-dimensional multivariate ANOVA via bootstrapping max statistics that involve the differences of sample mean vectors. The proposed method proceeds via the construction of simultaneous confidence regions for the differences of population mean vectors. It is suited to simultaneously test the equality of several pairs of mean vectors of potentially more than two populations. By exploiting the variance decay property that is a natural feature in relevant applications, we are able to provide dimension-free and nearly-parametric convergence rates for Gaussian approximation, bootstrap approximation, and the size of the test. We demonstrate the proposed approach with ANOVA problems for functional data and sparse count data. The proposed methodology is shown to work well in 
		simulations and several real data applications.
\end{abstract}

\noindent%
{\it Keywords:}   Bootstrap methods, Functional data analysis, Hypothesis testing, Gaussian approximation, Mean function, Physical activity, Poisson data, Simultaneous confidence intervals.
\vfill

\newpage
\spacingset{1.5} 

\section{Introduction}\label{sec:intro}
The MANOVA  problem of detecting significant differences among the means of multivariate populations is of central importance in a myriad of statistical applications.
%
However, the classical MANOVA  approaches  are only intended to handle low-dimensional settings where the number of covariates is much smaller than the sample size, which is a crucial limitation for modern high-dimensional data analysis.
%
Due to the demand for methodology that provides valid inference for  high-dimensional data,  the challenge of finding suitable new MANOVA methods has developed into a major line of research.
For example, the special case of  high-dimensional  two-sample testing has been investigated by \cite{Bai1996,Lopes2011,Cai2014,Thulin2014,Xu2016,ZHANG2016a,Zhang2019a} under the condition that populations share a common covariance matrix, while procedures designed by \cite{Chen2010,Feng2015,Feng2015a,Gregory2015,Staedler2016,Chang2017,Xue2020} do not require such a common covariance assumption. For the more general multiple-sample problem, methods and theory were studied by \citet{Fujikoshi2004,SRIVASTAVA2006,Schott2007,Yamada2012,SRIVASTAVA2013,Cai2014a,ZHANG2017,Bai2018,Li2020} when the populations share common covariance structure, while \citet{Zhang2009,YAMADA2015,Li2017,Hu2017,Zhou2017,Zhang2018b} eliminated the requirement of common covariance. Among these, \cite{Chang2017,Zhang2018b,Xue2020} adopt a bootstrap approach following  \cite{Chernozhukov2013,Chernozhukov2017}. 

{An important observation in this context is that  the variances of variables often exhibit a certain decay pattern.}  As an example, consider a multinomial model of $p$ categories.  Without loss of generality, assume that the probabilities of the $p$ categories are ordered as $\pi_1\geq\cdots\geq\pi_p$. Since the probabilities sum to one, it follows that the variance $\sigma_j^2=\pi_j(1-\pi_j)$ of the $j$th category must decay at least as fast as $j^{-1}$. Additional examples that arise in connection with principal component analysis and the Fourier coefficients of functional data may be found in~\cite{Lopes2020}.

When the structure of variance decay is available, \cite{Lopes2020} showed that near-parametric and dimension-free rates of Gaussian and bootstrap approximation can be established for max statistics of the form $\max_{1\leq j\leq p} \sqrt{n}\{\bar{X}-\mu\}(j)/\sigma_j^\tau$. In this expression,  $\bar{X}=(\bar X(1),\dots,\bar X(p))$ is the sample mean of $n$ independent and identically distributed random vectors with mean vector $\mu=(\mu(1),\dots,\mu(p))$ and coordinate-wise variances $\sigma_1^2,\dots,\sigma_p^2$, while the symbol $\tau$ denotes a tuning parameter in the interval $ [0,1)$. Remarkably, the near-parametric rates of approximation remain valid even when the decay is very weak, i.e., $\sigma_j\asymp j^{-\alpha}$ for an arbitrarily small $\alpha>0$. 
 In this paper, we {harness such decay patterns to develop promising bootstrap-based inference for the high-dimensional MANOVA problem.}

We consider a general setting with $\ngroup\geq 2$ populations having mean vectors \smash{$\mu_1,\ldots,\mu_\ngroup\in\R^p$.} For any collection of ordered pairs $\mathcal{P}$ taken from the set $\{(k,l): 1\leq k<l\leq \ngroup\}$, the hypothesis testing problem of interest is 
\begin{equation}\label{eq:test}
\mathbf{H}_0: \mu_k=\mu_l\,\,\text{  for all \ } (k,l)\in\mathcal{P}\quad\quad\text{ \ \ versus \ \ }\quad\quad \mathbf{H}_a: \mu_k\neq \mu_l \text{ \ for some \ } (k,l)\in\mathcal{P}.
\end{equation}
Note that this includes a very general class of null hypotheses of possible interest. The proposed  strategy is to construct simultaneous confidence regions for the differences $\mu_k-\mu_l$ for all pairs in $\mathcal{P}$ via bootstrapping a maximum-type statistic related to $\mu_k-\mu_l$ across all coordinates and all pairs. In addition, we adopt the idea of partial standardization developed in \cite{Lopes2020} to take advantage of the variance decay. This differs from the existing  bootstrap-based methods proposed in \cite{Chang2017,Xue2020,Zhang2018b} that do not exploit the decay. Furthermore, in the first two papers the authors consider only one- or two-sample problems, and in the last paper  only the standard global null hypothesis $\mu_1=\cdots=\mu_\ngroup$.

The proposed method has several favorable properties:
\begin{itemize}
	\item  There is flexibility in the choice of null hypothesis. In addition to the basic global null hypothesis  $\mu_1=\cdots=\mu_\ngroup$, which corresponds to choosing $\mathcal{P}=\{(k,l): 1\leq k<l\leq \ngroup\}$, we can also test more specific hypotheses. For instance, the null hypothesis $\mu_1=\mu_2$ and $\mu_3=\mu_4$ corresponds to $\mathcal{P}=\{(1,2),(3,4)\}$. In general, whenever $\mathcal{P}$ contains more than one pair, traditional methods often require that two or more separate tests are performed. This requires extra adjustments for multiple comparisons, which often have a negative impact on power. Indeed, the effect of multiplicity can be severe, because the number of pairs $|\mathcal{P}|$ may grow quadratically as a function of $\ngroup$, as in the case of the global null hypothesis with $|\mathcal{P}|=\ngroup(\ngroup-1)/2$.
	\item  The proposed method performs the test via constructing simultaneous confidence regions (SCR) for the differences $\mu_k-\mu_l$  indexed by $(k,l)\in\mathcal{P}$. Such SCRs are also valuable in their own right {(in addition to their utility for hypothesis testing)},  as they provide quantitative information about the separation of the mean vectors $\mu_1,\dots,\mu_\ngroup$ {that is often of interest in applications.} 
	\item When the null hypothesis is rejected, the proposed approach makes it possible to  immediately identify pairs of populations that have significantly different means without performing additional tests. By contrast, additional testing is often necessary when one adopts and extends traditional MANOVA approaches. 
	\item  Like \cite{Chang2017,Zhang2018b,Xue2020},  who essentially propose two-sample {or multiple-sample} 
	comparisons  based on bootstrapping, we do not require that the ratio of the sample sizes of any pair of populations converges to a specific limit. 
	\item In contrast to the testing procedures of \cite{Chang2017,Zhang2018b} (where the convergence rates for the size of the test are not established), and the method of \cite{Xue2020} (for which the convergence rate is at most $\sqrt{\log p}/n^{1/6}$), the proposed approach is shown to enjoy a near-parametric rate of convergence. Furthermore, this near-parametric rate is free of the dimension $p$ and holds under mild assumptions. These improvements are achieved by exploiting variance  decay. 
\end{itemize}


To demonstrate the usefulness of the proposed approach, we apply our procedure to perform ANOVA for functional data and sparse count data. Functional data are commonly encountered in many types of statistical analysis, as surveyed in the monographs \citet{Ramsay2005,Ferraty2006,Horvath2012,Zhang2013,Hsing2015,Kokoszka2017} and review papers \citet{WangCM16,Aneiros:2019aa}. 
{Previous examples of methods for functional ANOVA are} 
pointwise $F$-tests \citep[p.227,][]{Ramsay2005}, an integrated $F$-test and its variants \citep{Shen2004,Zhang2011,Zhang2013}, globalization of pointwise $F$-tests 
\citep{Zhang2014}, \lin{a test based on the maximum of pointwise $F$-statistics \citep{Zhang2019},} the HANOVA method \citep{Fan1998}, $L^{2}$ 
norm based methods \citep{Faraway1997,Zhang2007}, \lin{random projection based test \citep{Cuesta-Albertos2010}, a global envelope test with graphical interpretation \citep{Mrkvicka2020}}, and an empirical likelihood ratio approach  \citep{Chang2020}, in addition to resampling methods 
\citep{Zhang2013,Paparoditis2016}. 

While the proposed approach makes use of the techniques and some results developed in \cite{Lopes2020}, adapting  these results to the multiple-sample setting is a major challenge. The key obstacle is that, in contrast to the situation studied in \cite{Lopes2020}, the max statistic \eqref{eq:M} in the MANOVA  setting is not the maximum of an average of independent vectors.
Overcoming this difficulty requires a delicate transformation of the statistic to represent it as the maximum of the average of independent random vectors that are further transformations of the data; see Proposition \ref{prop:gau-approx-II} in the Supplement.  
In addition, the theory here is more comprehensive in the way that it accounts for the effect using estimated standard deviations $\hat\sigma_j$ in the SCR. This is done by establishing a uniform bound on the estimation error of $\hat\sigma_j$ over all coordinates and groups,  which holds when the data satisfy a basic continuity assumption; see Lemma \ref{lem:bound-on-sigma-hat} in the Supplement.


The  rest of the paper is structured as follows. In Section \ref{sec:method} we present the details of the proposed method. In Section \ref{sec:theory} we establish theoretical guarantees for bootstrapping max statistics under a multiple-sample setting, including a result on the convergence rate of the empirical size of the proposed test. Our signature application to functional ANOVA  is given in Section \ref{sec:functional-anova} and a second application to sparse count data is given in Section \ref{sec:count-data}. We conclude the paper in Section \ref{sec:conclusion}.

\section{High-dimensional multiple-sample test}\label{sec:method}

Consider $\ngroup$ independent groups of observations, where we assume that for the $k$th group one has  $n_k$ i.i.d. (independently and identically distributed) $p$-dimensional observations $X_{k,1},\ldots,X_{k,n_k}$ with mean $\mu_k\in\real^p$. Our goal is to test any of the null hypotheses in \eqref{eq:test} based on these data.

To motivate our approach, consider a two-sample test in the classical setting that corresponds to the special case $p=1$ and $\ngroup=2$ with $(k,l)=(1,2)$. The common statistic  $T=\{(\bar{X}_k-\mu_k)-(\bar{X}_l-\mu_l)\}/\sqrt{\mathrm{var}(\bar{X}_k-\bar{X}_l)}$ asymptotically follows a standard Gaussian distribution, where $\bar{X}_k=n_k^{-1}\sum_{i=1}^{n_k}X_{k,i}$ denotes the sample mean of the $k$th group for $k=1,2$. This statistic can be used to construct  a confidence interval of level $1-\varrho$ for the difference $\mu_k-\mu_l$, which can then be used to implement the standard  two-sample test at level $\varrho$.  When $p>1$, one can construct a simultaneous confidence region for $\mu_k-\mu_l\in\real^p$ in terms of the distribution of the max statistic
\[
M^\prime(k,l)=\max_{1\leq j\leq p}\frac{\{\bar X_{k}(j)-\mu_k(j)\}-\{\bar X_l(j)-\mu_l(j)\}}{\sqrt{\mathrm{var}(\bar X_{k}(j)-\bar X_{l}(j))}}.
\]
For the general case when  $\ngroup\geq 2$, it is natural to consider the  max statistic $M^\prime=\max_{(k,l)\in\mathcal{P}} M^\prime(k,l).$ 
One may equivalently rewrite the statistic $M^\prime(k,l)$ as
 \[
M^\prime(k,l)=\max_{1\leq j\leq p}\left(\sqrt{\ts\frac{n_l}{n_k+n_l}}\ts\frac{S_{k,j}}{\sigma_{k,l,j}}-\sqrt{\ts\frac{n_k}{n_k+n_l}}\ts\frac{S_{l,j}}{\sigma_{k,l,j}}\right),
\]
where 
$S_{k}=n_{k}^{-1/2}\sum_{i=1}^{n_{k}}(X_{k,i}-\mu_{k})$,  $S_{k,j}=S_k(j)$ denotes the $j$th coordinate, and $\sigma_{k,l,j}^2=\{n_l\mathrm{var}(X_k(j))+n_k\mathrm{var}(X_l(j))\}/(n_k+n_l)$. 
As shown in \cite{Lopes2020}, when the variances $\sigma_{k,l,j}^2$ exhibit a decay pattern, it is beneficial to use \emph{partial standardization}, 
\begin{equation}\label{eq:M}
M(k,l)=\max_{1\leq j\leq p}\left(\sqrt{\ts\frac{n_l}{n_k+n_l}}\ts\frac{S_{k,j}}{\sigma_{k,l,j}^{\tau}}-\sqrt{\ts\frac{n_k}{n_k+n_l}}\ts\frac{S_{l,j}}{\sigma_{k,l,j}^{\tau}}\right)\quad\text{and}\quad
M=\max_{(k,l)\in\mathcal{P}} M(k,l),
\end{equation}
where $\tau\in[0,1)$ is a parameter that may be tuned to maximize power. 

\begin{rem*} 
	To intuitively understand the role of $\tau$, it is helpful to consider the extreme cases of $\tau=1$ (ordinary standardization) and $\tau=0$ (no standardization). In the case of $\tau=1$,  the $j$th difference in \eqref{eq:M} has variance equal to 1 for every $j=1,\dots,p$, and hence, the ``low-dimensional structure'' of variance decay is eliminated. Likewise, in this situation, all of the $p$ coordinates are ``equally important'', which makes the problem genuinely high-dimensional --- and hence, makes bootstrap approximation more difficult. In the opposite case when $\tau=0$, a different issue arises. It can be seen from equation \eqref{eq:sci-two-sided} {below} that all of the $p$ simultaneous confidence regions will have the same width. 
This is undesirable, as the widths of the intervals should be adapted to the variance of each coordinate. In view of these undesirable effects when choosing the endpoints $\tau=1$ or $\tau=0$,   the proposed partial standardization seeks a tradeoff by allowing for intermediate values of $\tau$ between $0$ and $1$.\end{rem*}

As $M$ is the maximum of random variables that are in turn coordinate-wise maxima of a random vector, it is difficult to derive its distribution.\footnote{Note that $M$ itself is not a test statistic since it involves unknown parameters, but being able to estimate the quantiles of $M$ will enable our testing procedure based on SCRs.} This difficulty, fortunately, can be circumvented efficiently by bootstrapping, as follows.
Let $\hat{\Sigma}_{k}=n_{k}^{-1}\sum_{i=1}^{n_{k}}(X_{k,i}-\bar{X}_{k})(X_{k,i}-\bar{X}_{k})^{\transpose}$ be the sample covariance of the $k$th group. Define the bootstrap version of $S_k$ by $S_{k}^{\star}\sim N(0,\hat{\Sigma}_{k})$. (An equivalent definition is $S_k^\star=n_k^{-1/2}\sum_{i=1}^{n_k}X_{k,i}^\star$ with $X_{k,i}^\star$ i.i.d. sampled from $N(0,\hat\Sigma_k)$.) Likewise, the bootstrap version of  $M(k,l)$ is  defined by
\[
M^{\star}(k,l)=\max_{1\leq j\leq p}\left(\sqrt{\ts\frac{n_l}{n_k+n_l}} \ts\frac{S_{k,j}^{\star}}{\hat{\sigma}_{k,l,j}^{\tau}}-\sqrt{\ts\frac{n_k}{n_k+n_l}}\ts\frac{S_{l,j}^{\star}}{\hat{\sigma}_{k,l,j}^{\tau}}\right),
\]
where $\hat{\sigma}_{k,l,j}^{2}$ are diagonal elements of $\hat{\Sigma}_{k,l}=\frac{n_l}{n_k+n_l}\hat{\Sigma}_{k}+\frac{n_k}{n_k+n_l}\hat{\Sigma}_{l}$, and altogether, 
the bootstrap version of $M$ is defined by
$$M^\star =\max_{(k,l)\in\mathcal{P}} M^\star(k,l).$$
%
%
For a given dataset $X=\{X_{k,i}:\,1\leq k\leq \ngroup,\, 1\leq i \leq n_k\}$, we generate $B\geq 1$ independent samples of $(S^\star_1,\ldots,S^\star_{\ngroup})$, which yield $B$ independent samples of $M^\star$. Then, the empirical quantile function of these samples of $M^{\star}$, denoted by $\hat{q}_M(\cdot)$,  serves as an estimate of the quantile function $q_M(\cdot)$ of $M$.

Analogously, we define the min statistic \[
L(k,l)=\min_{1\leq j\leq p}\left(\sqrt{\ts\frac{n_l}{n_k+n_l}}\ts\frac{S_{k,j}}{\sigma_{k,l,j}^{\tau}}-\sqrt{\ts\frac{n_k}{n_k+n_l}}\ts\frac{S_{l,j}}{\sigma_{k,l,j}^{\tau}}\right)
\quad\text{and} 
\quad
L=\min_{(k,l)\in\mathcal{P}} M(k,l),
\]
as well as  their bootstrap counterparts,
\[
L^{\star}(k,l)=\min_{1\leq j\leq p}\left(\sqrt{\ts\frac{n_l}{n_k+n_l}}\ts\frac{S_{k,j}^{\star}}{\hat{\sigma}_{k,l,j}^{\tau}}-\sqrt{\ts\frac{n_k}{n_k+n_l}}\frac{S_{l,j}^{\star}}{\hat{\sigma}_{k,l,j}^{\tau}}\right)
\quad\text{and}\quad 
L^\star =\min_{(k,l)\in\mathcal{P}} L^\star(k,l).\]
Similarly, the  quantile function of $L^\star$ can be obtained by drawing samples from the distributions $N(0,\hat\Sigma_k)$.

Finally, the $1-\varrho$ two-sided simultaneous confidence regions (SCR) for the $j$th coordinates of $\mu_k-\mu_l$ for $j=1,\ldots,p$, $(k,l)\in\mathcal{P}$,  are given by 
\begin{align}
\sci{(k,l,j)}=\big[\bar{X}_k(j)-\bar{X}_l(j)-\ts\frac{\hat{q}_M(1-\varrho/2)\hat\sigma_{k,l,j}^\tau}{\sqrt{n_{k,l}}}\ , \ \,\,\bar{X}_k(j)-\bar{X}_l(j)-\ts\frac{\hat{q}_L(\varrho/2)\hat\sigma_{k,l,j}^\tau}{\sqrt{n_{k,l}}}\big],\label{eq:sci-two-sided}
\end{align} 
where $n_{k,l}:={n_kn_l/(n_k+n_l)}$ denotes the harmonic sample size of the $k$th and $l$th groups.  {With these SCRs in hand,} we perform the test in \eqref{eq:test} {by rejecting the null hypothesis  at the significance level $\varrho$ if  $0\notin \sci(k,l,j)$ for some $(k,l)\in\mathcal{P}$ and $j=1,\dots,p$}.   One-sided SCRs can be constructed and one-sided hypothesis tests can be conducted in a similar fashion. For the testing problem \eqref{eq:test}, it is often desirable to obtain the $p$-value, which corresponds to the largest value of $\varrho$ such that all SCRs in \eqref{eq:sci-two-sided} contain zero and can easily be found numerically. 

In practical applications,  one needs to determine a value for the parameter $\tau$.  Although in the next section it is shown that any fixed value in $[0,1)$ gives rise to the same asymptotic behavior of the proposed test, a data-driven method to optimize the empirical power is desirable. 
We propose to select the value of $\tau$ that yields the smallest $p$-value while keeping the size at the nominal level $\varrho$. We first observe that for a given value of $\tau$, the above bootstrap test provides a corresponding $p$-value. {It remains to estimate the empirical size for a given value of $\tau$. To this end, we propose the following resampling approach.} First, the data are centered within each group, so that the null hypothesis holds for the centered data. For each group, a new sample of the same size is generated by resampling the original dataset with replacement. Then, the proposed test is applied on the new samples {with the nominal significance level $\varrho$.} This process is repeated several times, for example, 100 times, and the empirical size is estimated by the proportion of the resampled datasets that lead to rejecting the null hypothesis.  {If a value of $\tau$ yields an empirical size that is bounded by the nominal level $\varrho$, then it is retained, and from these retained values of $\tau$, the one corresponding to the smallest $p$-value is selected.}

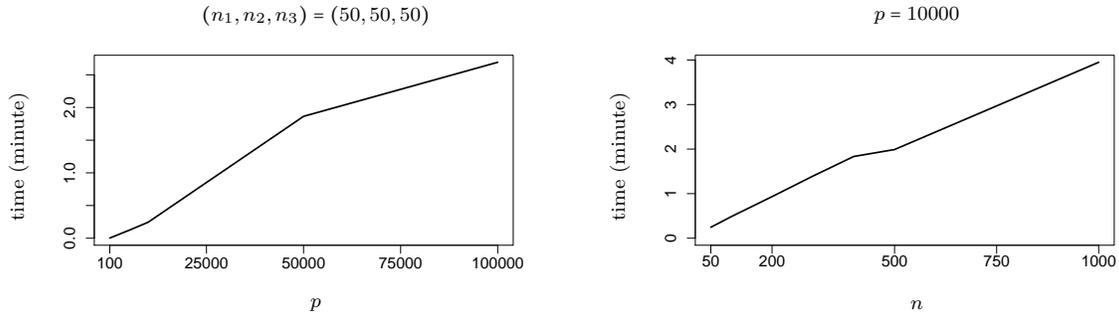
\begin{figure}[t]
	\begin{center}
	\begin{tikzpicture}[scale=0.94, every node/.style={scale=0.94}]
		\newcommand\y{0.8}
		\newcommand\dx{8.5}
		\newcommand\x{-1}

		\node at (\x,\y) {\includegraphics[width=0.45\textwidth]{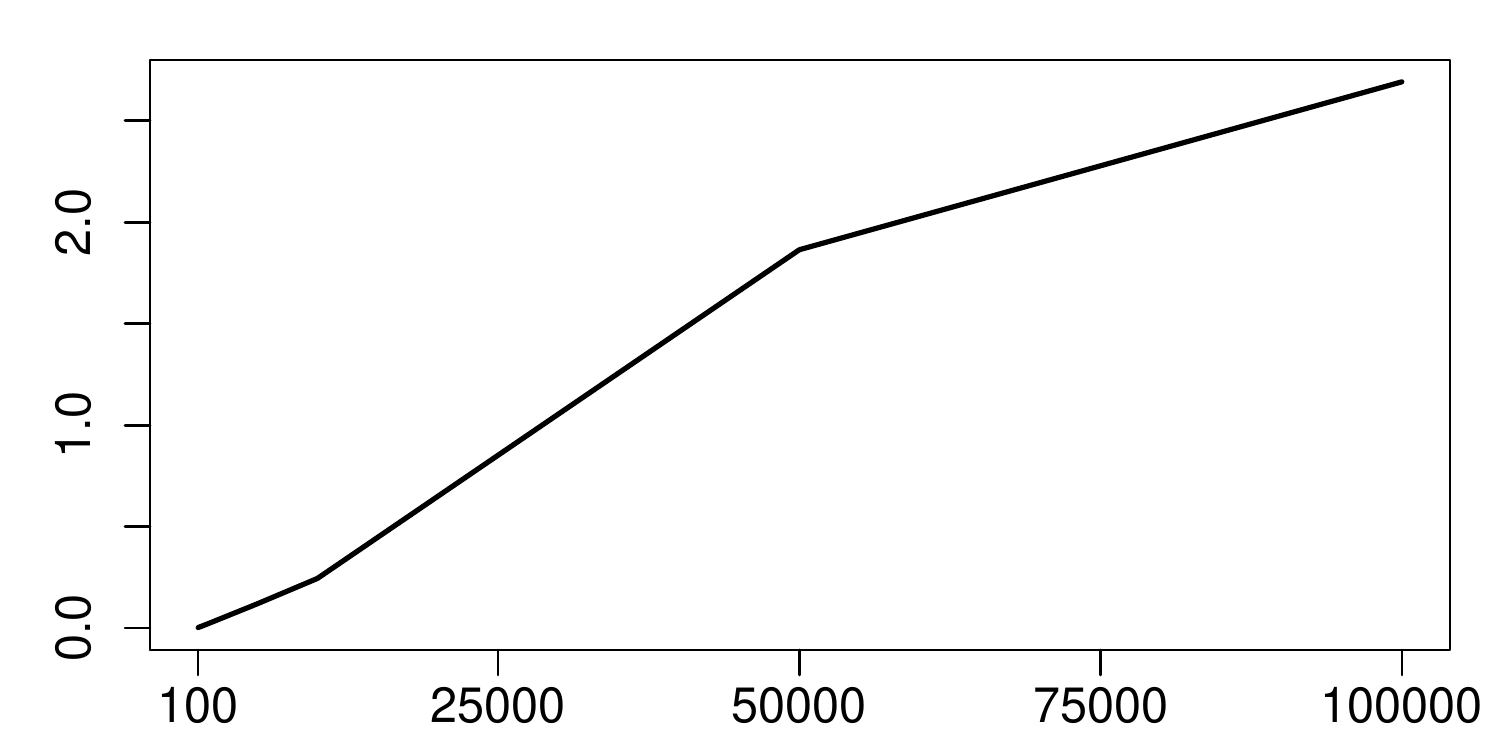}};
		\node at (\x+\dx,\y) {\includegraphics[width=0.45\textwidth]{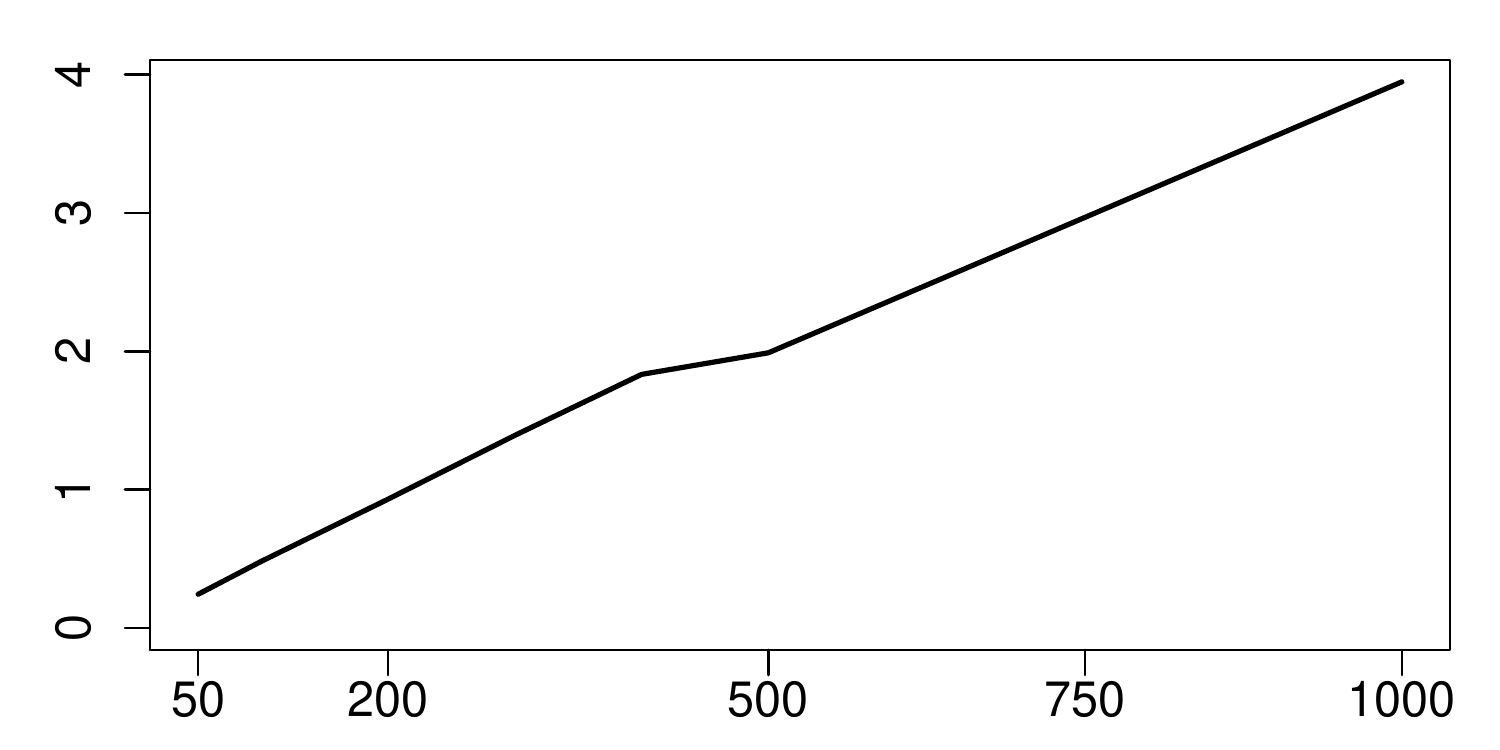}};	
		
		\pgfmathsetmacro{\x}{-0.6}
		\newcommand\dy{-2.1}
		\node at (\x,\y+\dy) {\scriptsize $p$};
		\node at (\x+\dx,\y+\dy) {\scriptsize $n$};
		
		\pgfmathsetmacro{\x}{-4.8}
		\node[rotate=90] at (\x,\y) {\scriptsize time (minute)};
		\node[rotate=90] at (\x+\dx,\y) {\scriptsize time (minute)};

		\pgfmathsetmacro{\x}{-0.6}
		\pgfmathsetmacro\dy{2}
		\node at (\x,\y+\dy) {\scriptsize $(n_1,n_2,n_3)=(50,50,50)$};
		\node at (\x+\dx,\y+\dy) {\scriptsize $p=10000$};
		
	\end{tikzpicture} 
\end{center}
\vspace{-0.2in}
	\caption{Computation time in a server with an NVIDIA Quadro P400 graphics card.
		 \label{fig:ct}}
\end{figure}

To tackle the additional {computational burden that this incurs,}  one can leverage the two levels of parallelism of the proposed algorithm: Each candidate value of $\tau$ in a grid can be examined in parallel, and for a given $\tau$, all the subsequent computations are parallel. Therefore, the proposed method is scalable with modern cloud,  cluster or GPU (graphics processing unit) based computing. 
For illustration, {we created an \texttt{R} software to implement the above parallel algorithm  for a GPU based platform.}  Figure \ref{fig:ct} shows the computation time that includes selecting a value for $\tau$ from $11$ candidate values, constructing the SCRs and performing the test, for datasets of $K=3$ groups, $(n_1,n_2,n_3)=(n,n,n)$ samples and $p$ dimensions. It is observed that the computation time scales efficiently in both $n$ and $p$. 

\section{Theory}\label{sec:theory}
\subsection{Bootstrapping max statistics for multiple samples}
\emph{Notation.} The identity matrix of size $p\times p$ is denoted by $I_p$. For a deterministic vector $v\in\real^p$ and $r>0$, let $\|v\|_r=(\sum_{j=1}^p|v_j|^r)^{1/r}$, and for a scalar random variable $\xi$, let  $\|\xi\|_r=\expect(|\xi|^r)^{1/r}$. The $\psi_1$-Orlicz norm of a random variable $\xi$ is denoted and defined by $\|\xi\|_{\psi_1}=\inf\{t>0:\expect[\exp(|\xi|/t)]\leq 2\}$. If $a$ and $b$ are real numbers, then we write $a\wedge b=\min\{a,b\}$ and $a\vee b=\max\{a,b\}$.

\noindent \emph{Conventions.} The main results are formulated in terms of a sequence of models indexed by the integer {$n=\min\{n_1,\dots,n_\ngroup\}$.}  All aspects of these models may depend on $n$, except where stated otherwise. Likewise, the following numbers may depend on $n$: the dimension $p$, the number of groups $K$, the group sample sizes $n_1,\dots,n_K$,\footnote{i.e.~for each $n=1,2,\dots$, the equation $n=\min\{n_1(n),\dots,n_K(n)\}$ is satisfied.} and the tuning parameter $\tau$. The set of pairs $\mathcal{P}$, as well as the population distributions of the groups may also depend on $n$. Accordingly, if it is stated that a constant $c$ does not depend on $n$, then it is understood that $c$ does not depend on any of these other numbers or objects. For constants of this type, the symbol $c$ will often be re-used with different values at each occurrence. If $a_n$ and $b_n$ are two sequences of non-negative real numbers, then $a_n\asymplt b_n$ means that there is a constant $c>0$ not depending on $n$, such that $a_n\leq cb_n$ holds for all large $n$. If both of the conditions $a_n\asymplt b_n$ and $b_n\asymplt a_n$ hold, then we write $a_n\asympeq b_n$.

\begin{assumption}[Data-generating model]
	\label{assu:1}~
	\begin{enumerate}[label=(\roman*)]
		\item  For each $k \in \{1,\ldots,\ngroup\}$, there exists a vector $\mu_k\in\real^{p}$
		and a positive semi-definite matrix $\Sigma_{k}\in\real^{p\times p}$, such that the
		observations $X_{k,1},\ldots X_{k,n_{k}}\in\real^{p}$ are generated
		as $X_{k,i}=\mu_{k}+\Sigma_{k}^{1/2}Z_{k,i}$ for each $1\leq i\leq n_{k}$, where $Z_{k,1},\ldots,Z_{k,n_{k}}\in\real^{p}$
		are i.i.d.~random vectors.
		\item  There is a constant $c_{0}>0$ not depending on $n$, such that for each $k\in\{1,\dots,\ngroup\}$, the random vector $Z_{k,1}$ satisfies $\sup_{\|u\|_{2}=1}\|Z_{k,1}^{\transpose}u\|_{\psi_{1}}\leq c_{0}$, as well as
		$\expect Z_{k,1}=0$ and $\expect(Z_{k,1}Z_{k,1}^{\transpose})={I}_{p}$.
	\end{enumerate}
\end{assumption}
In the above assumption, the mean vectors $\mu_k$ and covariance matrices $\Sigma_k$ are allowed to vary with the sample size $n_k$. Also, the random vectors $Z_{1,1},\ldots,Z_{1,n_1},\ldots,Z_{\ngroup,1},\ldots,Z_{\ngroup,n_\ngroup}$ across different populations are independent, and $Z_{1,1},\ldots,Z_{\ngroup,1}$ may have different distributions.

To state the next assumption, for $d\in\{1,\ldots,p\}$, we use $\mathcal{J}_{k}(d)$ to denote a set
of indices corresponding to the $d$ largest values among $\sigma_{k,1},\ldots,\sigma_{k,p}$.
In addition, let $R_{k}(d)\in\real^{d\times d}$ denote the correlation
matrix of the random variables $\{X_{k,1}(j):\, j\in\mathcal{J}_{k}(d)\}$.
Lastly, let $a\in(0,1/2)$ be a fixed constant, 
and define the integers $\ell_{k}$ and $m_{k}$ according to
\[
\ell_{k}=\lceil(1\vee\log^{3}n_{k})\wedge p\rceil,
\]
\[
m_{k}=\lceil(\ell_{k}\vee n_{k}^{\frac{1}{\log(n_{k})^{a}}})\wedge p\rceil.
\]

\begin{assumption}[Structural assumptions]
	\label{assu:2}~
	\begin{enumerate}[label=(\roman*)]
		\item The parameters $\sigma_{k,1},\ldots,\sigma_{k,p}$ are positive, and
		there are positive constants $\alpha$, $c_1$, and $c_{\circ}\in(0,1)$,
		\notdependon, such that for each $k\in \{1,\ldots,\ngroup\}$,
		\[
		\sigma_{k,(j)}\leq c_1 j^{-\alpha}\quad\text{ for all }j\in\{m_{k},\ldots,p\},
		\]
		\[
		\sigma_{k,(j)}\geq c_{\circ}j^{-\alpha}\quad\text{ for all }j\in\{1,\ldots,m_{k}\},
		\]
		where $\sigma_{k,(j)}$ denotes the $j$th largest value of $\sigma_{k,1},\ldots,\sigma_{k,p}$.
		\item There exists a constant $\epsilon_{0}\in(0,1)$, \notdependon, such that for $k=1,\ldots,\ngroup$,
		\[
		\max_{i\neq j}R_{k,i,j}(\ell_{k})\leq1-\epsilon_{0},
		\]
		where $R_{k,i,j}(\ell_k)$ denotes the $(i,j)$ entry of the matrix $R_k(\ell_k)$. 
		Also, for $k=1,\ldots,\ngroup$, the matrix $R_{k}^{+}(\ell_{k})$ with $(i,j)$
		entry given by $\max\{R_{k,i,j}(\ell_{k}),0\}$ is positive semi-definite.
		Moreover, there is a constant $C_0>0$, \notdependon, such that for each $k=1,\ldots,\ngroup$, we have
		\[
		\sum_{1\leq i<j\leq \ell_{k}}R_{k,i,j}^{+}(\ell_{k})\leq C_0\ell_{k}.
		\]
	\end{enumerate}
\end{assumption}

The above two assumptions are  multiple-sample analogs  of  assumptions in \cite{Lopes2020}, where examples of correlation matrices satisfying the above conditions are 
given. The following assumption imposes  constraints on $\tau$ in conjunction with $n$  and on the sample sizes $n_1,\ldots,n_\ngroup$. 
\begin{assumption}
	\label{assu:parameters} There exist positive constants $c_2$ and $c_3$ not depending on $n$ such that the bounds $c_2\leq\frac{n_k}{n_k+n_l}\leq c_3$ hold for all $k,l\in\{1,\ldots,\ngroup\}$. Also, the conditions $(1-\tau)\sqrt{\log \nmin}\gtrsim 1$ and $\max\{\ngroup,|\mathcal P|\}\asymplt e^{\sqrt{\log \nmin}}$ hold.
\end{assumption}

	In the last assumption, note that $\tau$ is allowed to approach to 1 at a slow rate. 
	Although $n_{1},\ldots,n_\ngroup$ are required to be of the same order, their ratios do not have to converge to certain limits. Such convergence conditions are required by some of  the test procedures   surveyed in Section \ref{sec:intro} that are based on asymptotic  limit distributions of test statistics rather than  bootstrap.  Also, it is notable that the current setting allows $\ngroup\to\infty$ and $|\mathcal{P}|\to\infty $ as $n\to\infty$.
{Overall, Assumptions \ref{assu:1}--\ref{assu:parameters}  are quite mild and are satisfied for many relevant applications, with examples in Sections \ref{sec:functional-anova} and \ref{sec:count-data}.}

 Let $\tilde{S}_{k}\sim N(0,\Sigma_{k})$ for each $k=1,\dots,K$, and define the Gaussian counterparts
of the partially standardized statistics $M(k,l)$ and $M$, 
\[
\tilde{M}(k,l)=\max_{1\leq j\leq p}\left(\sqrt{\ts\frac{n_l}{n_k+n_l}}\ts\frac{\tilde{S}_{k,j}}{\sigma_{k,l,j}^{\tau}}-\sqrt{\ts\frac{n_k}{n_k+n_l}}\ts\frac{\tilde{S}_{l,j}}{\sigma_{k,l,j}^{\tau}}\right)
\quad
\text{and} \quad \tilde M=\max_{(k,l)\in \mathcal{P}} \tilde M(k,l).\]
The following two theorems,  with proofs provided in the Supplement, extend the Gaussian and bootstrap approximation results in \cite{Lopes2020} to the multiple-sample setting as encountered in MANOVA, where $\dk$ denotes the Kolmogorov distance, defined by $\dk(\mathcal{L}(U),\mathcal{L}(V))=\sup_{t\in\real}|\P(U\leq t)-\P(V\leq t)|$ for generic random variables $U$ and $V$ with probability distributions $\mathcal{L}(U)$ and $\mathcal L(V)$. As discussed in the introduction, this extension from the one- to the multi-sample case is nontrivial. The key theoretical results are the following Theorems \ref{thm:gauss-appox} and \ref{thm:boot-appox}, which provide theoretical justifications for the proposed bootstrap procedure. In these theorems, the constant $\delta$ may be taken to be arbitrarily small, and so the {convergence rates are} nearly parametric.

\begin{thm}[Gaussian approximation]
	\label{thm:gauss-appox}Fix any small $\delta>0$, and suppose that
	Assumptions \ref{assu:1}--\ref{assu:parameters}
	hold. Then,
	\[
	\dk\left(\law(M),\law(\tilde{M})\right) \ \asymplt \  \nmin^{-\frac{1}{2}+\delta}.
	\]
\end{thm}

\begin{thm}[Bootstrap approximation]
	\label{thm:boot-appox}Fix any small $\delta>0$, and suppose that
	Assumptions \ref{assu:1}--\ref{assu:parameters}
	hold. Then there is a constant $c>0$, \notdependon,
	such that the event 
	\[
	\dk\left(\law(\tilde{M}),\law(M^{\star}|X)\right) \ \leq \ c\nmin^{-\frac{1}{2}+\delta}
	\]
	occurs with probability at least $1-c\nmin^{-1}$, where $\law(M^{\star}|X)$ represents the distribution of $M^\star$ conditional on the observed data.
\end{thm}




\subsection{High-dimensional MANOVA}\label{sec:manova}

We first analyze the power of the proposed method in Section \ref{sec:method}.
 All proofs are deferred to  the Supplement.

\begin{thm}\label{prop:quantile-bound}
	If Assumptions \ref{assu:1}--\ref{assu:parameters} hold and the number of bootstrap samples satisfies {$B\asympgt \log^2 n$}, then the following statements are true. 
	\begin{enumerate}[label={(\roman*)}]
		\item \label{enu:power-a} For any fixed $\varrho\in (0,1)$, we have $|\hat q_{M}(\varrho)|\leq c\log^{1/2}\nmin$
		with probability at least $1-c\nmin^{-1}$, where $c$ is a constant \notdependon. 
		\item \label{enu:power-b} For some constant $c>0$ \notdependon, we have
		\[
		\prob\left(\max_{(k,l)\in\mathcal{P}}\max_{1\leq j\leq p}\hat{\sigma}_{k,l,j}^{2}<2\sigma_{\max}^{2}\right)\geq1-c\nmin^{-1},
		\]
		where $\sigma_{\max}=\max\{\sigma_{k,j}:1\leq j\leq p,1\leq k\leq \ngroup\}$.
	\end{enumerate}
	  Consequently, if $\max_{(k,l)\in\mathcal{P}}\max_{1\leq j\leq p}|\mu_{k}(j)-\mu_{l}(j)|\geq c \sigma_{\max}\nmin^{-1/2} \log^{1/2}\nmin$ for a sufficiently large positive constant $c$ \notdependon, then for any choice of $\mathcal{P}$, the null hypothesis will be rejected with probability tending to one as $n\to\infty$.
\end{thm}

To analyze the size of the proposed test, we observe that when we construct the SCRs, we use $\hat\sigma_{k,l,j}$ instead of $\sigma_{k,l,j}$. This requires us to quantify the Kolmogorov distance between the distributions of $M$ and \begin{equation}
\hat{M}=\max_{(k,l)\in\mathcal{P}}\hat{M}(k,l),\label{eq:hatm}
\end{equation} where 
\begin{equation}
\hat{M}(k,l)=\max_{1\leq j\leq p}\left(\sqrt{\ts\frac{n_l}{n_k+n_l}}\ts\frac{S_{k,j}}{\hat{\sigma}_{k,l,j}^{\tau}}-\sqrt{\ts\frac{n_k}{n_k+n_l}}\ts \frac{S_{l,j}}{\hat{\sigma}_{k,l,j}^{\tau}}\right).\label{eq:hatm-g-h}
\end{equation}
Note that like $M$ defined in \eqref{eq:M}, the random variable $\hat M$ itself is not a test statistic. 
With $F_{k,j}$ denoting the cumulative distribution function of the standardized random variable $\{X_{k,1}(j)-\mu_{k}(j)\}/\sigma_{k,j}$, we  require the following mild condition on the distribution of the standardized observations.

\begin{assumption} 
	\label{assu:X}  There are positive constants $\nu$, $r_0$, and $c$ not depending on $n$ such that $\max_{1\leq k\leq K}\max_{1\leq j\leq p}\sup_{x\in\real}\sup_{r\in(0,r_0)}r^{-\nu}\Big(F_{k,j}(x+r)-F_{k,j}(x-r)\Big)\leq c$.
\end{assumption}

The above condition is essentially equivalent to common H\"older continuity of the distribution functions $F_{k,j}$, i.e., there is  a common H\"older constant $\nu$ that is fixed but could be arbitrarily small. The assumption is satisfied if each of the distributions $F_{k,j}$ has a density function $f_{k,j}$ such that $\max_{1\leq k\leq K}\max_{1\leq j\leq p}\|f_{k,j}\|_{\infty}\lesssim 1$, where $\|\cdot\|_{\infty}$ is the supremum norm. 
However, the condition  is much weaker than this, as it may hold even when the distributions do not have densities, or the densities are unbounded.


\begin{thm}
	\label{thm:from-hat-to-truth-2}Fix any small $\delta>0$, and suppose that
	Assumptions \ref{assu:1}--\ref{assu:X}
	hold. Then, 
	\[
	\dk(\law(\hat{M}),\law(M))\asymplt\nmin^{-\frac{1}{2}+\delta}.
	\]
\end{thm}
With the triangle inequality, the above theorem together with Theorem \ref{thm:gauss-appox} and \ref{thm:boot-appox} implies that, with probability at least $1-c\nmin^{-1}$, we have $\dk(\law(\hat{M}),\law(M^\star\mid X))\leq c\nmin^{-\frac{1}{2}+\delta}$, for some constant $c>0$ \notdependon. This allows us to quantify the convergence rate of the size of the test, as follows.  Let $\size(\varrho)$ be the probability that $\mathbf{H}_0$ is rejected at the level $\varrho$ when it is true. {When  $B\asympgt {n}$, the Dvoretzky--Kiefer--Wolfowitz--Massart inequality \citep{Dvoretzky1956,Massart1990} implies that the empirical distribution of $B$ independent samples of $M^\star$ uniformly converges to the distribution of $M^\star$ at the rate $n^{-1/2+\delta}$ with probability at least $1-cn^{-1}$.}  The following result  is then a direct consequence of Theorems \ref{thm:gauss-appox}--\ref{thm:from-hat-to-truth-2} and it  asserts that the size of the test is asymptotically correctly controlled at the rate $\nmin^{-1/2+\delta}$.
\begin{thm}\label{thm:size}Fix any small $\delta>0$, and fix any $\varrho\in (0,1)$. If
	Assumptions \ref{assu:1}--\ref{assu:X} hold, {with $B\asympgt {n}$}, then 
		$$|\size(\varrho)-\varrho| \ \lesssim \ \nmin^{-1/2+\delta}.$$
\end{thm}

We note that in Theorems \ref{thm:from-hat-to-truth-2} and \ref{thm:size}, Assumption \ref{assu:X} can be replaced with the condition {$\nmin^{-1/2}\log^{3}p\ll1$} which then imposes an upper bound on the growth rate of $p$ relative to $\nmin$. In conjunction with the consistency of the general test as in Theorem~\ref{prop:quantile-bound}, Theorem~\ref{thm:size} provides strong justification for the application of the proposed test for a large class of null hypotheses that are typically all of interest 
 in MANOVA in addition to the main global null hypothesis that all means are equal.

\section{Application to functional ANOVA}\label{sec:functional-anova}

Consider a separable Hilbert space   $\hilbert$  and
a second-order random element $Y$ with mean element $\mu\in\hilbert$,
i.e., $\expect\|Y\|_{\hilbert}^{2}<\infty$, where $\|\cdot\|_{\hilbert}$ denotes the norm of the Hilbert space. In our context, the random
element $Y$ represents an observed functional data atom drawn from a population of functional data. 
Commonly considered Hilbert spaces in the area of functional data
analysis include reproducing kernel Hilbert spaces and the space 
$L^2(\tdomain)$ of squared integrable functions defined on a domain $\tdomain$. In one-way functional ANOVA, one aims to test the hypothesis
\begin{equation}
\mathbf H_0:\,\mu_{1}=\cdots=\mu_{\ngroup},\label{eq:null}
\end{equation}
given $K$ independent groups of  i.i.d.~elements $Y_{k,1},\ldots,Y_{k,n_{k}}\in\hilbert$ with common mean element $\mu_k\in\hilbert$, with $k=1,\dots,K$.

Given an orthonormal basis $\phi_{1},\phi_{2},\ldots$ of $\hilbert$,  each $\mu_{k}$ may be represented  in terms of this basis, i.e., $\mu_{k}=\sum_{j=1}^{\infty}u_{kj}\phi_{j}$,
where $u_{k,j}$ are generalized Fourier coefficients. Then the null  hypothesis
(\ref{eq:null}) is equivalent to the statement that $u_{k,j}=u_{l,j}$
for all $j\geq1$ and all $1\leq k<l\leq \ngroup$. This suggests that in empirical situations we choose a large integer $p\geq 1$ and test whether the vectors $u_k\equiv(u_{k,1},\ldots,u_{k,p})$ are equal for $k=1,\ldots,\ngroup$, which is precisely the hypothesis testing problem introduced in Section \ref{sec:method}. \lin{This idea of transforming a functional ANOVA problem into a MANOVA problem has been proposed by \cite{Gorecki2015} with a classic standard MANOVA method. Here we modify this idea with the proposed MANOVA method to exploit the inherited decay in variances for functional data. We first observe that}
 each $Y_k$ admits the Karhunen--Lo\`eve expansion $Y_k=\mu_k+\sum_{j=1}^\infty \xi_{k,j}\varphi_j$, where $\varphi_1,\varphi_2,\ldots$ are orthonormal elements of $\mathcal H$, and $\xi_{kj}$ are uncorrelated random variables such that $\mathbb E\xi_{kj}=0$ and $\sum_{j=1}^\infty\mathrm{var}(\xi_{kj})<\infty$. This {implies}  that $\mathrm{var}(\xi_{kj})$ decays to zero at a rate faster than $j^{-1}$. Consequently, Proposition 2.1 of \cite{Lopes2020} asserts that the variance of the (random) generalized Fourier coefficient of $Y_k$ with respect to the basis element $\phi_j$ also decays, which allows us to adopt the test proposed in Section \ref{sec:method}. 
 
\subsection{Simulation studies}
We assess the above method in terms of its finite sample performance by  numerical simulations and compare it with three popular methods in the literature, namely, the $L^2$  based method (L2) \citep{Faraway1997,Zhang2007}, the $F$-statistic based method (F) \citep{Shen2004,Zhang2011} and the global pointwise $F$ test (GPF) \citep{Zhang2014}. These were briefly  reviewed in the introduction and numerical implementations are available from \cite{Gorecki2019}, see also 
\cite{Gorecki2015}. We also compare it \lin{with the random projection  based method (RP)  \citep{Cuesta-Albertos2010}, the global envelope test (GET) \citep{Mrkvicka2020}} and  a method (MPF) recently developed by \cite{Zhang2019} that takes the maximum of the pointwise $F$-statistics as a test statistic and also leverages bootstrapping to approximate the critical value of the test.

\begin{figure}[t]
	\begin{minipage}{0.24\textwidth}
		\includegraphics[scale=0.38]{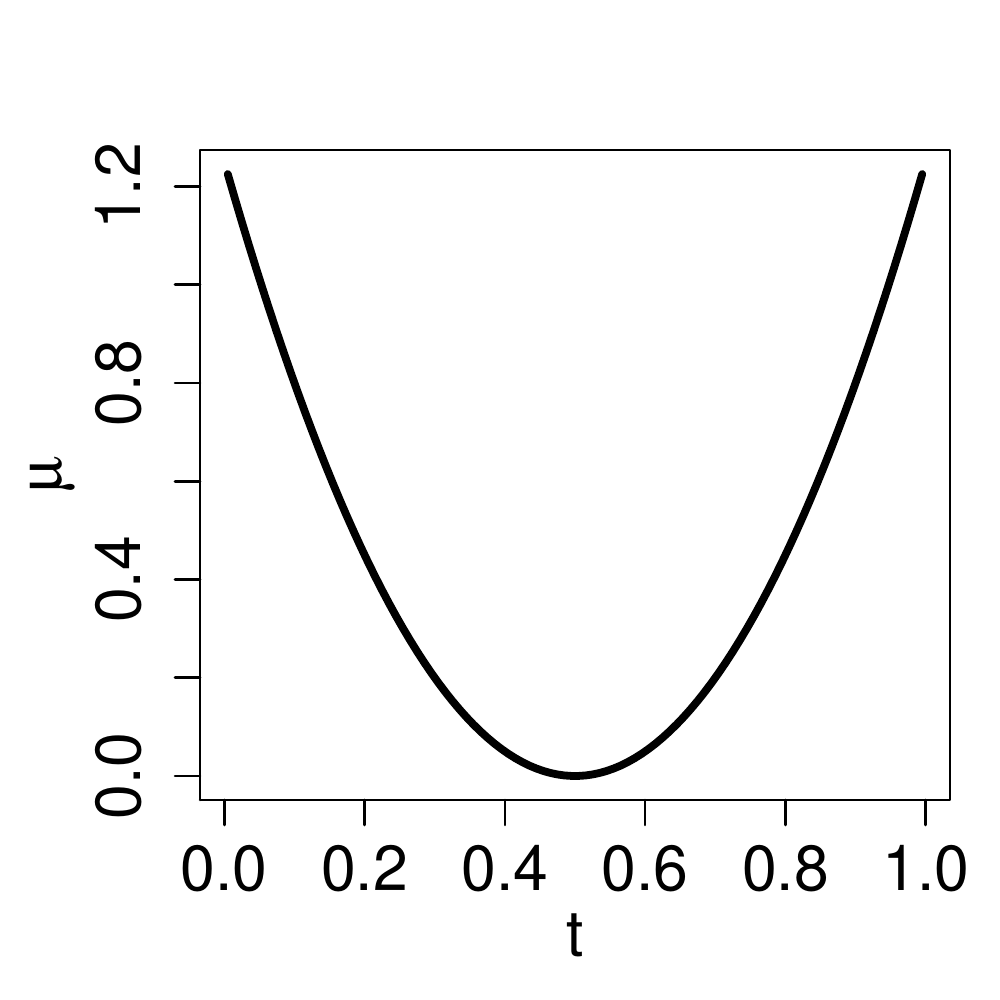}
	\end{minipage}
	\begin{minipage}{0.24\textwidth}
		\includegraphics[scale=0.38]{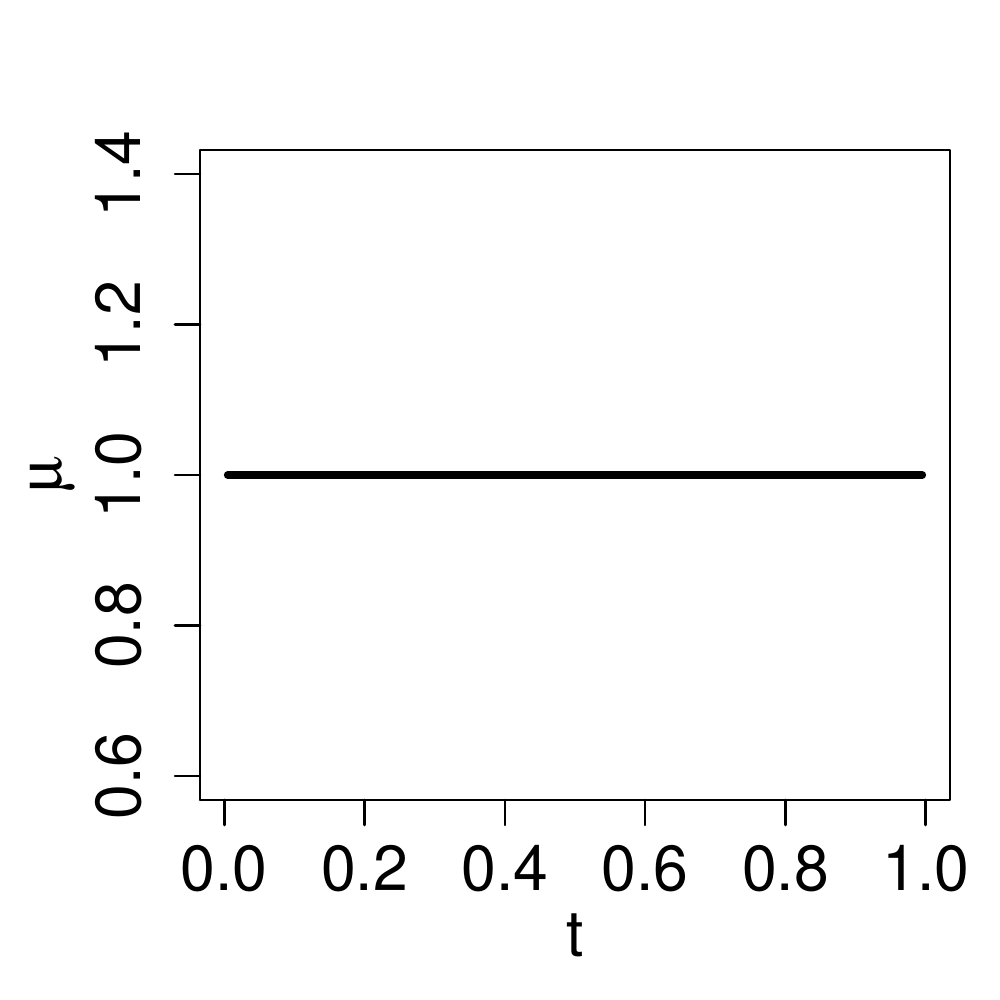}
	\end{minipage}
	\begin{minipage}{0.24\textwidth}
		\includegraphics[scale=0.38]{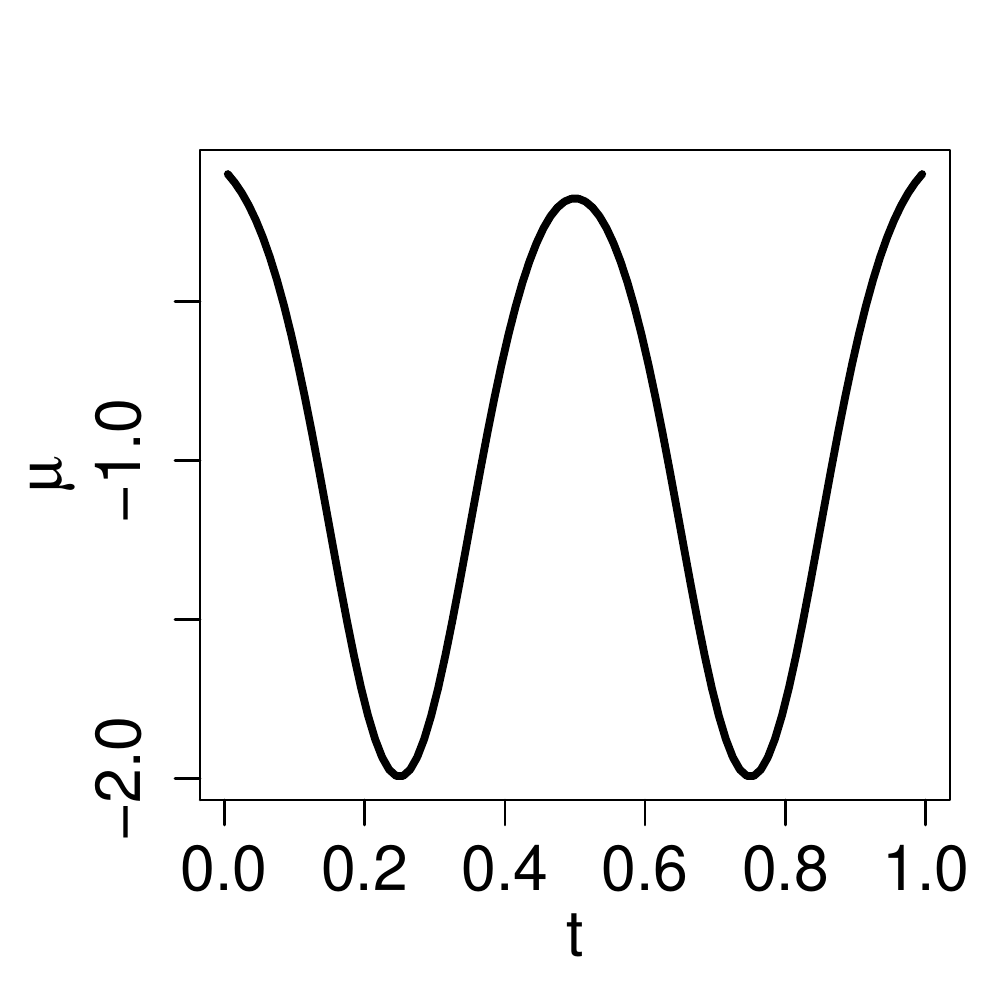}
	\end{minipage}
	\begin{minipage}{0.24\textwidth}
		\includegraphics[scale=0.38]{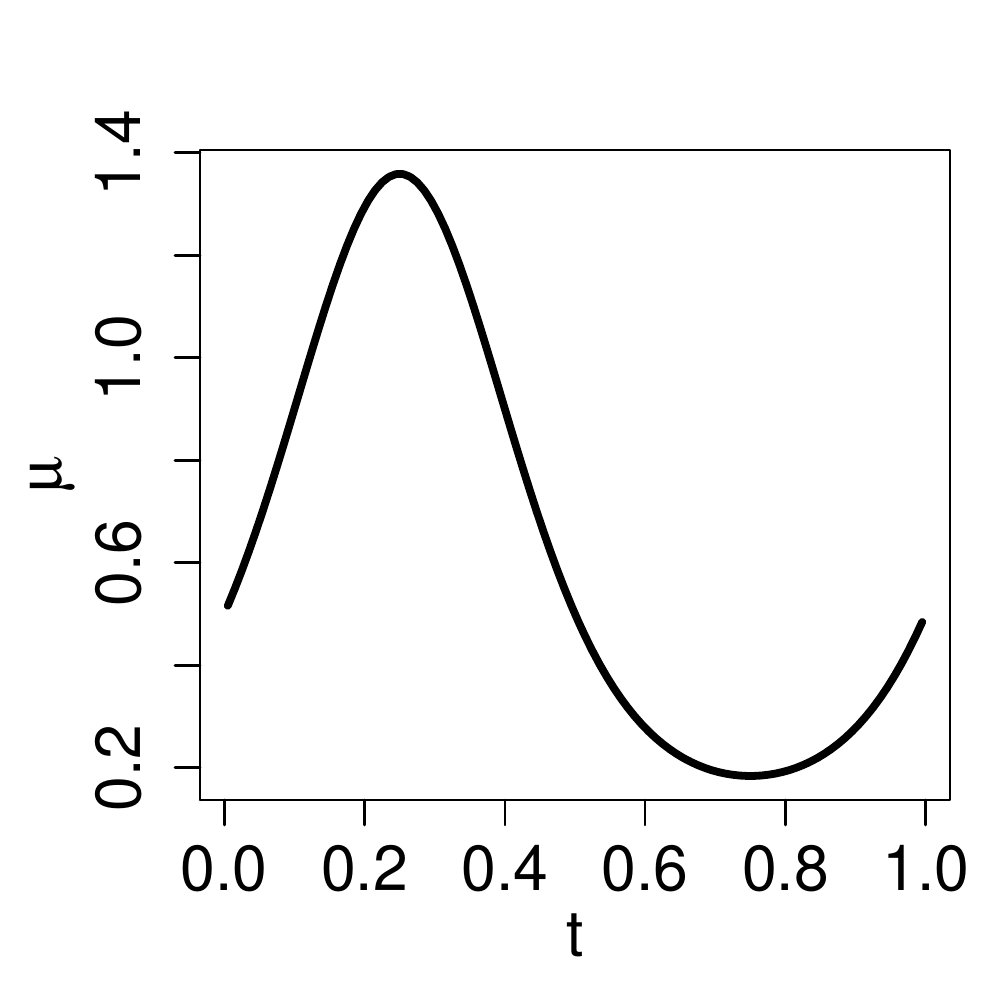}
	\end{minipage}

	\begin{minipage}{0.24\textwidth}
		\includegraphics[scale=0.38]{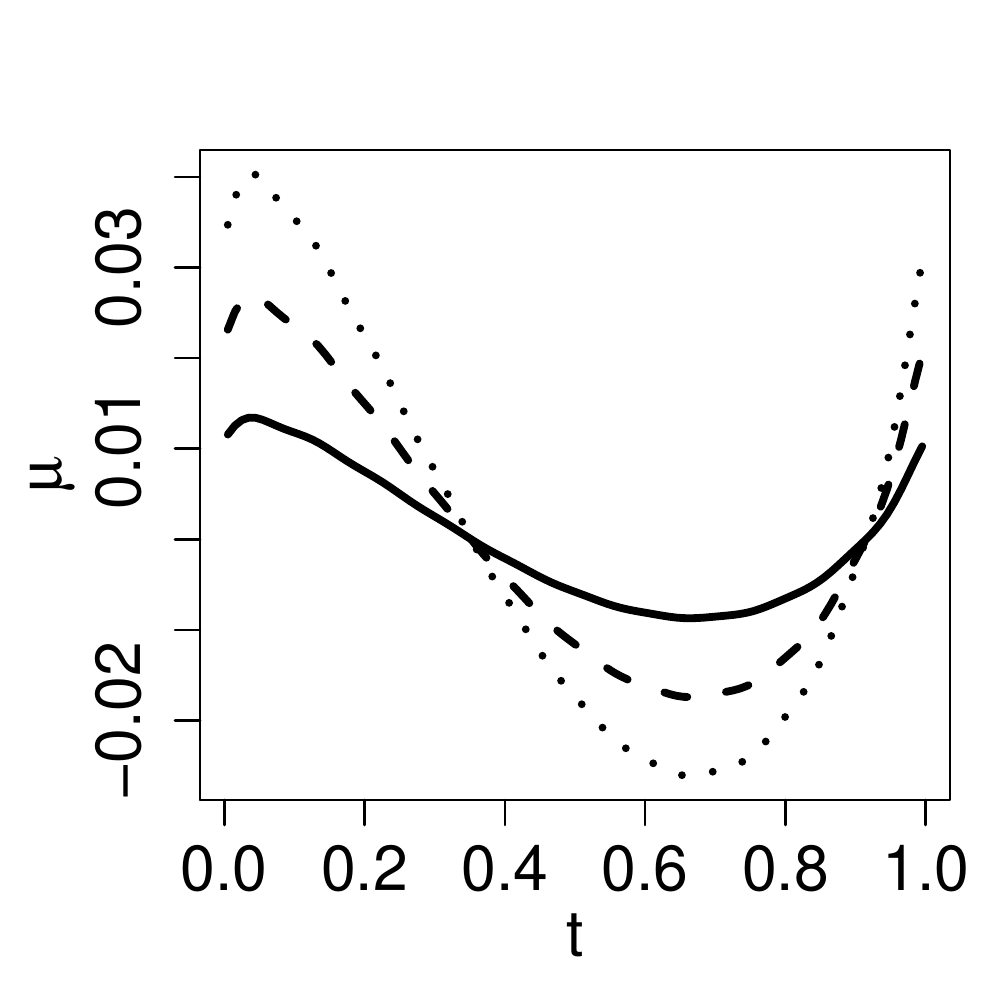}
	\end{minipage}
	\begin{minipage}{0.24\textwidth}
		\includegraphics[scale=0.38]{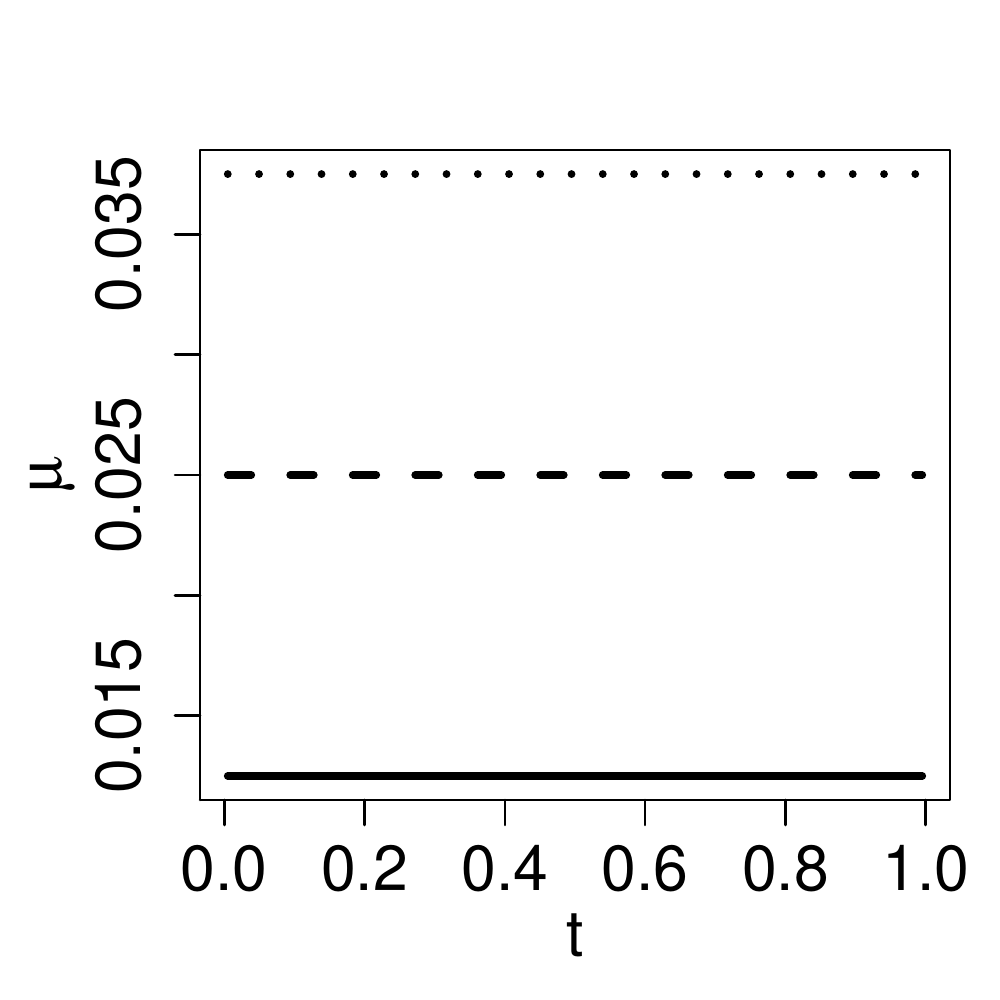}
	\end{minipage}
	\begin{minipage}{0.24\textwidth}
		\includegraphics[scale=0.38]{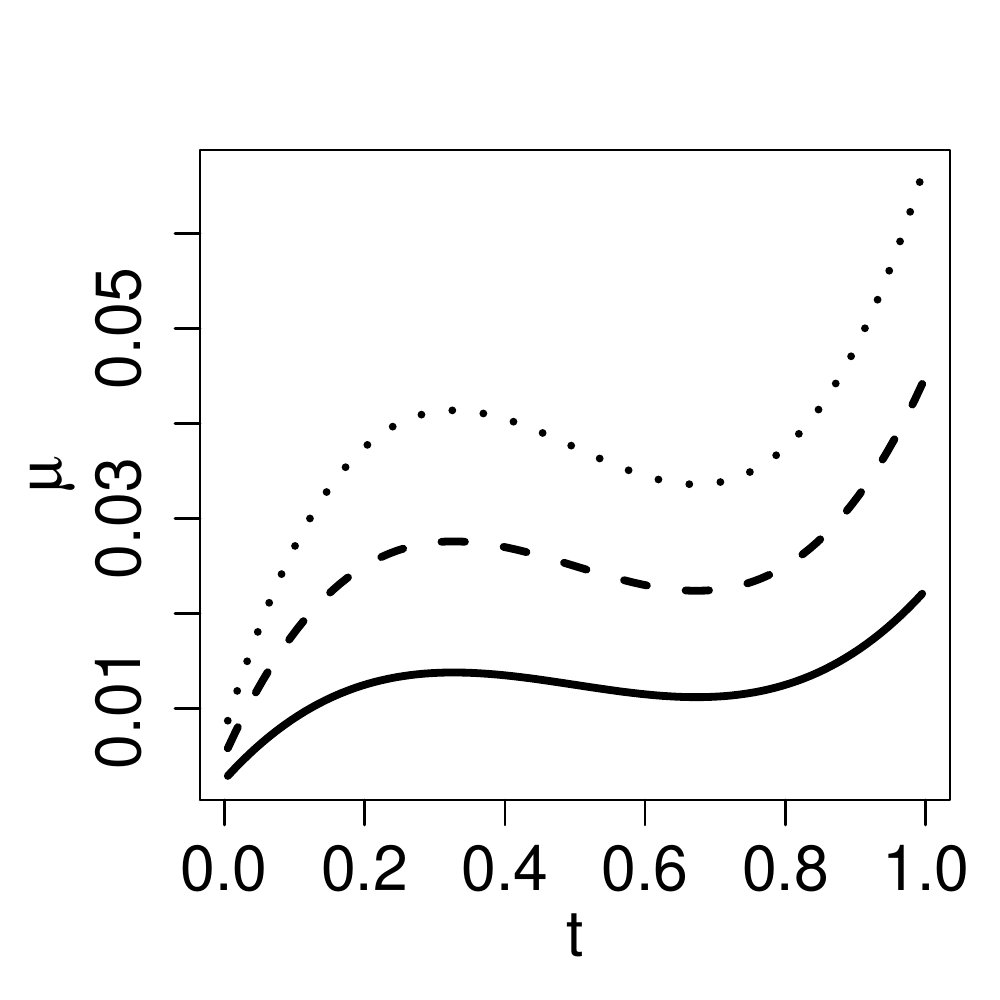}
	\end{minipage}
	\begin{minipage}{0.24\textwidth}
		\includegraphics[scale=0.38]{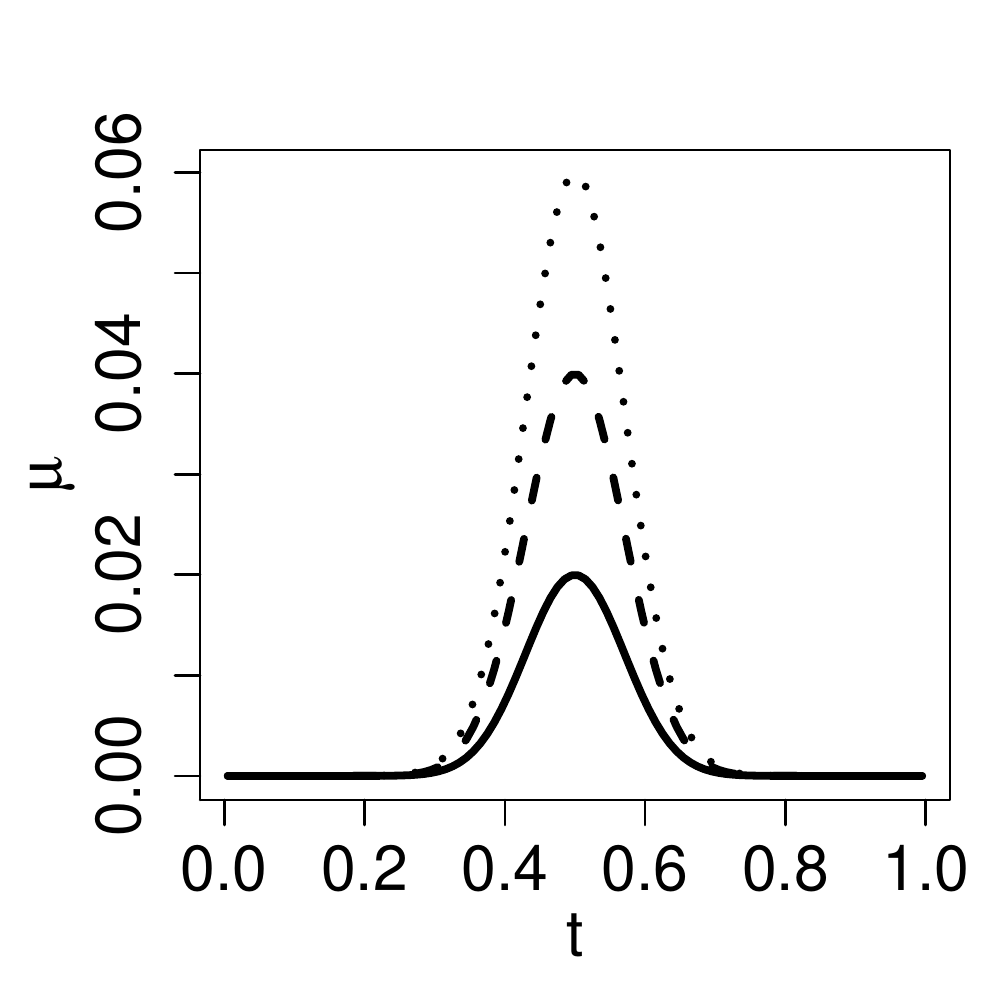}
	\end{minipage}
	\caption{Mean functions. The first row shows the functions $\mu_0$ employed for families (M1)--(M4), respectively, and the second row displays the functions $\mu_1-\mu_0$ (solid), $\mu_2-\mu_0$ (dashed) and $\mu_3-\mu_0$ (dotted) with $\theta=0.5$ in the families (M1)--(M4), respectively, from left to right.  \label{fig:mean}}
\end{figure}

In the simulation study, we set $\hilbert=L^2([0,1])$, and consider four families of mean functions, parameterized by $\theta\in[0,1]$, as follows, 
\begin{description}
\item [{(M1)}] $\mu_{k}(t)=\mu_0(t)+\theta k\sum_{j=1}^{10}j^{-2}\{\sin(2j\pi t) + \cos(2j\pi t)\}/50$ with $\mu_0(t)=5(t-1/2)^2$,
\item [{(M2)}] $\mu_{k}(t)=\mu_0(t)+\theta k/40$ with $\mu_0(t)\equiv 1$,
\item [{(M3)}] $\mu_{k}(t)=\mu_0(t)+\theta k\{1+(10t-2)(10t-5)(10t-8)\}/40$ with $\mu_0(t)=-(f_{1/4,1/10}(t)+f_{3/4,1/10}(t))$,
\item [{(M4)}] $\mu_{k}(t)=\mu_0(t)+\theta k \exp\{-(t-1/2)^2/100\}/25$ with $\mu_0(t)=\exp\{\sin(2\pi t)\}/2$,
\end{description}
for $k=1,2,3$, where $f_{a,b}$ denotes the probability density function of the normal distribution with mean $a$ and variance $b^2$. Obviously $\mu_1,\mu_2,\mu_3$ are identical and equal to $\mu_0$ when $\theta=0$, and differ from each other when $\theta\neq 0$. These families are shown in Figure \ref{fig:mean}. Mean function families (M1) and (M2) represent ``sparse alternatives'' in the frequency domain in the sense that the Fourier coefficients of the mean functions differ most in the first few leading terms under the alternative when $\theta \neq 0$, while the function family (M3) represents  a ``dense alternative'' in the frequency domain. When $\theta\neq 0$,  the families (M1)--(M3) are ``dense'' in the time domain. 
In particular, the alternatives in (M2) are uniformly dense in the time domain, in the sense that the differences of the mean functions between the groups are nonzero and uniform in $t\in\tdomain=[0,1]$.  Thus, families (M1)--(M3) favor the integral-based methods such as the L2, F and GPF tests, as these methods integrate certain statistics over the time domain. In contrast, the alternatives in the last family (M4) are ``sparse'' in the time domain. 

We sample functional data of the form $\mu_k(\cdot)+W_k(\cdot)$, for certain choices of centered random processes $W_k(\cdot)$ in two different settings. In the first ``common covariance'' setting,  the random processes of all groups are Gaussian with the following common  Mat\'ern  covariance function
\begin{equation}\label{eq:matern-cov}
\mathcal{C}(s,t)=\ts\frac{\sigma^2}{16}\frac{2^{1-\nu}}{\Gamma(\nu)}\left(\ts\frac{\sqrt{2\nu}|s-t|}{\eta}\right)^{\nu}B_{\nu}\left(\frac{\sqrt{2\nu}|s-t|}{\eta}\right),
\end{equation}
where $\Gamma$ is the gamma function, $B_{\nu}$ is the modified
Bessel function of the second kind, $\sigma^2$ is set to $2.5$, $\eta$ is set to $1$, and $\nu$ is set to $1/2$. In the  ``group-specific covariance'' setting,  the groups have different covariance functions, as follows. For the first group, the random process is the Gaussian process with the Mat\'ern covariance function \eqref{eq:matern-cov}. For the second group, the process is the Wiener process with dispersion $\sigma=0.1$, i.e., the Gaussian process with the  covariance function $\mathcal C(s,t)=\sigma^2 \min(s,t)$. For the third group, we set $W_3(\cdot)=\sum_{j=1}^{51} \xi_j\phi_j(\cdot)/20$, where $\phi_1(t)\equiv 1$, $\phi_{2j}=\sin(2j\pi t)$ and $\phi_{2j+1}=\cos(2j\pi t)$, and $\xi_j$ follows a uniform distribution on $[-j^{-2}\sqrt{3},j^{-2}\sqrt{3}]$, providing a non-Gaussian case. All sampled functions are observed at $m=100$ equally spaced points on the interval $[0,1]$. Using larger values of $m$ does not have much effect on the performance; this is in agreement  with the findings in \cite{Zhang2019}.

\begin{figure}[t]
	\begin{minipage}{0.24\textwidth}
		\includegraphics[scale=0.38]{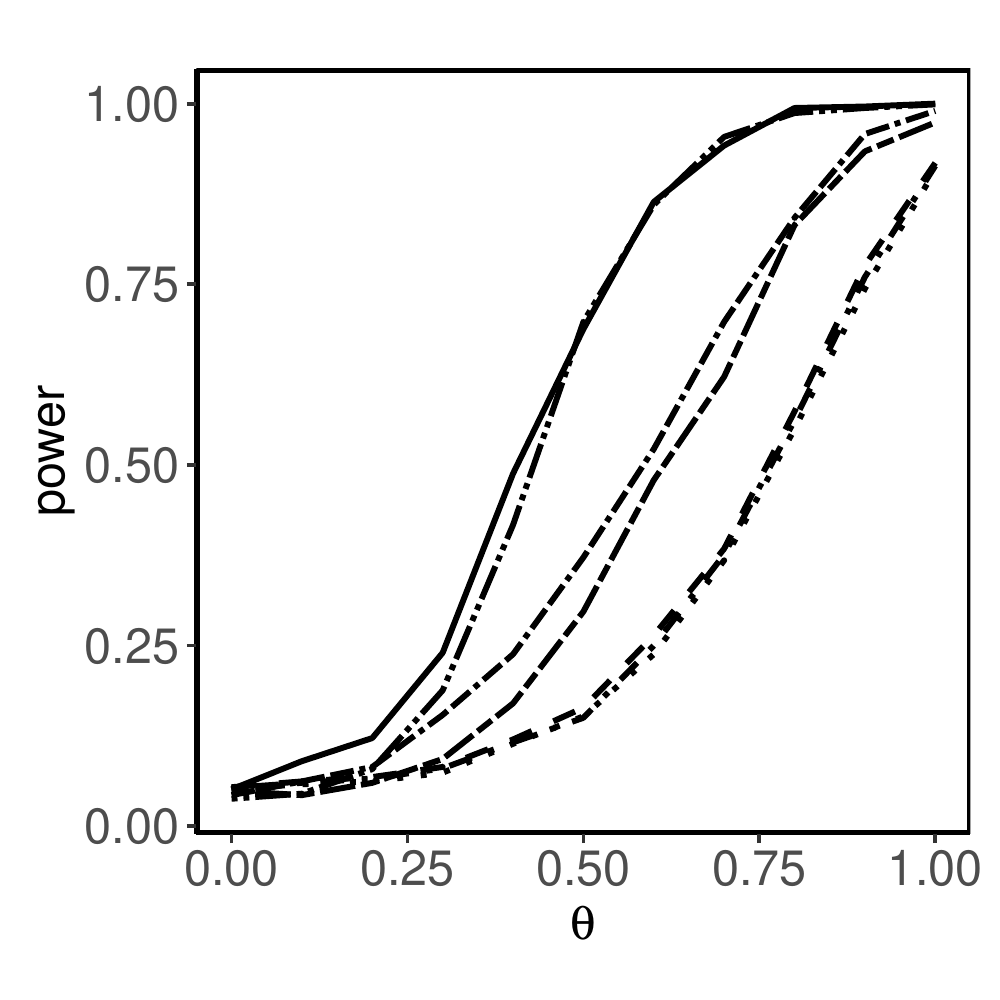}
	\end{minipage}
	\begin{minipage}{0.24\textwidth}
		\includegraphics[scale=0.38]{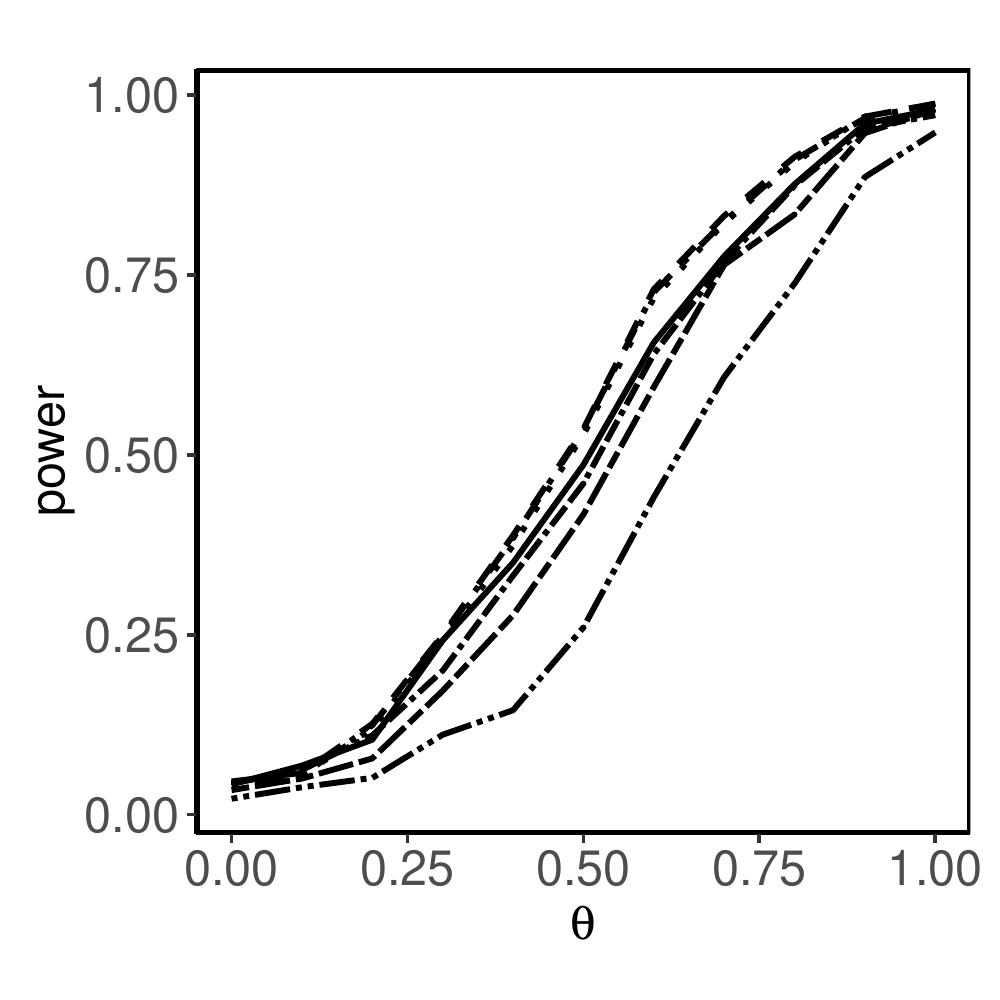}
	\end{minipage}
	\begin{minipage}{0.24\textwidth}
		\includegraphics[scale=0.38]{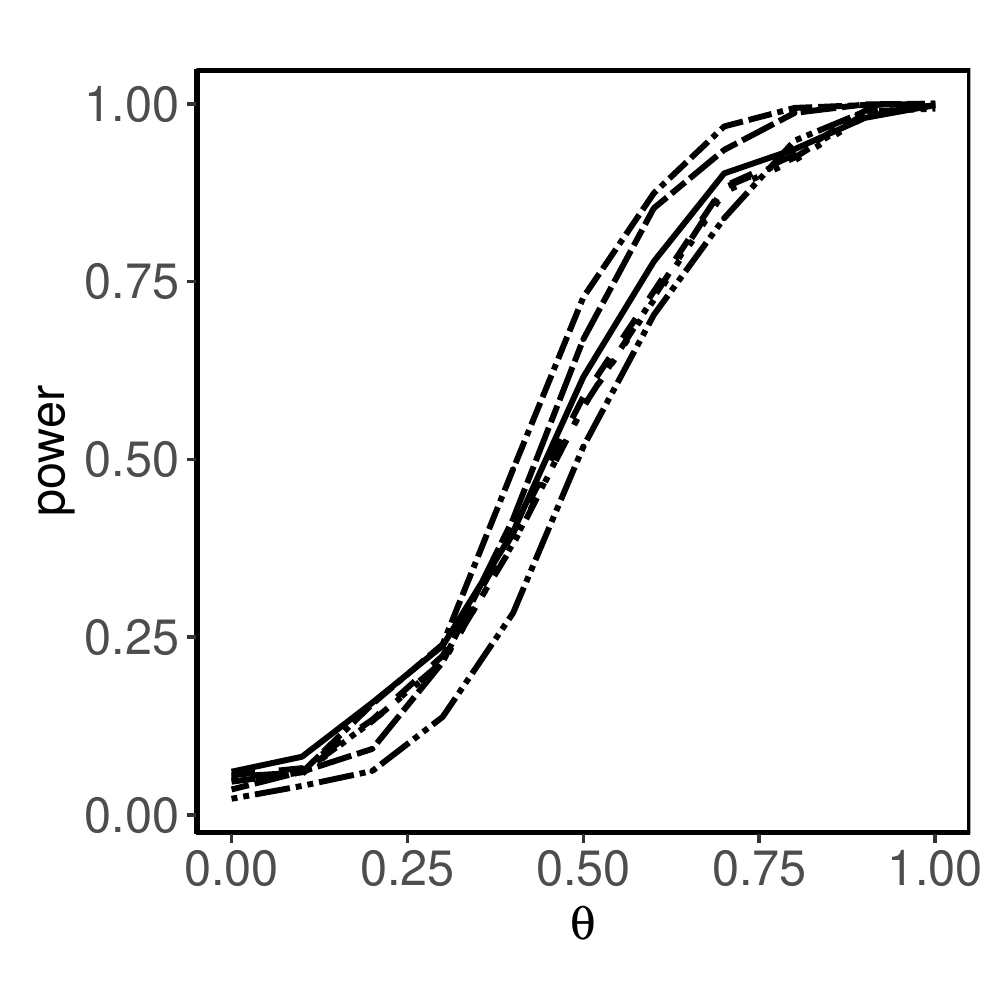}
	\end{minipage}
	\begin{minipage}{0.24\textwidth}
		\includegraphics[scale=0.38]{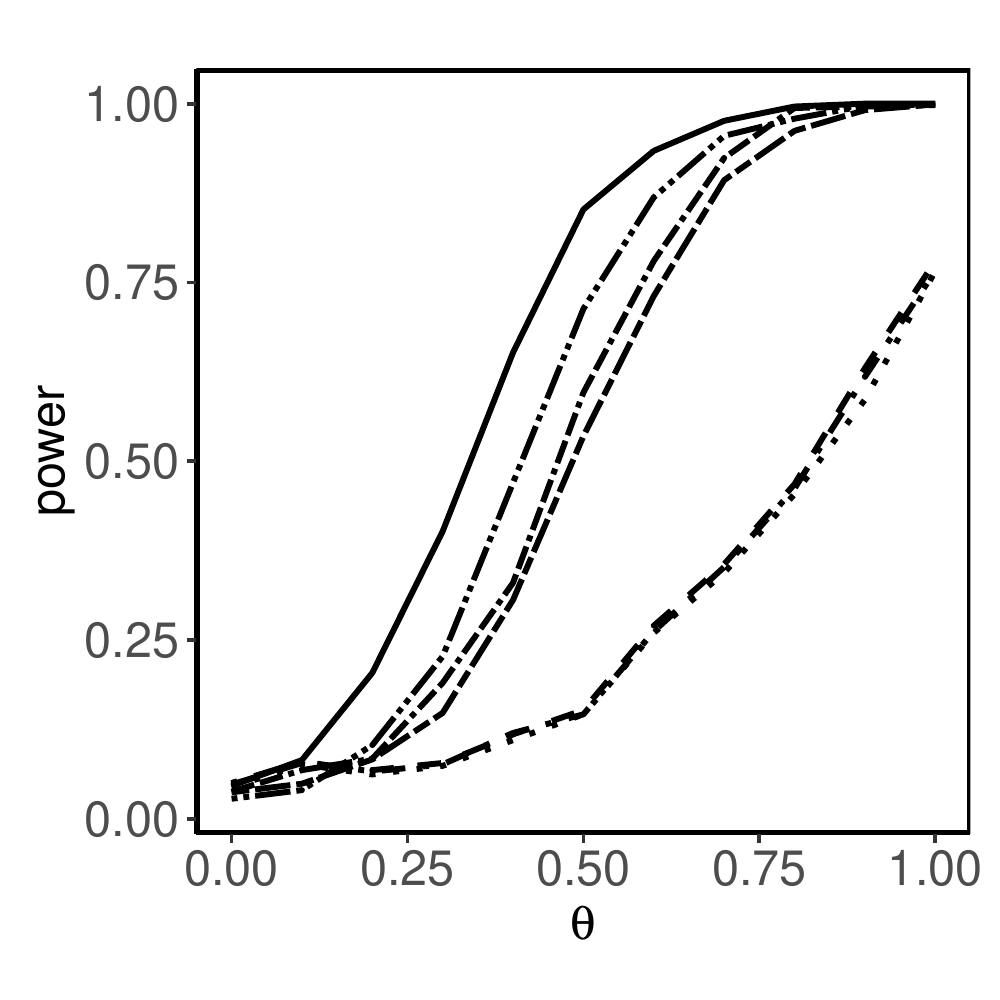}
	\end{minipage}
\\
	\begin{minipage}{0.24\textwidth}
		\includegraphics[scale=0.38]{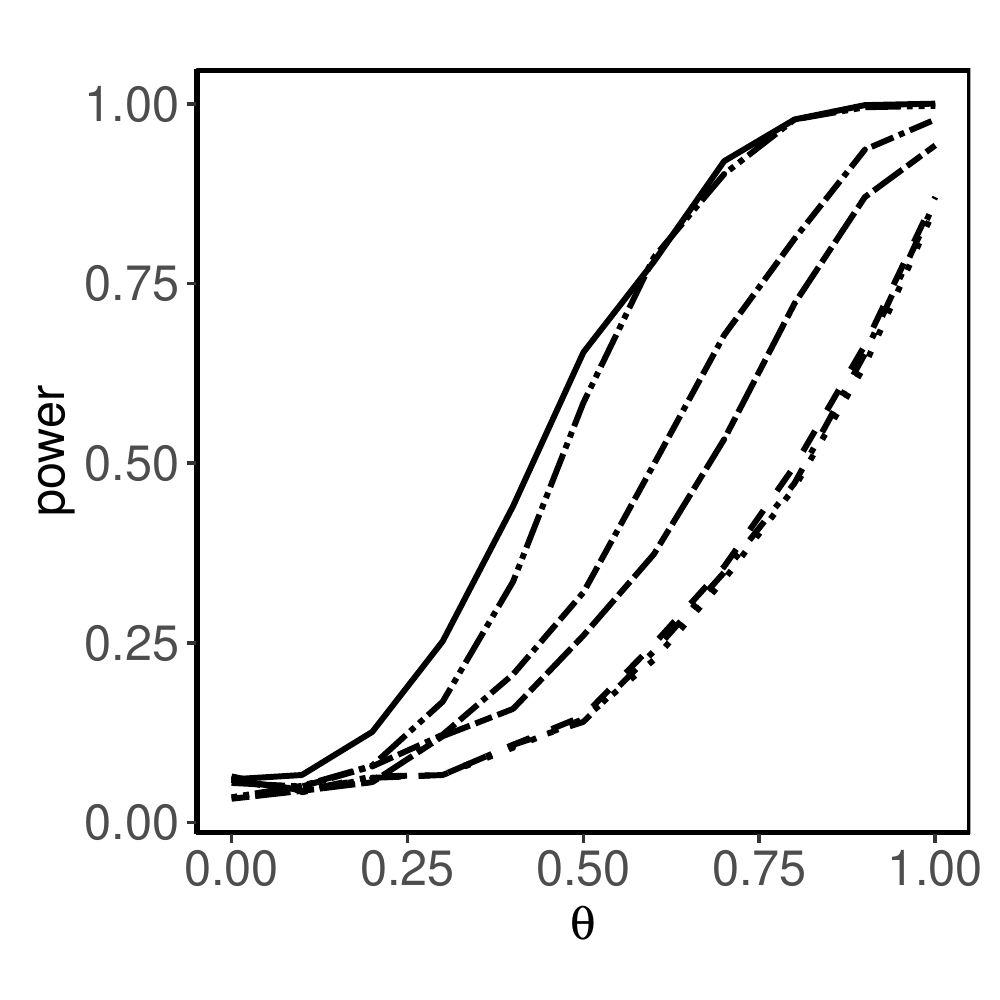}
	\end{minipage}
	\begin{minipage}{0.24\textwidth}
		\includegraphics[scale=0.38]{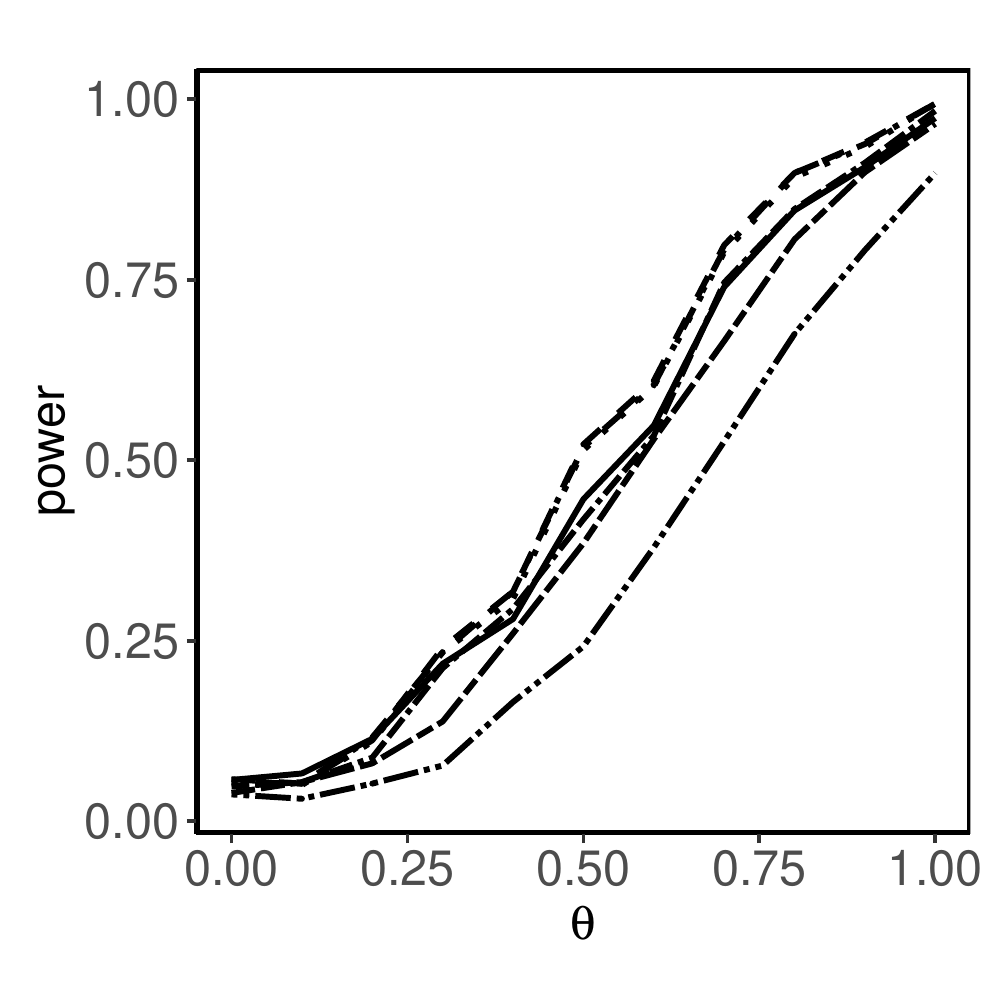}
	\end{minipage}
	\begin{minipage}{0.24\textwidth}
		\includegraphics[scale=0.38]{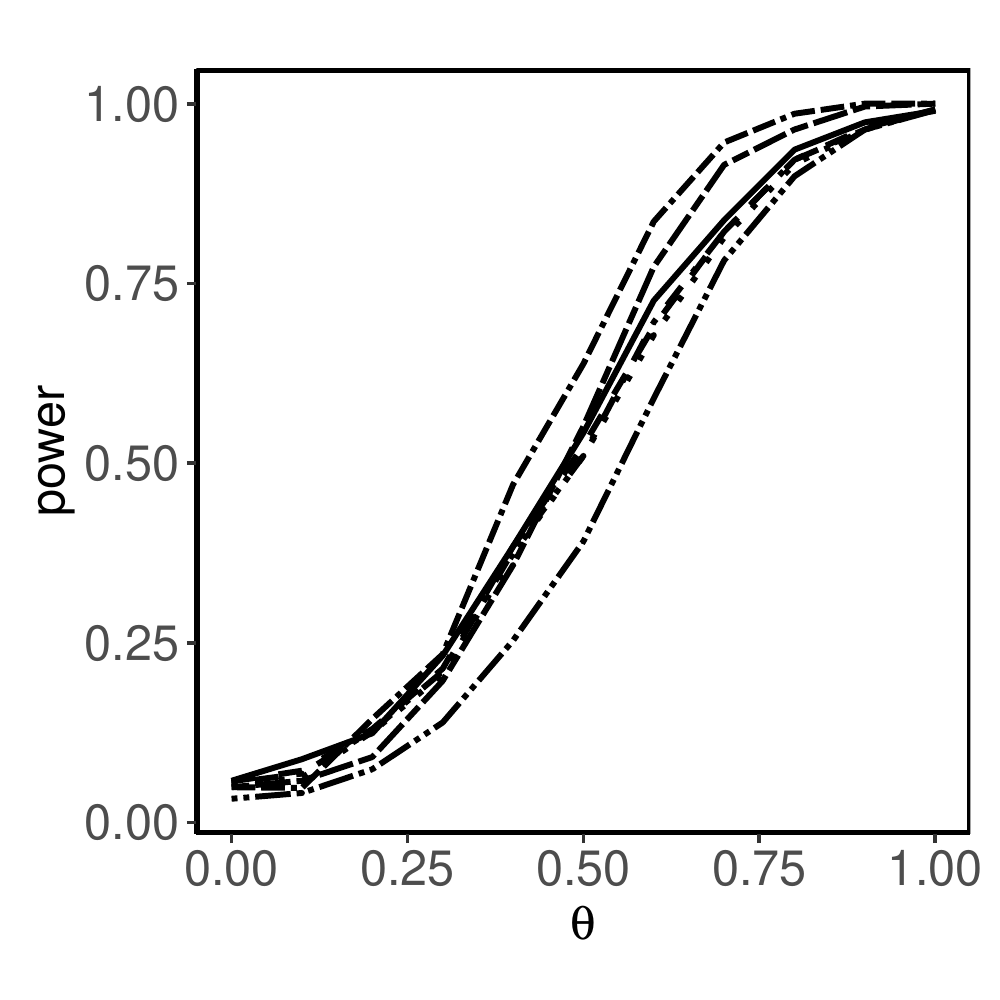}
	\end{minipage}
	\begin{minipage}{0.24\textwidth}
		\includegraphics[scale=0.38]{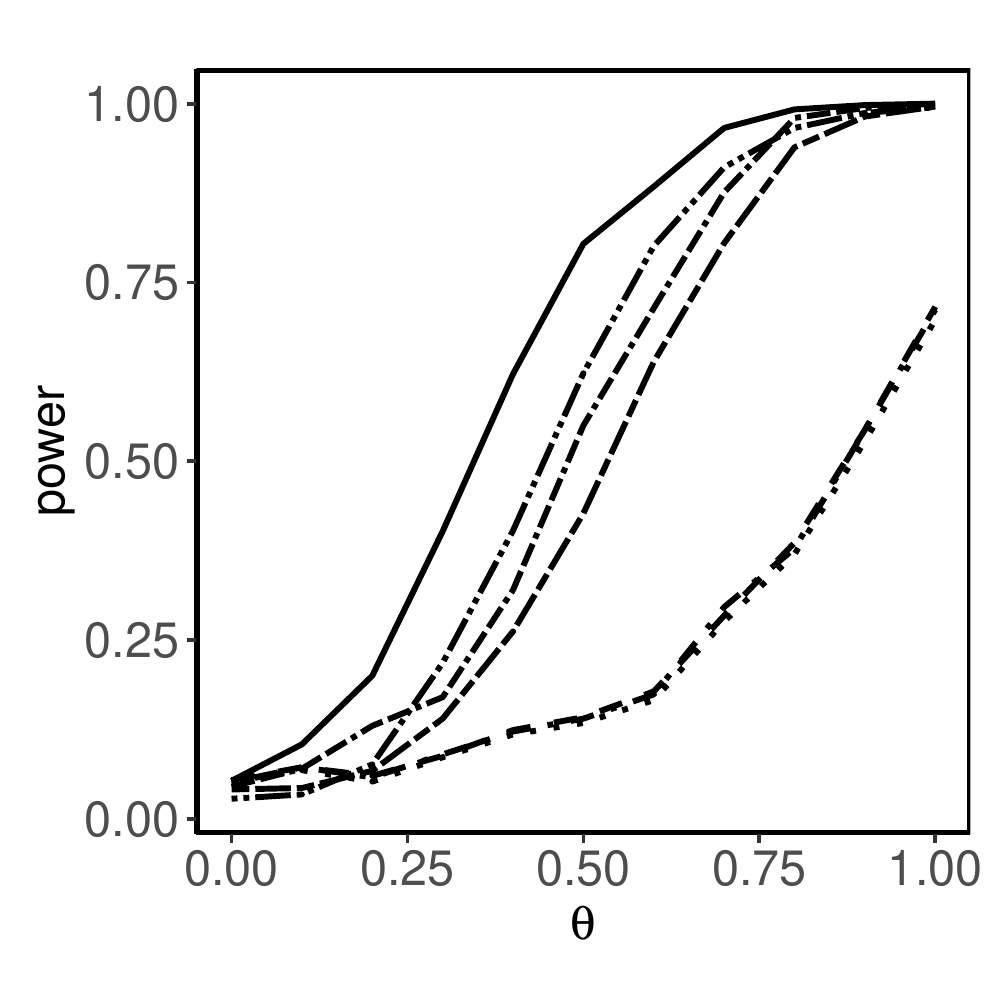}
	\end{minipage}
	\caption{Empirical power of the proposed functional ANOVA (solid), L2 (dashed), F (dotted), GPF (dot-dashed), {MPF (dot-dash-dashed), GET (short-long-dashed) and RP (dot-dot-dashed)} in the ``common covariance'' setting. Top: from left to right the panels display the  empirical power functions for families (M1), (M2), (M3) and (M4), when $n_1=n_2=n_3=50$. Bottom: from left to right the panels display the  empirical power functions for families (M1), (M2), (M3) and (M4) for unbalanced designs  when $n_1=30, n_2=50$ and $n_3=70$. The power functions  of L2, F and GPF  are nearly indistinguishable.\label{fig:power-fda-common}}
\end{figure}

 We set the significance level at $\varrho=0.05$, consider balanced sampling with  $n_{1}=n_{2}=n_{3}=50$ and also unbalanced sampling with $(n_1,n_2,n_3)=(30,50,70)$,  and use the aforementioned basis $\phi_1(t),\ldots,\phi_p(t)$ with $p=51$. The parameter $\tau$ is selected by the method described in Section \ref{sec:method} from 11 candidate values, namely, $0,0.1,\ldots,0.9,0.99$. Each simulation setup is replicated 1000 
times independently. The results for the size of the global test  are summarized in Table \ref{tab:size-fda}, showing that the proposed method and most of the other methods have an empirical size that is reasonably close to the nominal level. 
The performance in terms of power is depicted in Figure \ref{fig:power-fda-common} for the scenario with common covariance structure. The average of the selected values for $\tau$ is $0.713\pm0.155$ and $0.754\pm 0.172$ for the scenarios with common covariance structure and group-specific covariance structure, respectively.

\begin{figure}
	\begin{minipage}{0.24\textwidth}
		\includegraphics[scale=0.38]{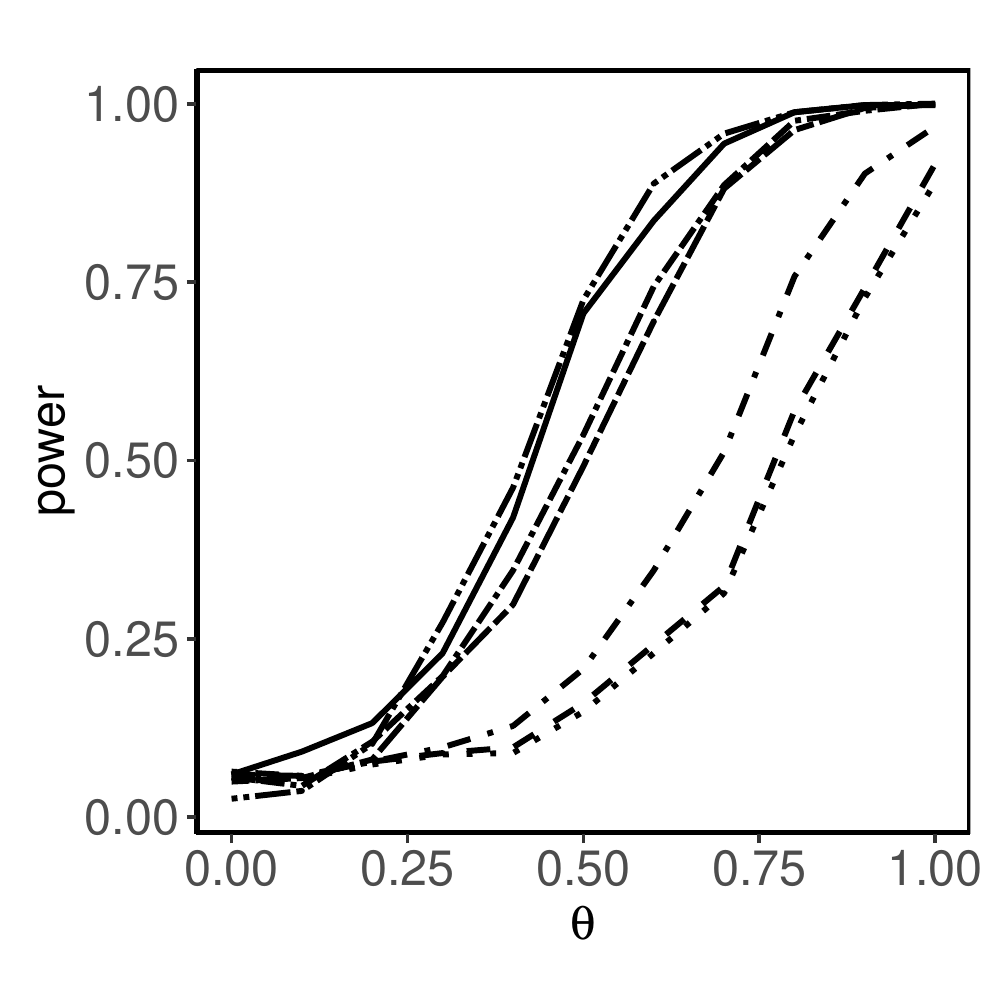}
	\end{minipage}
	\begin{minipage}{0.24\textwidth}
		\includegraphics[scale=0.38]{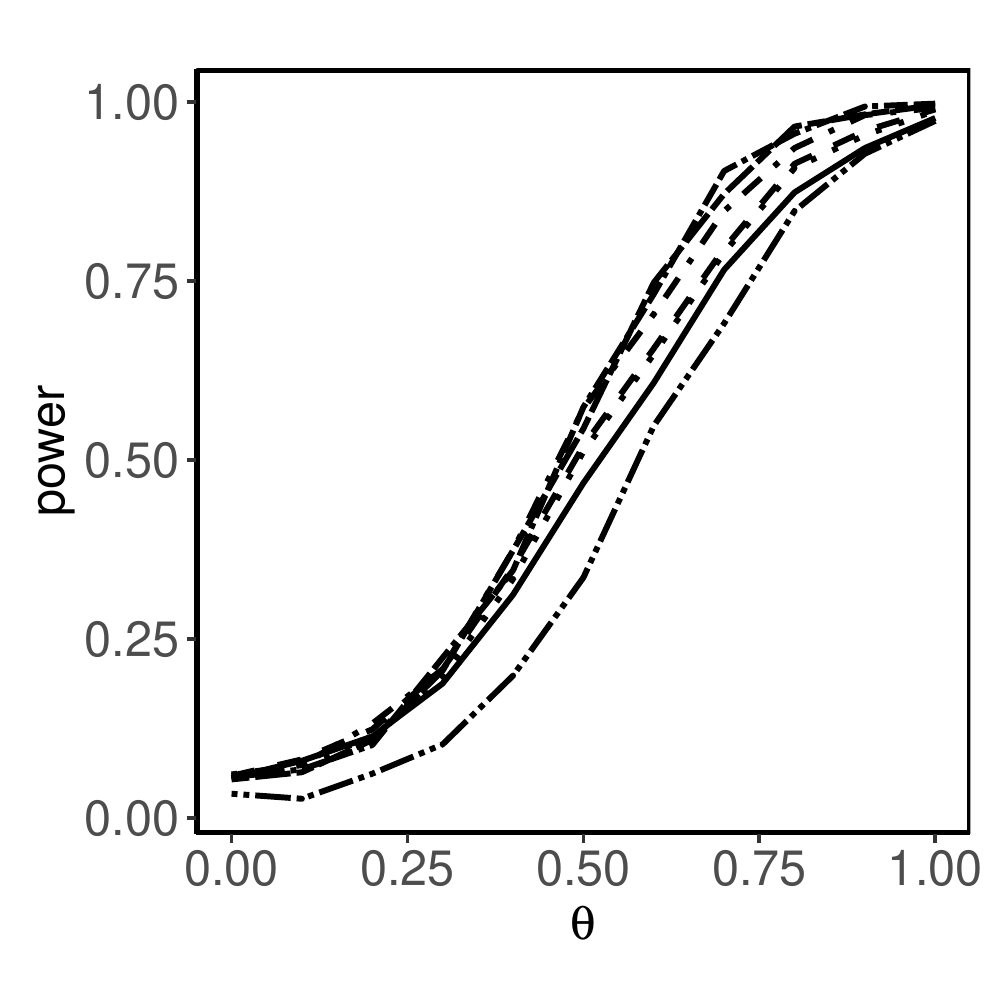}
	\end{minipage}
	\begin{minipage}{0.24\textwidth}
		\includegraphics[scale=0.38]{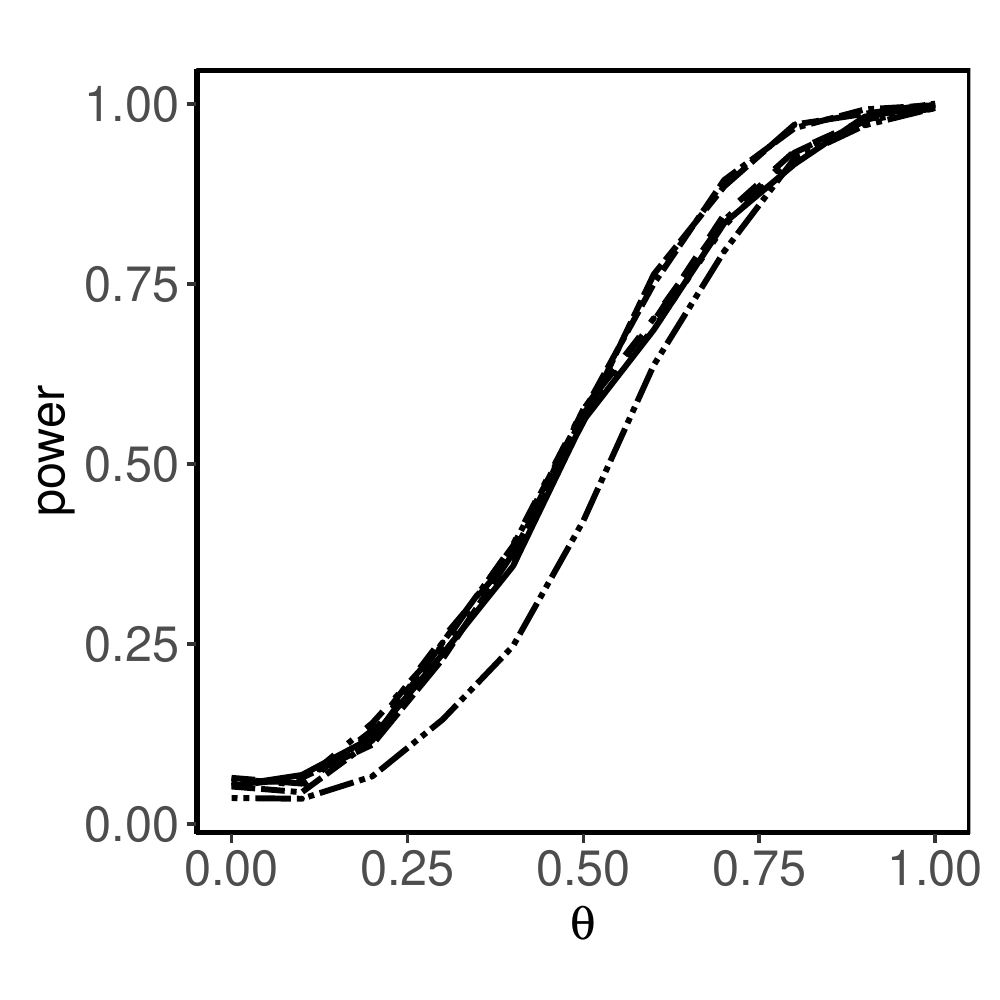}
	\end{minipage}
	\begin{minipage}{0.24\textwidth}
		\includegraphics[scale=0.38]{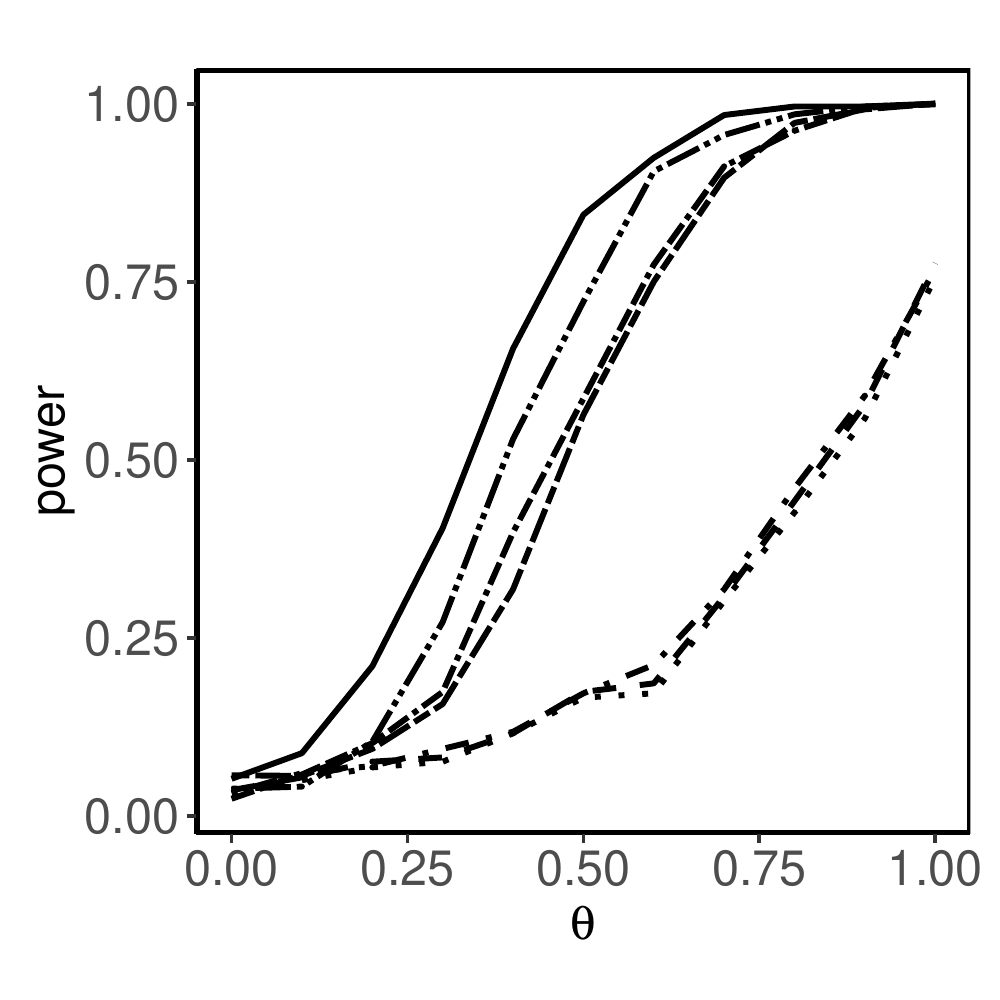}
	\end{minipage}
\\
	\begin{minipage}{0.24\textwidth}
		\includegraphics[scale=0.38]{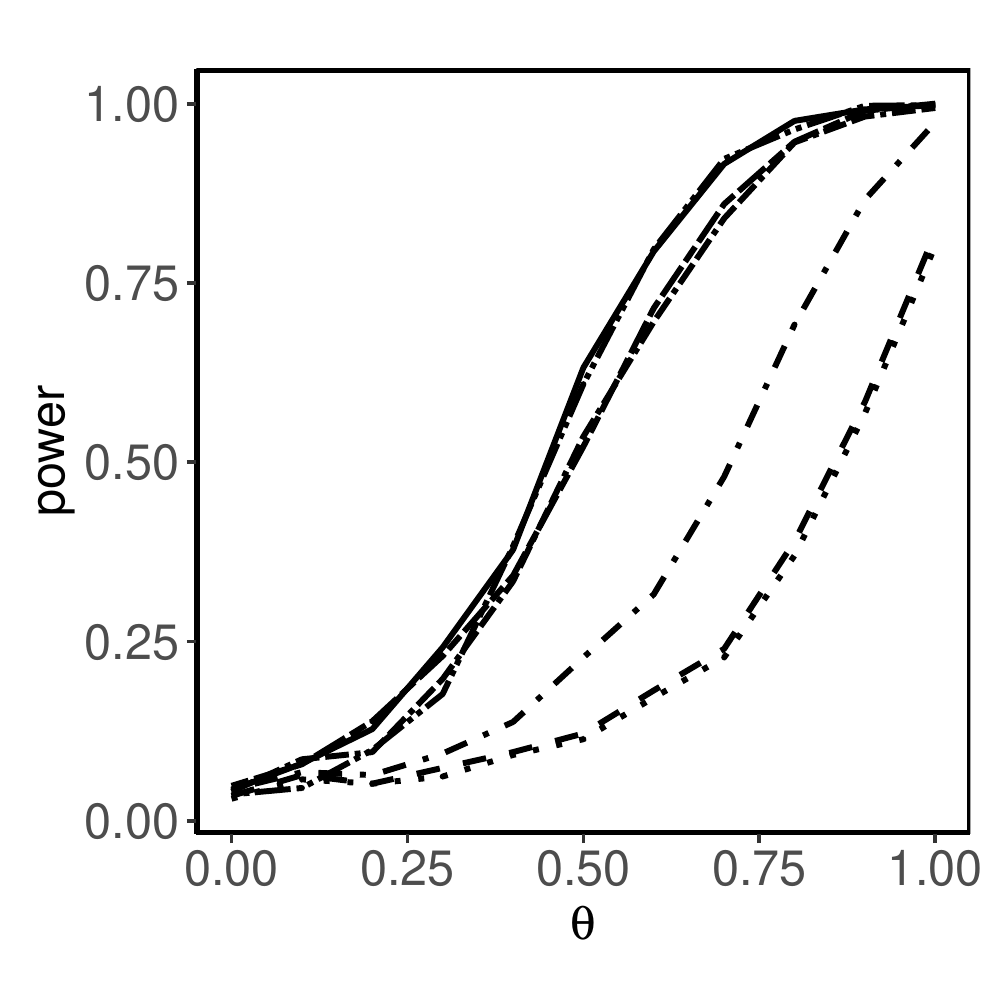}
	\end{minipage}
	\begin{minipage}{0.24\textwidth}
		\includegraphics[scale=0.38]{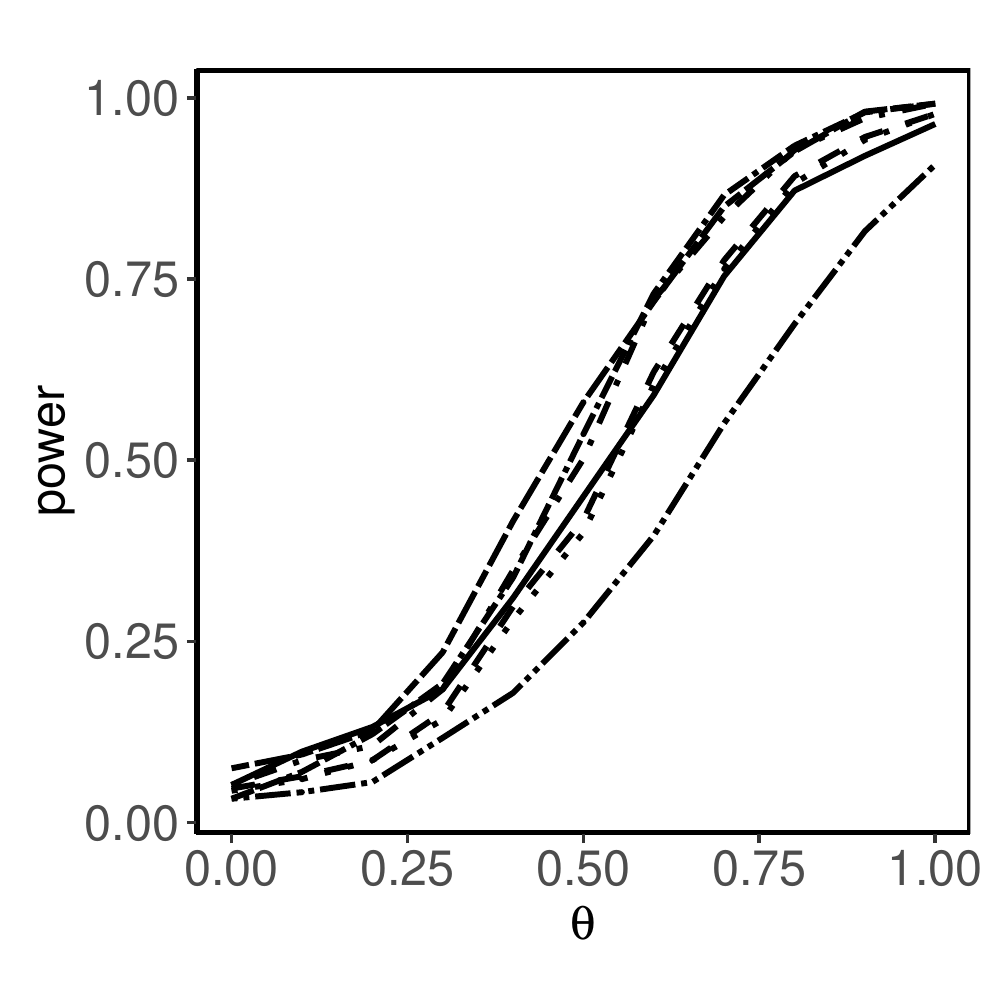}
	\end{minipage}
	\begin{minipage}{0.24\textwidth}
		\includegraphics[scale=0.38]{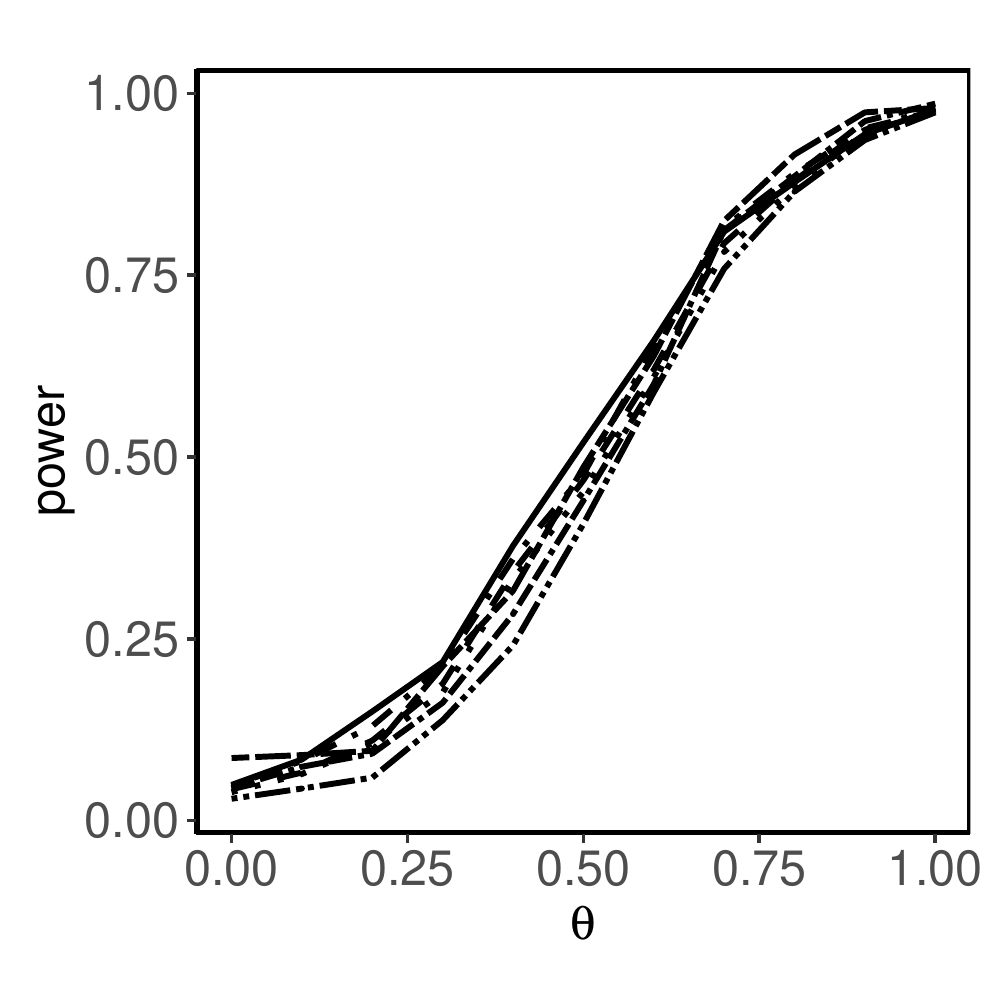}
	\end{minipage}
	\begin{minipage}{0.24\textwidth}
		\includegraphics[scale=0.38]{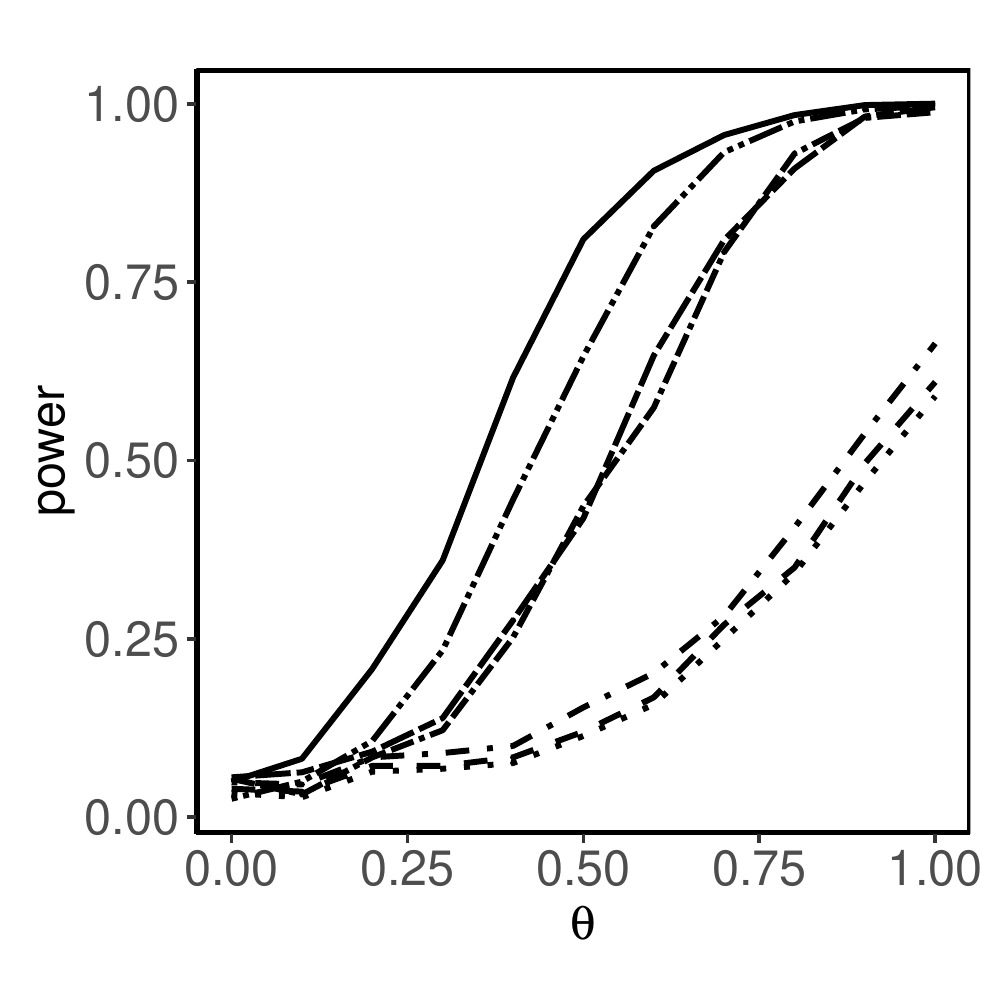}
	\end{minipage}
	\caption{Same as Figure~\ref{fig:power-fda-common} but for the case of covariance functions that differ between groups.} 
	\label{fig:power-fda-diff}
\end{figure}

When the alternatives are sparse in the frequency domain but not uniformly dense in the time domain (as in (M1)), or when the alternatives are sparse in the time domain (as in (M4)), the proposed method clearly outperforms most existing methods in terms of power  by a large margin. \lin{ The only exception is the RP method, which has similar power in the case of (M1). For the family (M2), all methods have nearly indistinguishable power, except for the RP method, which has substantially lower power.} For the family (M3), the power of MPF is slightly larger in relation to the other methods. Similar observations emerge for the scenario of group-specific covariance functions with results shown in Figure \ref{fig:power-fda-diff}, except that the power of GPF and MPF is slightly larger when the family is (M2), where the alternatives are uniformly dense in the time domain. In the group-specific context, the power of MPF is closer to the power of the proposed method  for (M1), while the power of all methods {except the RP method} is nearly indistinguishable for (M3). In conclusion, the proposed test is powerful against both dense and sparse alternatives in either time or frequency domain, and provides strong improvements over existing methods in the important case where  the alternative is sparse in the time domain or in the frequency domain (but not uniformly dense in the time domain).

The average computation time to complete a single Monte Carlo simulation replicate in seconds, {including selecting the parameter $\tau$ by the proposed data-driven procedure in Section \ref{sec:method}} is presented in Table \ref{tab:ct-fda}. \lin{It shows that  a single simulation replicate can be completed within 5 seconds without GPU acceleration and within only 0.1 seconds when  utilizing an NVIDIA Quadro P400 graphics card}. Following the suggestion of a reviewer, we also investigated the impact of within-function correlation on the power by using the simulation models from \cite{Zhang2019} and found that the proposed method is preferred when the within-function correlation is strong; see Section \ref{sec:additional-simulation-fda} of the Supplement for details, {where we also examined the effectiveness of the data-driven selection procedure for $\tau$ proposed in Section \ref{sec:method}.}

\begin{table}[t]
	\caption{Empirical size of functional ANOVA\label{tab:size-fda}}
	\begin{center}
		\renewcommand*{\arraystretch}{1.2}
		\begin{tabular}{|c|c|c|c|c|c|c|c|c|c|}
			\hline
			Covariance & M & $(n_1,n_2,n_3)$ & proposed & L2 & F & GPF & MPF & GET & RP \tabularnewline
			\hline
			\multirow{8}{*}{common} & \multirow{2}{*}{M1} & 50,50,50 & .051 & .054 & .052 & .053 & .043 & .049 & .038 \tabularnewline
			\cline{3-10}
			\cline{3-10}
			& & 30,50,70 & .053 & .056 & .057 & .056 & .055  & .033 & .035 \tabularnewline
			\cline{3-10}
			\cline{2-10}
			& \multirow{2}{*}{M2} & 50,50,50 & .042 & .046 & .041 & .044 & .043 & .034 & .022\tabularnewline
			\cline{3-10}
			\cline{3-10}
			& & 30,50,70 & .057 & .058 & .052 & .054 & .039 & .048 & .037\tabularnewline
			\cline{3-10}
			\cline{2-10}
			& \multirow{2}{*}{M3} & 50,50,50 & .057 & .056 & .050 & .054 & .047  & .036 & .023 \tabularnewline
			\cline{3-10}
			\cline{3-10}
			& & 30,50,70 & .056 & .057 & .053 & .055 & .049  & .049 & .033 \tabularnewline
			\cline{3-10}
			\cline{2-10}
			& \multirow{2}{*}{M4} & 50,50,50 & .046 & .048 & .044 & .050 & .038 & .037 & .028 \tabularnewline
			\cline{3-10}
			\cline{3-10}
			& & 30,50,70 & .053 & .054 & .052 & .051 & .045 & .041 & .028 \tabularnewline
			\cline{3-10}
			\hline
			\multirow{8}{*}{group-specific} & \multirow{2}{*}{M1} & 50,50,50 & .055 & .055 & .052 & .058 & .056 & .050 & .026  \tabularnewline
			\cline{3-10}
			\cline{3-10}
			& & 30,50,70 & .043 & .035 & .031 & .044 & .041 & .049 & .037  \tabularnewline
			\cline{3-10}
			\cline{2-10}
			& \multirow{2}{*}{M2} & 50,50,50 & .056 & .059 & .056 & .061 & .057 & .054 & .034 \tabularnewline
			\cline{3-10}
			\cline{3-10}
			& & 30,50,70 & .052 & .047 & .044 & .052 & .039 & .055 & .033 \tabularnewline
			\cline{3-10}
			\cline{2-10}
			& \multirow{2}{*}{M3} & 50,50,50 & .051 & .054 & .053 & .055 & .052 & .053 & .036 \tabularnewline
			\cline{3-10}
			\cline{3-10}
			& & 30,50,70 & .049 & .043 & .039 & .048 & .045  & .066 & .030 \tabularnewline
			\cline{3-10}
			\cline{2-10}
			& \multirow{2}{*}{M4} & 50,50,50 & .052 & .041 & .040 & .042 & .044 & .057 & .038 \tabularnewline
			\cline{3-10}
			\cline{3-10}
			& & 30,50,70 & .050 & .040 & .039 & .049 & .054 & .056 & .026 \tabularnewline
			\cline{3-10}
			\hline
		\end{tabular}
	\end{center}
\end{table}

\begin{table}[t]
	\caption{Computation times for functional ANOVA\label{tab:ct-fda} (in seconds)}
	\vspace{-0.1in}
	\begin{center}
		\renewcommand*{\arraystretch}{1.2}
			\begin{tabular}{|c|c|c|c|c|c|c|c|c|}
				\hline
				proposed (no GPU) & proposed (GPU) & L2 & F & GPF & MPF & GET & RP \tabularnewline
				\hline
				4.792 & .085 & .002 & .002 & .005 & 3.602 & 1.629 & .877 \tabularnewline
				\hline
		\end{tabular}
	\end{center}
\end{table}

\subsection{Data application}\label{sec:app:fda}

We apply the proposed  method to analyze the functional  data described in \cite{Carey2008} concerning egg-laying trajectories for  Mexican fruit flies ({\it Anastrepha ludens}) under various diets, with further perspective and background provided in \cite{Carey1998,Carey2002}. In this study, newly merged  female flies  were placed in individual glass cages  and during their entire lifespan were fed different diets. The number of eggs laid by each individual fly on each day was recorded and the resulting trajectories of daily egg-laying were then viewed as functional data. Since flies started egg-laying only around day 10 after emergence and to avoid selection effects due to individually varying age-at-death, we considered the trajectories on a domain $[10,50]$ days
and included only those flies that were still alive at the right endpoint at age 50 days.

\begin{figure}[t]
	\begin{center}
		\includegraphics[scale=0.8]{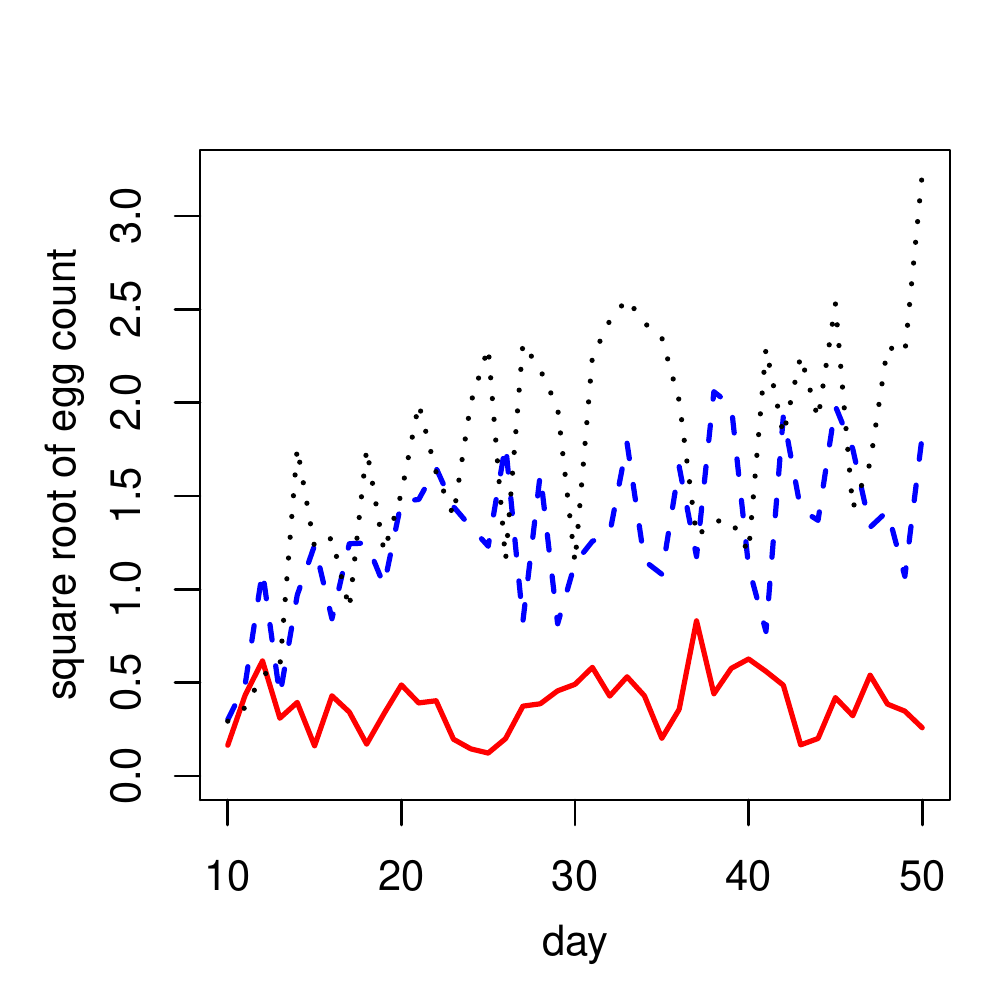}
	\end{center}
	\caption{Sample mean trajectories of the number of eggs laid between age 10 and 50 days by female fruit flies under three different diets, where the dotted curve corresponds to a
	cohort of $n_1=25$ flies receiving a  
	diet with 25\% protein, the dashed curve to a cohort of $n_2=41$ flies under a  diet with 10\% protein, and the solid curve to a cohort of $n=50$ flies under a diet with 4\% protein. 
	\label{fig:medfly}}
\end{figure}

Of interest is the effect of the amount of protein in the diet 
on the egg-laying trajectory, as female flies require protein to produce eggs. We compare three cohorts of fruit flies which all received an overall reduced diet at 25\% of full level and
three different protein levels, with sugar-to-protein ratios of 3:1, 9:1, and 24:1, corresponding to fractions of 25\%, 10\% and 4\% of protein in the diet. The cohorts  consist of $n_1=25, n_2=41$
and $n_3=50$ flies, respectively and are thus unbalanced. The sample mean functions for the three cohorts are depicted in 
Figure~\ref{fig:medfly}, where the noisy character of the data is reflected in the fluctuations of the functions. The mean of the cohort under a 4\% protein diet is
seen to be substantially smaller than the means for the other two groups, indicating that egg production is severely impeded if flies receive only 4\% protein. The mean functions for the cohorts receiving 10\% and 25\% are much closer, indicating that protein levels above 10\% have a relatively much smaller impact on egg-laying trajectories than 
protein levels declining below 10\%.

These visual impressions are confirmed when applying the proposed functional ANOVA approach. The selected value for $\tau$  was $\tau=0.4$ and 51 Fourier 
basis functions are used to represent the data. The  
overall $p$ value for the null hypothesis that the three mean functions are the same is $p< 10^{-7}$ {from Table \ref{tab:fda:app:pvalue}}. The pairwise comparisons between the groups with  
25\% protein and the 4\% protein as well as between the 10\% protein group and the 4\% protein group show significant differences, while this is not the case
for the comparison between the 25\% protein and the 10\% protein group. This confirms that there is a minimum protein level that needs to be maintained 
as otherwise egg-laying is impeded over the entire lifespan, while more than 10\% protein does not lead to major changes in the expected egg-laying trajectory.
This valuable extra information is obtained without performing additional hypothesis tests and thus no requirement for adjustments for multiple comparisons that might lower the power of the test.

\begin{table}[t]
	\caption{$p$-values for the study on the egg-laying trajectories\label{tab:fda:app:pvalue}}
	\vspace{-0.1in}
	\begin{center}
		\renewcommand*{\arraystretch}{1.2}
		\begin{tabular}{|c|c|c|c|c|c|c|c|}
			\hline
			 proposed & L2 & F & GPF & MPF & GET & RP \tabularnewline
			\hline
			 $<10^{-7}$ & $3.0\times 10^{-15}$ & $2.4\times 10^{-14}$ & $2.4\times  10^{-13}$  & .012 & .0005  & $.0007$ \tabularnewline
			\hline
		\end{tabular}
	\end{center}
\end{table}

\section{Application to sparse count data}\label{sec:count-data}
Count data, often modeled by multinomial or Poisson distributions, occur in many applications. For the multinomial model, the decay in variance is an inherent feature due to the requirement that the sum of the probabilities of all categories is one. For the Poisson distribution, since the variance is equal to the mean,  sparseness in the mean induces decay in the variance. Here, sparseness refers to situations where there are only a few nonzero coordinates, or where the ordered mean coordinates decrease to zero. For instance, in the field of text mining or information retrieval in which word frequency is an important feature, words in a vocabulary often have drastically different frequencies. In addition, the frequency of words decreases rapidly when moving from frequent to rare words. For example, for the English language, the ordered word frequency is found to approximately follow Zipf's law \citep{Zipf1949}. {Below we  assess the performance of the proposed method for sparse Poisson data via simulation studies and two real data applications.}

\subsection{Simulation studies}

We considered  three groups, represented by the $p$-dimensional random vectors $X_1$, $X_2$, and $X_3$. Each random vector $X_k$ follows a multivariate Poisson distribution \citep{Inouye2017} and is represented by $(W_{k0}+W_{k1},\ldots,W_{k0}+W_{kp})$, where for $k=1,2,3$, $W_{k0},\ldots,W_{kp}$ are independent Poisson random variables  with mean $\eta_{k0},\ldots,\eta_{kp}\in\real$, respectively. Then  the $j$th coordinate of $X_k$ follows also a Poisson distribution with mean $\eta_{k0}+\eta_{kj}$. In addition, all coordinates are correlated due to the shared  random variable $W_{k0}$. In our study, we set $\eta_{k0}=1$ for $k=1,2,3$, and consider two settings for $\eta_{k1},\ldots,\eta_{kp}$. In the first ``sparse'' setting, $\eta_{kj}=(1+\theta k)j^{-1}$ for $k=1,2,3$ and $j=1,\ldots,p$. In this setting, when $\theta\neq 0$, the difference of the mean in the $j$th coordinate decays as $j^{-1}$.  In the second ``dense'' setting, we set $\eta_{kj}=j^{-1}+\theta k/2$, so that the difference of the mean in each coordinate is  equal. Note that the setting with $\theta=0$ corresponds to the null hypothesis, under which the mean vectors of all groups are identical. For the dimension, we consider two cases, namely, $p=25$ and $p=100$, and for sample size  the balanced case $(n_1,n_2,n_3)=(50,50,50)$ and an unbalanced case with  $(n_1,n_2,n_3)=(30,50,70)$. The parameter $\tau$ is selected by the method described in Section \ref{sec:method}. Each simulation is repeated 1000 times. Across all settings, the average value of selected $\tau$ is $0.305\pm0.221$ and $0.341\pm 0.237$ for $p=25$ and $p=100$, respectively.

\begin{figure}
	\begin{minipage}{0.24\textwidth}
		\includegraphics[scale=0.4]{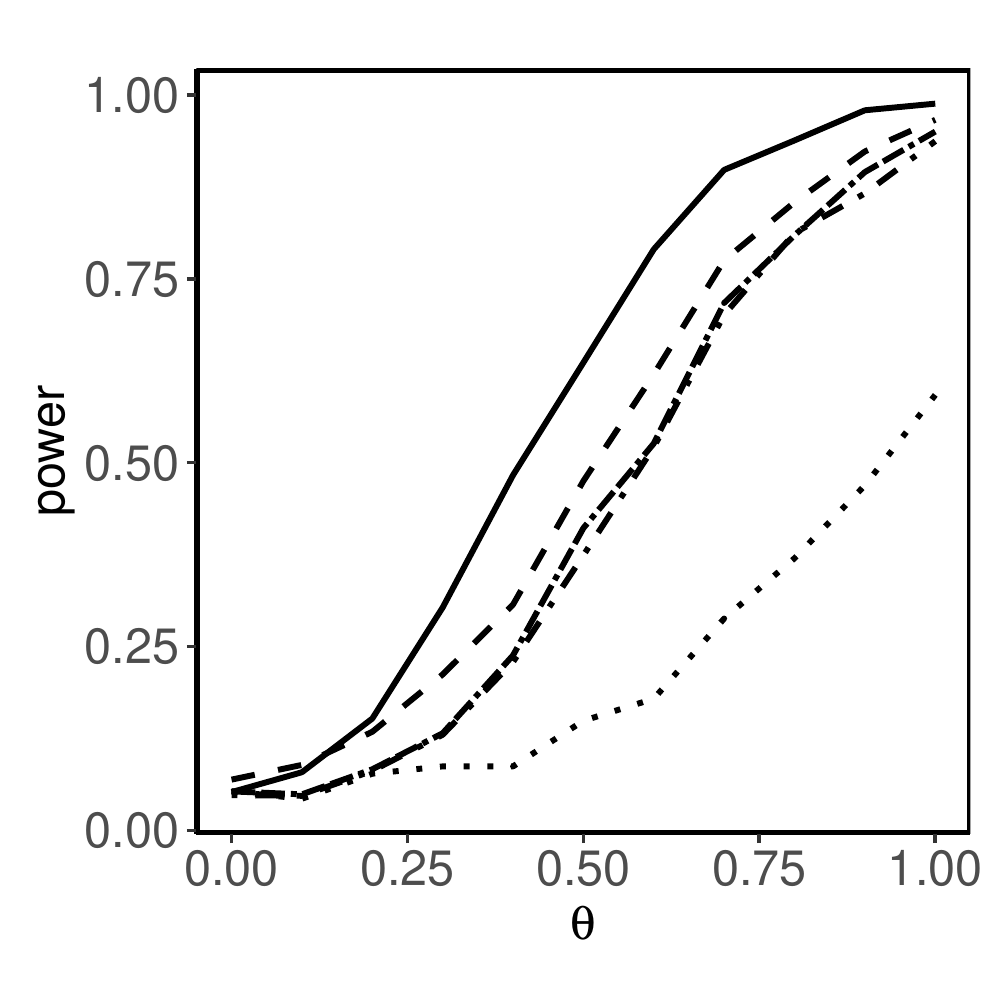}
	\end{minipage}
	\begin{minipage}{0.24\textwidth}
		\includegraphics[scale=0.4]{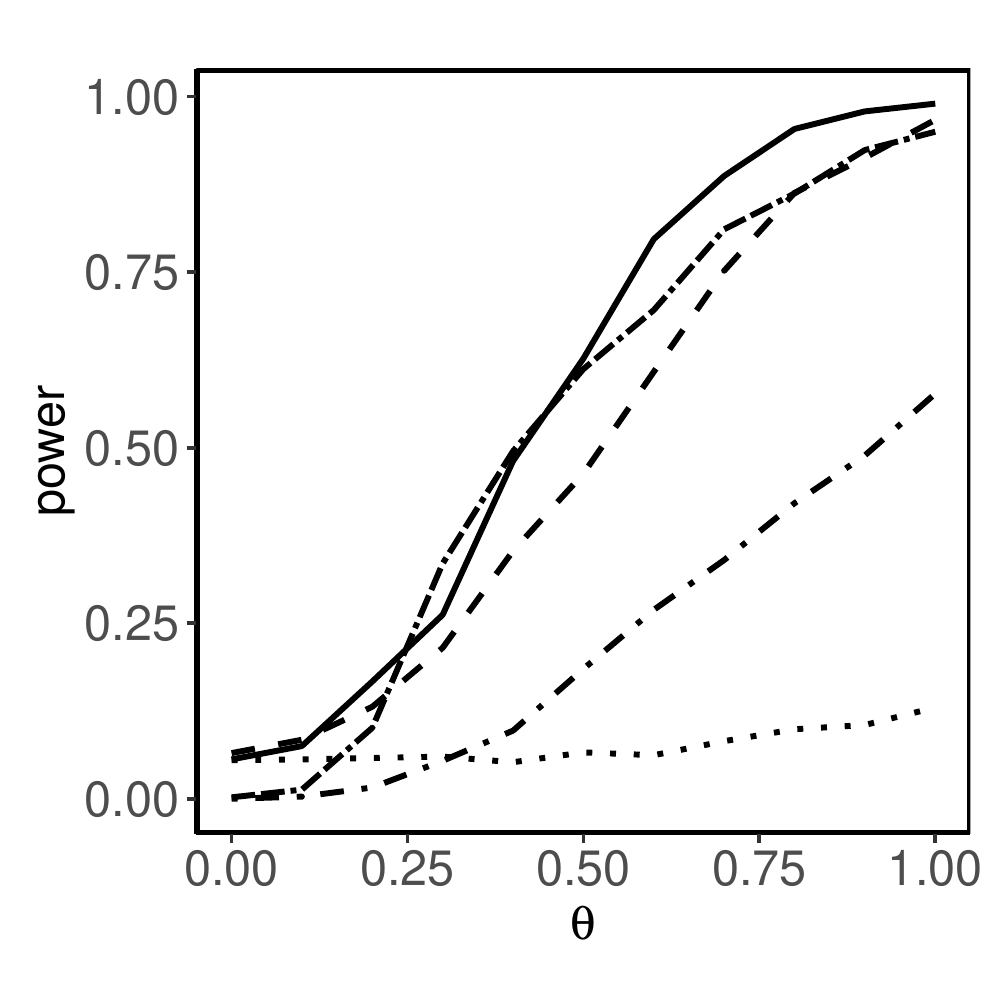}
	\end{minipage}
	\begin{minipage}{0.24\textwidth}
		\includegraphics[scale=0.4]{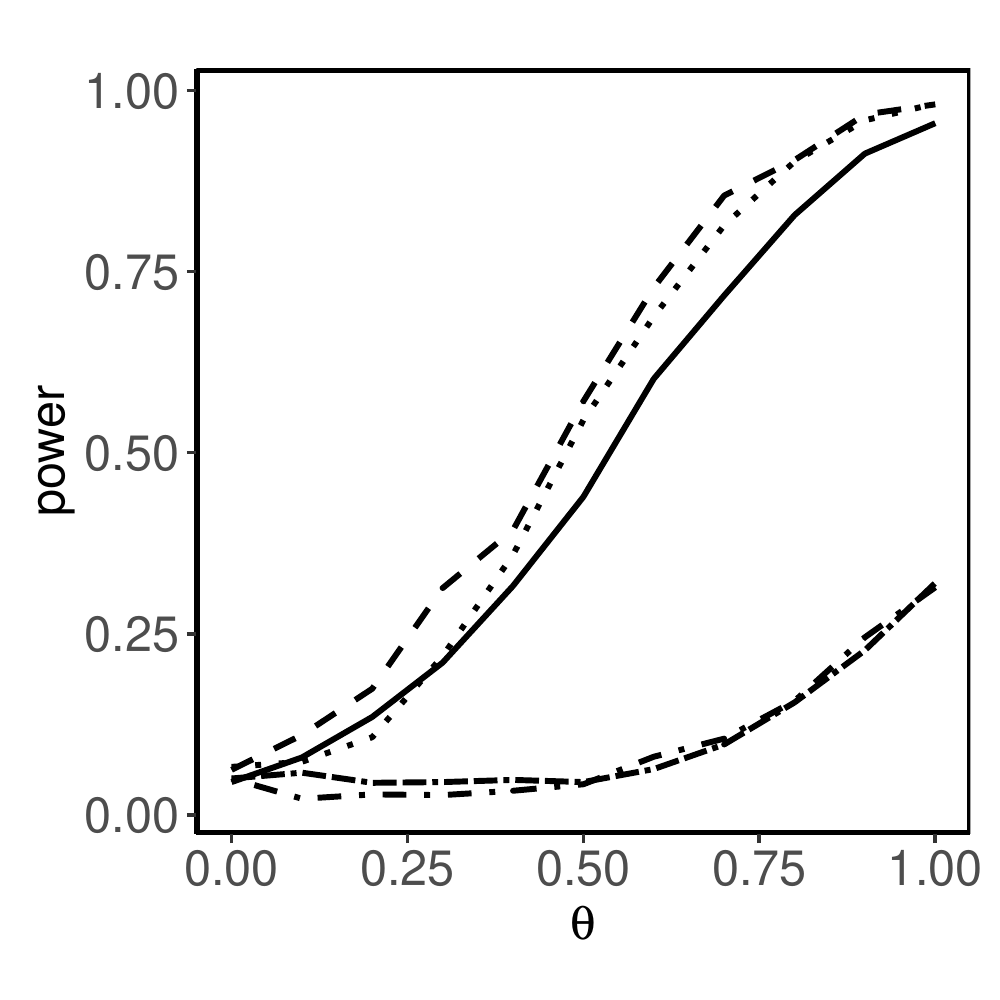}
	\end{minipage}
	\begin{minipage}{0.24\textwidth}
		\includegraphics[scale=0.4]{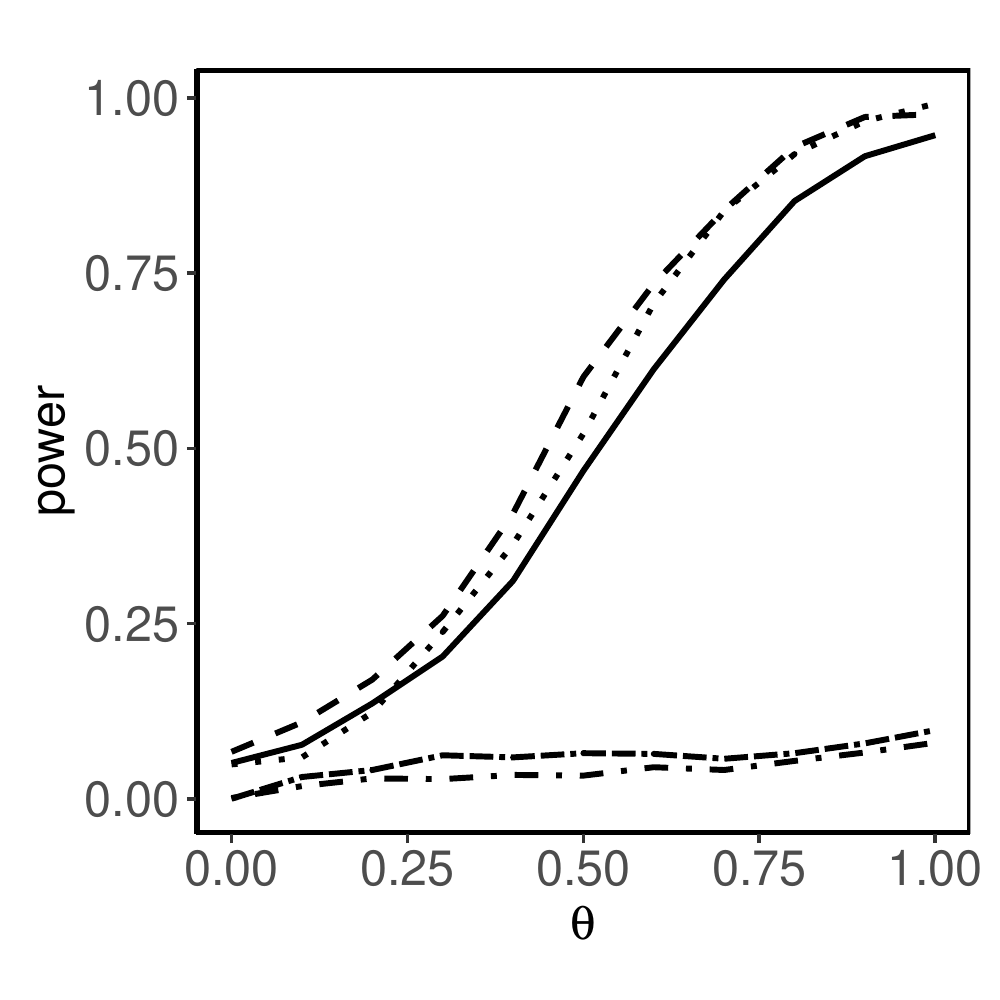}
	\end{minipage}
	\caption{Empirical power of the proposed high-dimensional ANOVA (solid), {DALp (dashed), S (dotted), LH (dot-dashed) and RRLH (dot-dash-dashed)}, when $(n_1,n_2,n_3)=(30,50,70)$, for the sparse setting with $p=25$ (first panel) and $p=100$ (second panel) and for the dense setting with $p=25$ (third panel) and $p=100$ (fourth panel).\label{fig:power-pois}}
\end{figure}

For comparison purposes, we implemented {the procedure (S) of \cite{Schott2007} and the data-adaptive $\ell_p$-norm-based test (DALp)  \citep{Zhang2018b}} that are reviewed in the introduction. The former is based on the limit distribution of a test statistic that is composed {of} inter-group and within-group {sums}  of squares,  
while the latter utilizes an adjusted $\ell_p$-norm-based test statistic whose distribution is approximated by a multiplier bootstrap. The former is favored for  testing problems with a dense alternative, while the latter has been reported to be powerful against different patterns of alternatives \citep{Zhang2018b}. We also include the classic Lawley--Hotelling trace test (LH) \citep{Lawley1938,Hotelling1947} as a baseline method which is not specifically designed for the high-dimensional setting, \lin{and its ridge-regularized version (RRLH) \citep{Li2020} targeting the high-dimensional scenario}. The empirical sizes  in Table \ref{tab:size-pois} demonstrate that those of the  proposed test  and the test of \cite{Schott2007} are quite close to the nominal level, while the size of the test of \cite{Zhang2018b} seems slightly inflated and the sizes of the Lawley--Hotelling trace test {and its regularized version are} rather conservative in the high-dimensional case $p=100$.  The power function  for the sparse case $(n_1,n_2,n_3)=(30,50,70)$ is shown in Figure \ref{fig:power-pois}, while the power function for $(n_1,n_2,n_3)=(50,50,50)$ is very similar (not shown).  One finds that in the sparse case, the proposed test has substantially more power   than the test of \cite{Zhang2018b}, while the latter in turn has more power  than the test of \cite{Schott2007}  and the Lawley--Hotelling trace tests. In the dense setting which does not favor the proposed test, it is seen to have  power behavior that is comparable with that of  the tests of  \cite{Schott2007} and \cite{Zhang2018b}, and all of these methods outperform the Lawley--Hotelling trace test whose performance substantially deteriorates for higher dimensions. \lin{The regularized Lawley--Hotelling trace test  substantially improves upon the classic version only in the sparse setting and when the dimension is relatively large, e.g., when $p=100$.}
The average computation time to complete a single Monte Carlo simulation replicate is presented in Table \ref{tab:ct-pois}, where $p=100$ and the parameter $\tau$ is selected from 11 candidate values by the data-driven procedure proposed in Section \ref{sec:method}. \lin{We observe that a single simulation replicate can be completed within 10 seconds without GPU acceleration and within 0.2 seconds by utilizing an NVIDIA Quadro P400 graphics card.} In addition to testing hypotheses, the proposed method can also simultaneously identify the pairs of groups, as well as  coordinates, that have significantly different means, as we demonstrate below for two real datasets. 
\begin{table}[t]
	\caption{Empirical size of ANOVA on Poisson data\label{tab:size-pois}}
	\vspace{-0.1in}
	\begin{center}
		\renewcommand*{\arraystretch}{1.2}
		\begin{tabular}{|c|c|c|c|c|c|c|c|}
			\hline
			 & $p$ & $n$ & proposed & S & DALp & LH & RRLH \tabularnewline
			\hline
			\multirow{4}{*}{sparse} & \multirow{2}{*}{25} & 50,50,50 & .055 & .042 & .065 & .045 & .051  \tabularnewline
			\cline{3-8}
			& & 30,50,70 & .052 & .053 & .069 & .048 & .053 \tabularnewline
			\cline{2-8}
			& \multirow{2}{*}{100} & 50,50,50 & .056 & .045 & .054 & .000 & .000 \tabularnewline
			\cline{3-8}
			& & 30,50,70 & .056 & .055 & .065 & .000 & .002 \tabularnewline
			\cline{2-8}
			\hline
			\multirow{4}{*}{dense} & \multirow{2}{*}{25} & 50,50,50 & .050 & .051 & .065 & .045  & .065 \tabularnewline
			\cline{3-8}
			& & 30,50,70 & .045 & .066 & .062 & .050  & .050 \tabularnewline
			\cline{2-8}
			& \multirow{2}{*}{100} & 50,50,50 & .057 & .054 & .064 & .001 & .004  \tabularnewline
			\cline{3-8}
			& & 30,50,70 & .051 & .049 & .067 & .001 & .000 \tabularnewline
			\cline{3-8}
			\hline
		\end{tabular}
	\end{center}
\end{table}

\begin{table}[t]
	\caption{Average computation time for ANOVA on Poisson data (in seconds)\label{tab:ct-pois}}
	\vspace{-0.1in}
	\begin{center}
		\renewcommand*{\arraystretch}{1.2}
			\begin{tabular}{|c|c|c|c|c|c|}
				\hline
				 proposed (no GPU) & proposed (GPU)  & S & DALp & LH & RRLH\tabularnewline
				\hline
				9.869 & .155  & .011 & .461 & .030 & .135 \tabularnewline
				\hline
		\end{tabular}
	\end{center}
\end{table}

\subsection{Data applications}

We apply the proposed method to analyze the CLASSIC3 dataset\footnote{Originally available from ftp://ftp.cs.cornell.edu/pub/smart, and now available publicly on the Internet, e.g., https://www.dataminingresearch.com/index.php/2010/09/classic3-classic4-datasets/} \citep{Dhillon2003} that has been studied in information retrieval. The data consist of 3891 document abstracts from three different domains, specifically, $n_1=1460$ from information retrieval (CISI), $n_2=1398$ from aeronautical systems (CRAN) and $n_3=1033$ from medical research (MED). Standard text preprocessing was applied to these abstracts, including removal of high-frequency common words (commonly referred to as stop words, such as ``the'', ``is'', ``and'', etc), punctuation and Arabic numbers. In addition, we follow common practice in the field of information retrieval to reduce inflected words to their word stem, base or root form by using a stemmer, such as the Krovetz stemmer \citep{Krovetz1993}.  Each document is then represented by a vector of word counts. These vectors are naturally sparse, as the number of distinct words appearing in a document is in general far less than the size of the vocabulary. Intuitively, vocabularies from different domains are different. Our goal is to examine this intuition and to find the words that are substantially different among the  three domains. To this end, we focus on words with at least 50 occurrences in total to eliminate the effects of rare words. This results in $p=1296$ distinct words under consideration. Then, we applied the proposed test to the processed data and found that the vocabularies used in these three domains are not the same among any pair of the domains, with $p$-value less than $10^{-7}$ where $\tau$ was selected as  $\tau=0.6$. In particular, the proposed  method simultaneously identifies the words that have significantly different frequency among the domains, which are shown in Table \ref{tab:classic3}, where the numbers represent the average frequency of the words within each domain. The results for CISI and CRAN match our intuition about these two domains. For the domain of medical research, the word ``normal'' is often used to refer to healthy patients or subjects, while the word ``increase'' is used to describe the change of certain {health}  
metrics, such as {blood pressure}. 

\begin{table}
	\caption{The average frequency of words that are significantly different among all categories\label{tab:classic3}}
	\vspace{-0.2in}
	\begin{center}
		\resizebox{\textwidth}{!}{
		\begin{tabular}{|c|c|c|c|c|c|c|c|c|c|c|}
			\hline
			     & use & data & pressure & effect & theory & problem & body & increase & normal & group \\
			\hline
			CISI & \textbf{0.715} & \textbf{0.401} &  0.011 &  0.060 & 0.167 & 0.301 &  0.017 &  0.089  & 0.007 & 0.129 \\
			\hline
			CRAN & 0.515 &  0.239 & \textbf{1.004} & \textbf{0.759} & \textbf{0.684} &  \textbf{0.456} &  \textbf{0.607} &  0.271  & 0.112 &  0.011 \\
			\hline 
			MED  & 0.265 &  0.082 &  0.139 & 0.338 & 0.024 & 0.069 & 0.162 & \textbf{0.437} & \textbf{0.351} & \textbf{0.304} \\
			\hline
		\end{tabular}}
	\end{center}
\end{table}

\begin{table}[t]
	\caption{$p$-values for studies on CLASSIC3 and NHANES datasets\label{tab:manova:app:pval}}
	\vspace{0.1in}
	\centering
	\begin{threeparttable}
		\renewcommand*{\arraystretch}{1.2}
		\begin{tabular}{|c|c|c|c|c|c|}
			\hline
			 & proposed  & S & DALp & LH & RRLH \tabularnewline
			\hline
			CLASSIC3 & $<10^{-7}$  & $0^\dagger$  & $<10^{-7}$  & $0^\dagger$  & $0^\dagger$ \tabularnewline
			\hline
			NHANES & .004  & .005 & .005 & .936 & .716 \tabularnewline
			\hline
		\end{tabular}
	\begin{tablenotes}\footnotesize
		\item[$\dagger$] The $p$-values are below machine precision.
	\end{tablenotes}
	\end{threeparttable}
\end{table}

Next, we apply the proposed  method to study physical activity using data collected by wearable devices, as available in the National Health and Nutrition Examination Survey (NHANES) 2005--2006. In the survey, each participant of age 6 years or above was asked to wear a physical activity monitor (Actigraph 7164) for seven consecutive days, with bedtime excluded. Also, as the device  is not waterproof, participants were advised to remove it during swimming or bathing. The monitor detected and recorded the magnitude of acceleration  of movement of the participant. For each minute, the readings were summarized to yield one single integer in the interval $[0,32767]$ that signifies the average intensity {of movement} within that minute. This results in $m=60\times 24\times 7=10080$ observations per participant. Demographic characteristics  of the participants are also available, and in our analysis  we focused on two age groups and two marital categories. The two age groups are young adulthood with age ranging from 18 to 44, and middle-age adulthood with age ranging from 45 to 65. The two marital  groups are ``single'' (including the widowed, divorced, separated and never-married categories in the original data) and ``non-single'' (including married and living-with-partner categories). These groups induce four cohorts: young non-single adults, young single adults, middle-age non-single adults and middle-age single adults. Our goal is to examine whether the physical activity patterns are different among these cohorts.

\begin{figure}
	\begin{minipage}{0.3\textwidth}
		\includegraphics[scale=0.5]{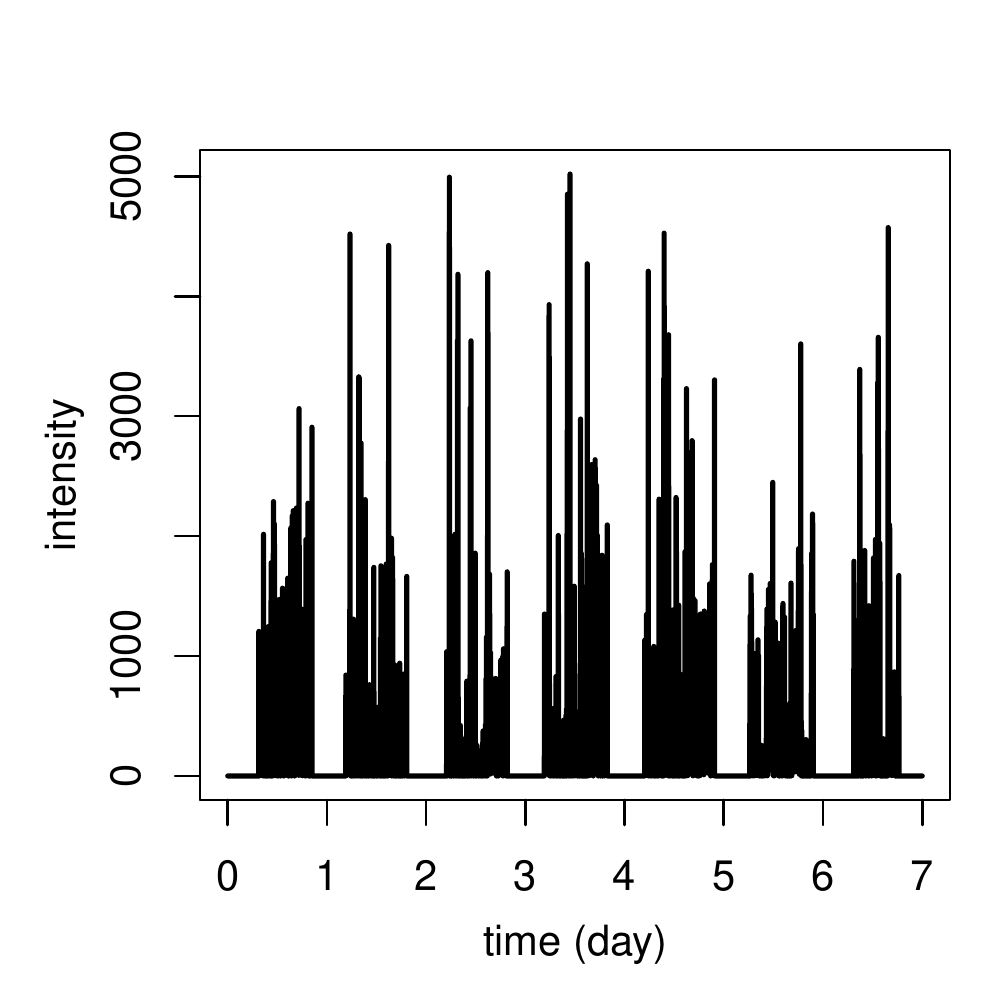}
	\end{minipage}
	\begin{minipage}{0.3\textwidth}
		\includegraphics[scale=0.5]{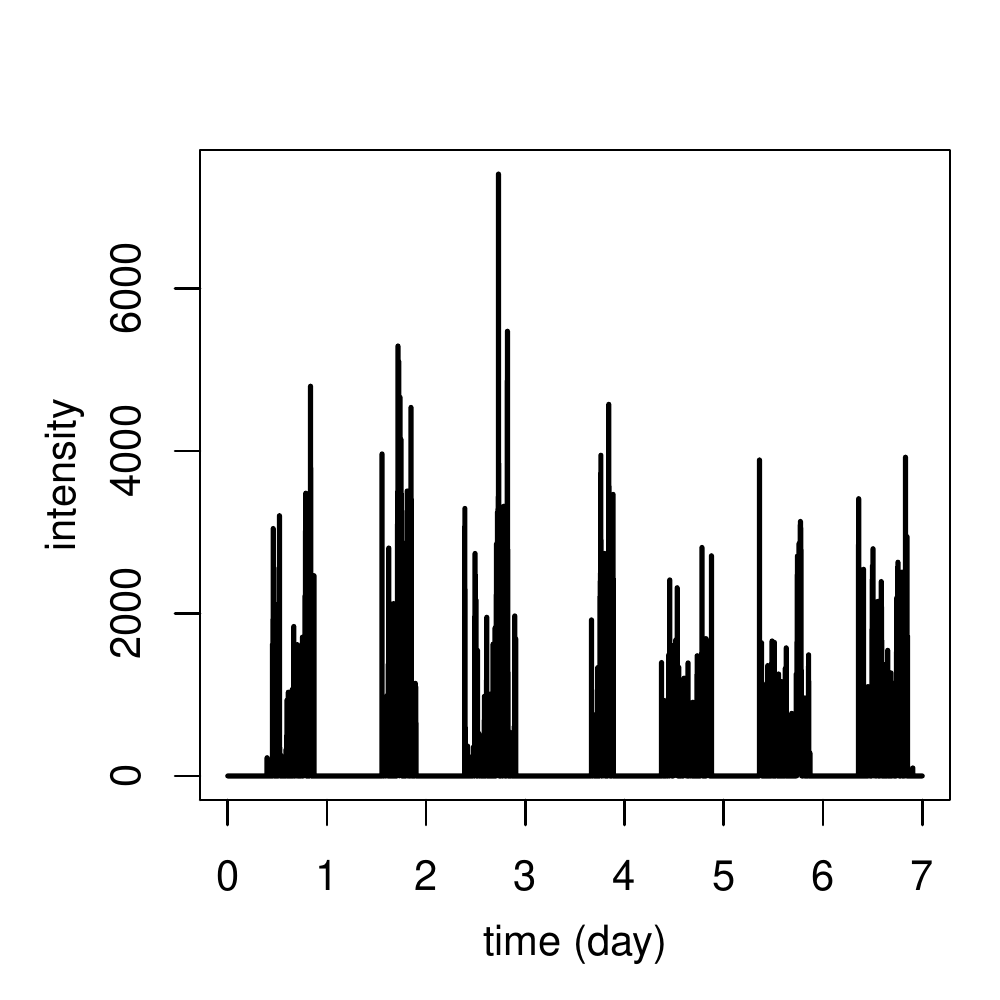}
	\end{minipage}
	\begin{minipage}{0.3\textwidth}
		\includegraphics[scale=0.5]{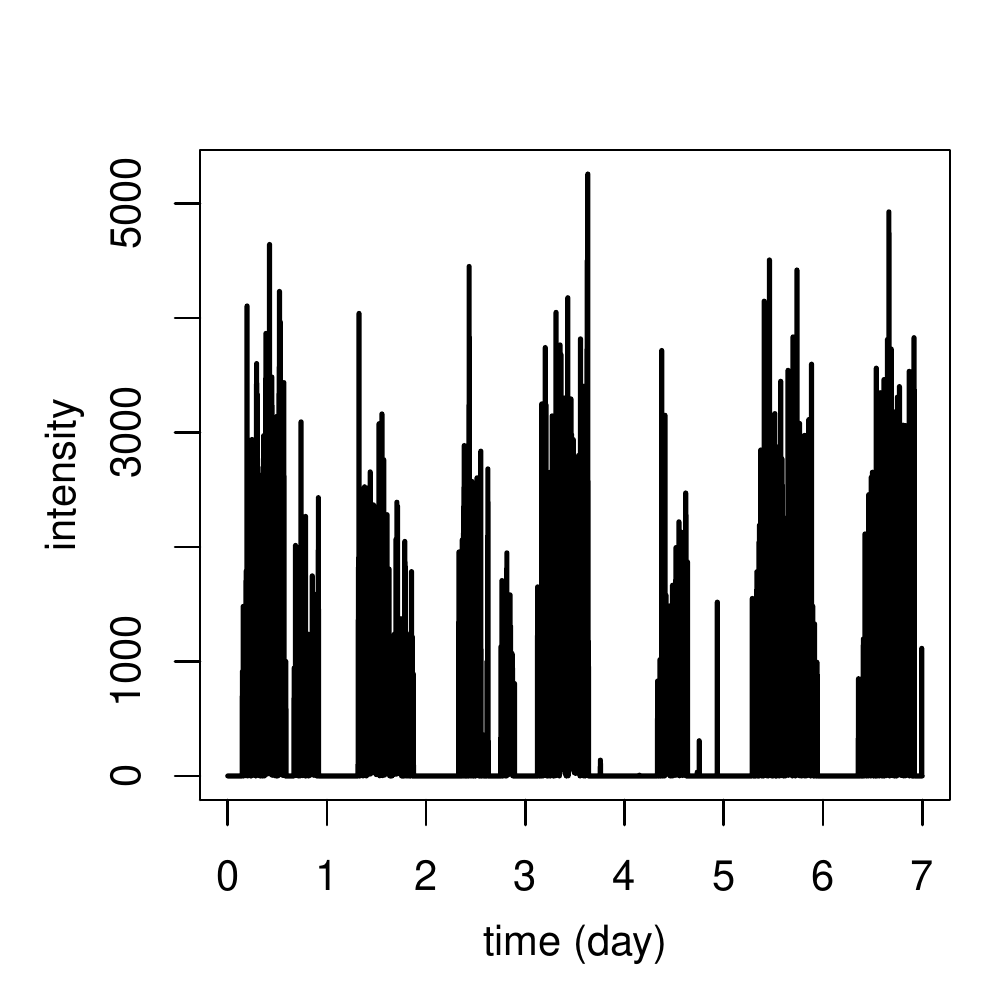}
	\end{minipage}
	\caption{Activity intensity trajectories of three randomly selected participants from the NHANES data 2005--2006.\label{fig:nhanes-raw}}
\end{figure}

Figure \ref{fig:nhanes-raw} presents the activity trajectories of three randomly selected participants, showing that they have different circadian rhythms. To address this problem, we adopt the strategy proposed by \cite{Chang2020}, who studied physical activity of elder veterans from the perspective of functional data analysis, by transforming each activity trajectory $A(t)$ into an activity profile $X(j)=\mathrm{Leb}(\{t\in[0,7]:A(t)\geq j\})$ for $j=1,\ldots,32767$, where $\mathrm{Leb}$ denotes the Lebesgue measure on $\real$. This is essentially equivalent to accumulated $F_A(j)/m$, where $F_A(j)$ denotes the frequency of $j$, i.e., the number of occurrences of the intensity value $j$, in the trajectory $A$. Therefore, the activity profile $X(j)$ can be viewed as count data normalized by $m$. As over 95\% of the  physical activity has low to moderate intensity, i.e., with intensity value below 1000, we focus on the intensity spectrum $[1,1000]$. {In addition, we exclude subjects with readings that are missing, unreliable or from a monitor not in calibration. This results in four cohorts of size $n_1=1027$, $n_2=891$, $n_3=610$ and $n_4=339$, respectively.}

\begin{figure}[H]
	\begin{minipage}{0.49\textwidth}
		\includegraphics[scale=0.6]{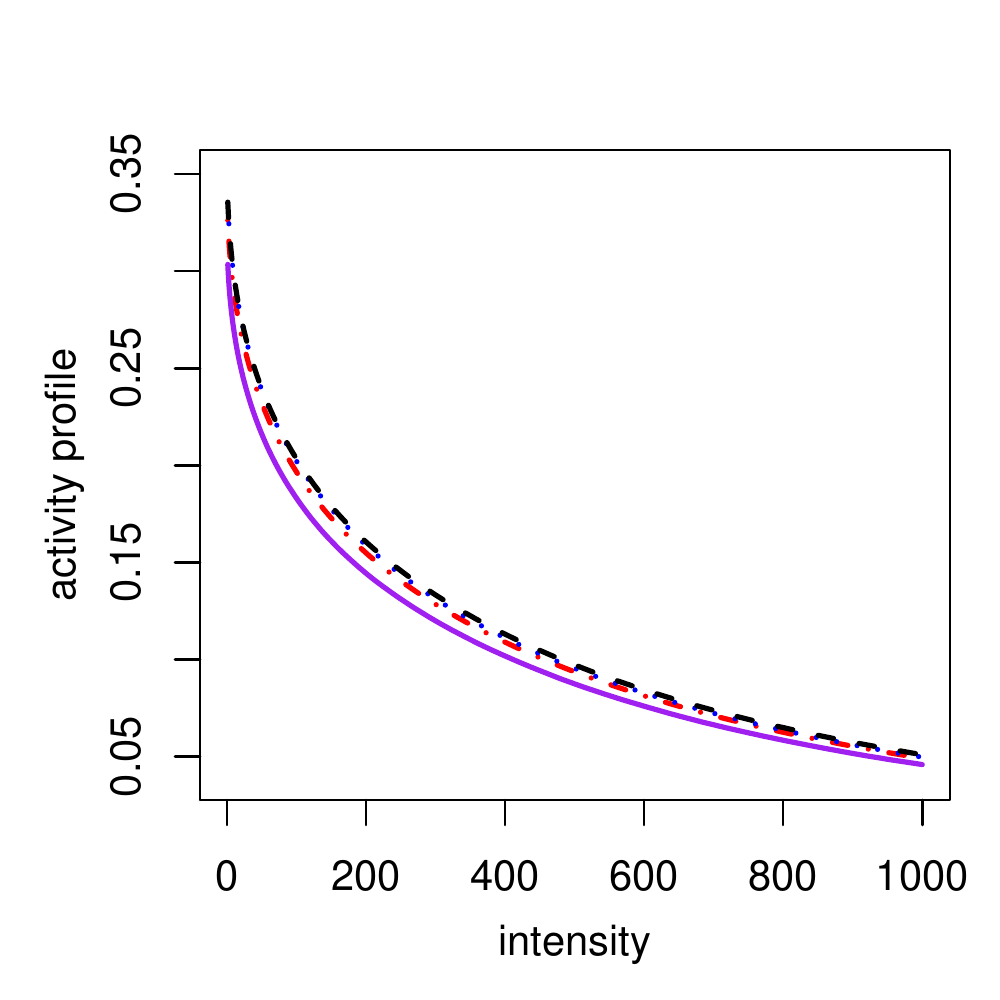}
	\end{minipage}
	\begin{minipage}{0.49\textwidth}
		\includegraphics[scale=0.6]{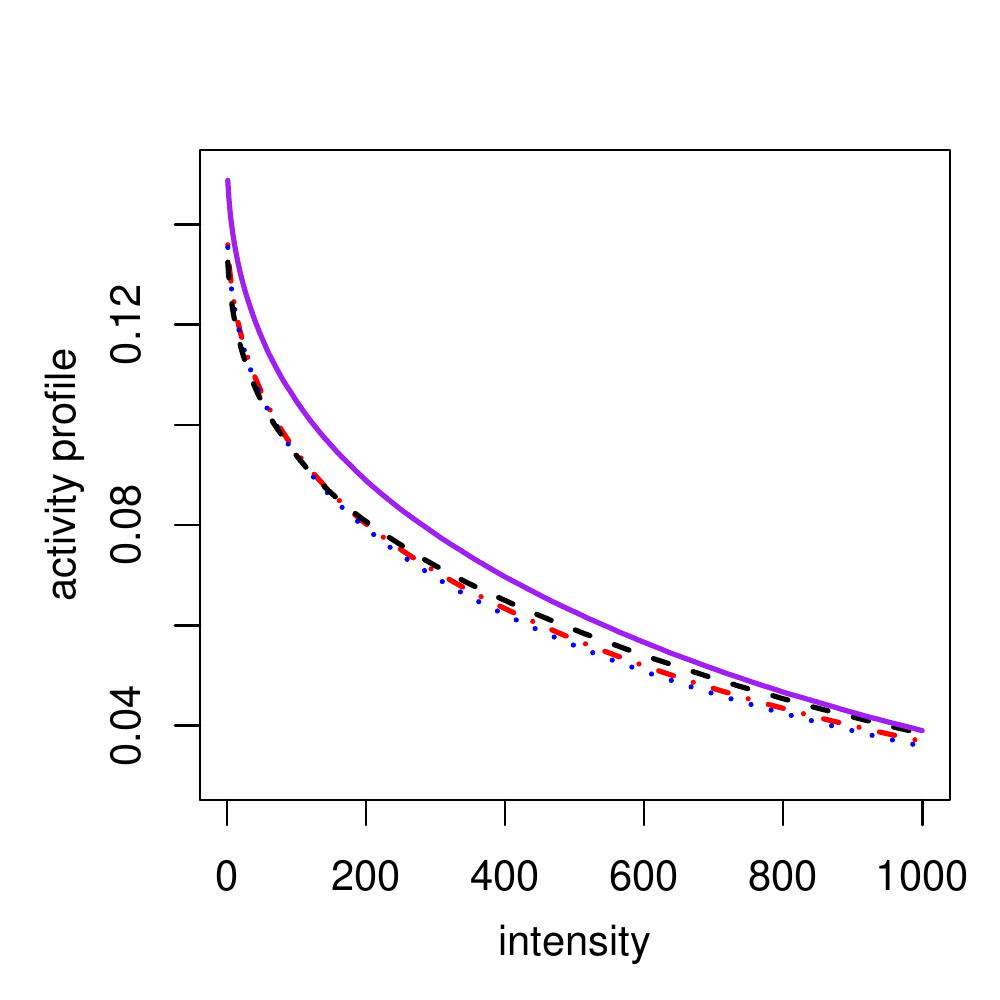}
	\end{minipage}
\\
	\begin{minipage}{0.49\textwidth}
		\includegraphics[scale=0.6]{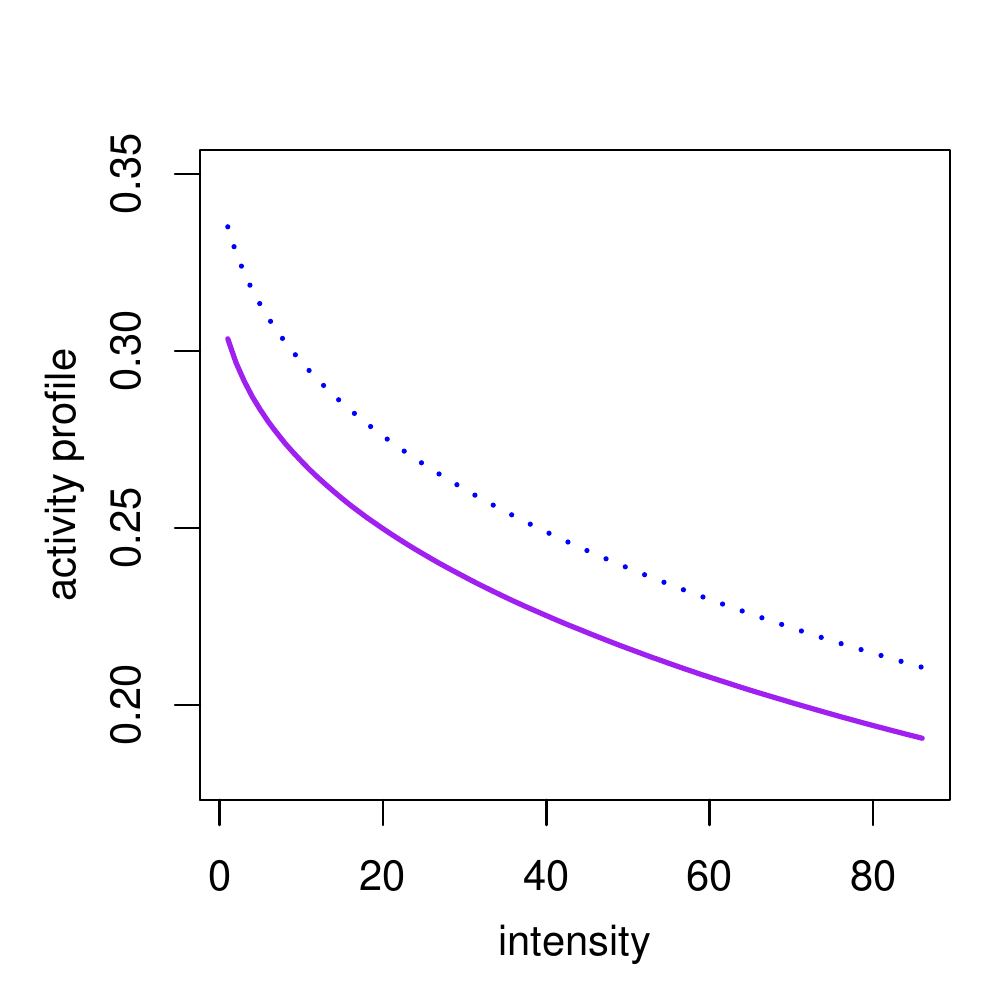} 
	\end{minipage}
	\begin{minipage}{0.49\textwidth}
		\includegraphics[scale=0.6]{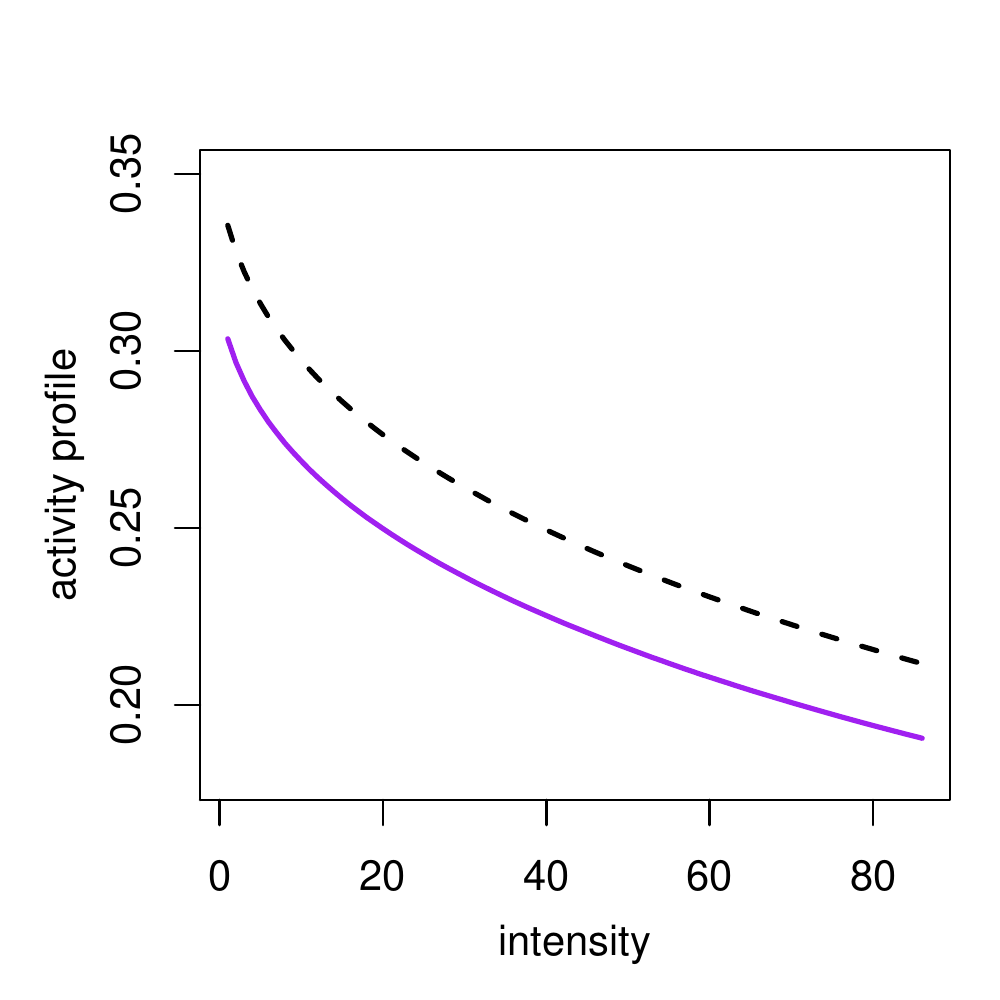} 
	\end{minipage}
	\caption{Top: the coordinate-wise mean activity (left) and its standard deviation (right) of young non-single cohort (dash-dotted), young single cohort (dotted), middle-age non-single cohort (dashed) and middle-age single cohort (solid); bottom-left: mean activity profiles of the young single cohort (dotted) and the middle-age single cohort (solid) shown for the  intensity spectrum on which the differences in the means are significant among the two cohorts; bottom-right:  mean activity profiles of the middle-age non-single cohort (dashed) and the middle-age single cohort (solid) over the spectrum on which the differences in the means are significant among the two cohorts.  \label{fig:nhanes-res}}
\end{figure}

\begin{figure}[H]
	\begin{minipage}{0.49\textwidth}
		\includegraphics[scale=0.7]{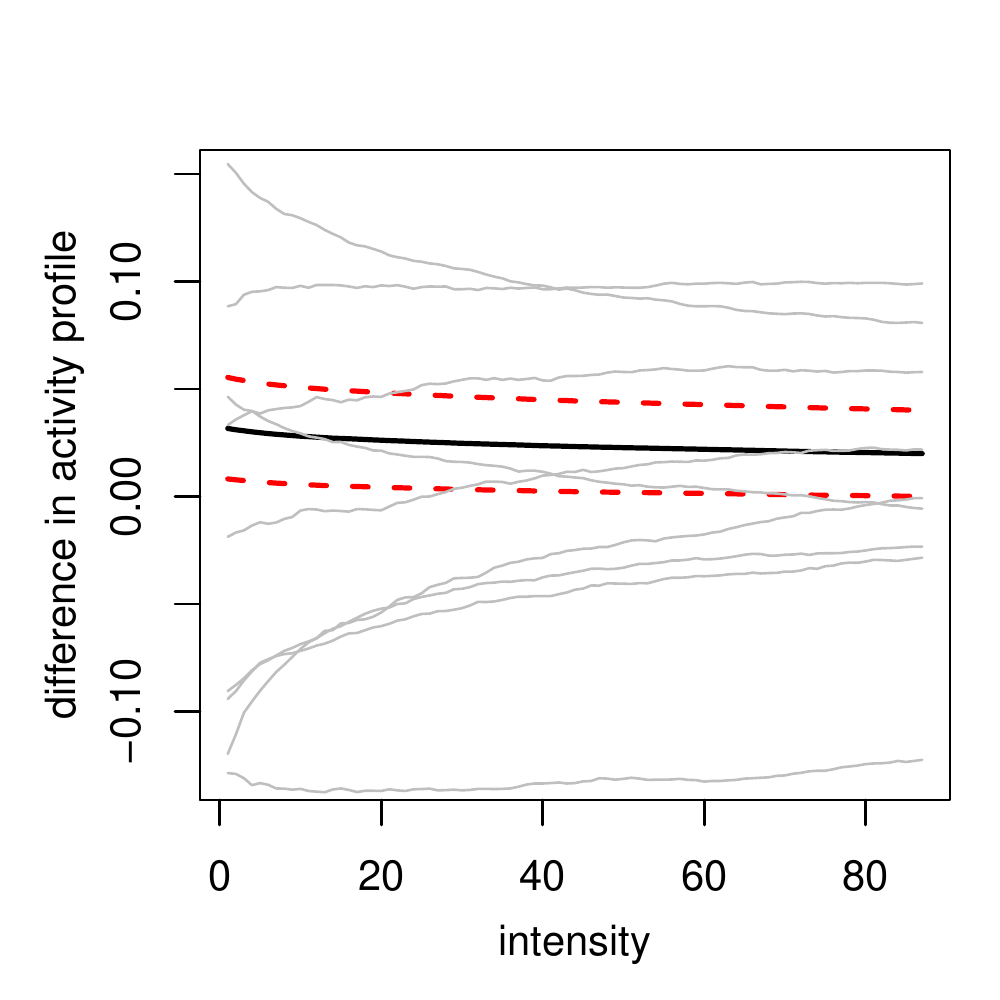}
	\end{minipage}
	\begin{minipage}{0.49\textwidth}
		\includegraphics[scale=0.7]{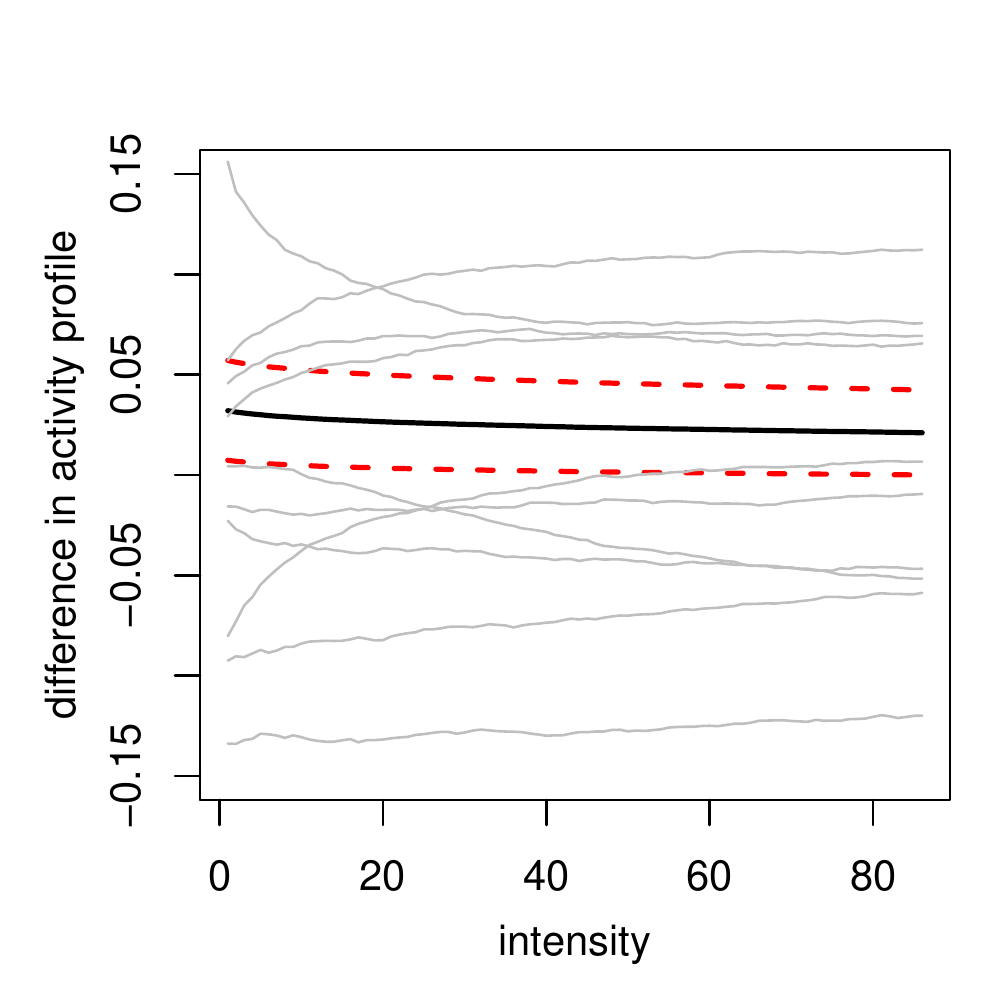}
	\end{minipage}
	\caption{The empirical simultaneous confidence regions (dashed) for the difference (solid) of mean activity profiles over $[1,87]$. The left panel corresponds to young single and middle-age single cohorts. The right panel corresponds to middle-age non-single and middle-age single cohorts. The light gray solid lines are differences of activity profiles of some pairs of participants from the corresponding pairs of cohorts, included to illustrate the variability of the differences in the individual level. \label{fig:nhanes-sci}}
\end{figure}

The  mean activity profiles and their standard deviations are depicted in the top panels of Figure \ref{fig:nhanes-res}, from which we observe that both the mean and standard deviation decay quite fast. In addition, the mean profiles from  the young single and middle-age non-single cohorts are almost indistinguishable in the plot, while the mean profile of the middle-age single cohort is visibly different from the others. These visual impressions are in line with the results obtained with the proposed test, which  rejects the global null hypothesis with an approximate $p$-value of  0.004 and thus suggests that some mean activity profiles are likely to be substantially different, where the selected value for $\tau$ is $0.5$. \lin{The methods of \cite{Schott2007} and \cite{Zhang2018b} also reject the null hypothesis with a similar $p$-value, while both  Lawley--Hotelling trace test and its regularized version do not; see Table \ref{tab:manova:app:pval} for the detailed $p$-values of these methods.} The proposed method also identifies two pairs of cohorts whose mean activity profiles are different and the intensity spectrum on which the differences are significant, namely, the young single cohort and the middle-age single cohort on the spectrum $[1,87]$, and the middle-age non-single cohort and middle-age single cohort on the spectrum $[1,86]$. These findings are visualized in the bottom panels of Figure \ref{fig:nhanes-res}. Furthermore, the proposed method provides SCRs for the differences of mean activity profiles among all pairs of cohorts. For instance, in Figure \ref{fig:nhanes-sci} we present the 95\% SCRs for the pairs with differences in the mean activity profiles over the spectrum on which the differences are statistically significant. In summary, comparing to the young single and middle-age non-single cohorts, the middle-age single cohort is found to have less activity on average in the low-intensity activity spectrum.

\section{Concluding Remarks}\label{sec:conclusion}

\lin{The proposed method for high-dimensional ANOVA via bootstrapping max statistics leads to the  construction of  simultaneous confidence regions for the differences of population mean vectors and is applicable for various statistical frameworks, including functional data analysis and multinomial and count data settings. The theoretical justifications rely on two key ingredients, variance decay and partial standardization, which imply near-parametric rates of convergence in high dimensions.   In simulations, the  resulting tests are shown to be highly competitive in terms of controlling the size of the tests and power in a variety of scenarios. It is notable that the proposed method can be completely parallelized which leads to very fast implementations on parallel processors.}

\lin{As predicted by theory,  performance of the proposed method is geared towards the case of sparse signals and in such scenarios it routinely outperforms competing methods in simulations. It is also found to be competitive for situations with dense  signals. Since it is often unknown whether signals are sparse or dense in practice, this makes the method
quite appealing for high-dimensional ANOVA in the presence of variance decay, notably for functional ANOVA problems where such variance decay is an inherent feature.}

\lin{The proposed method employs a parameter $\tau$ that controls  the partial standardization, which is chosen data-adaptively. The implementation can 
be further accelerated by choosing a fixed value, where  the choice $\tau=0.8$ was shown to be effective in simulation studies in Sections \ref{sec:additional-simulation-fda} and \ref{sec:additional-simulation-manova} of the Supplement. The principle of partial standardization may be of broader interest.}


\bigskip
\begin{center}
	{\large\bf SUPPLEMENTARY MATERIAL}
\end{center}

\begin{description}
	
	\item[Supplement:] The Supplement contains the proofs for the results in Section \ref{sec:theory}, and additional simulation studies for functional ANOVA and high-dimensional MANOVA. (PDF)
	
	\item[R-package:] The {\tt hdanova.cuda}  package\footnote{https://github.com/linulysses/hdanova.cuda}
	 implements the proposed method for the GPU based computing platform.

\end{description}

\bibliographystyle{asa}
\bibliography{hdanova}

\begin{thebibliography}{64}
\newcommand{\enquote}[1]{``#1''}
\expandafter\ifx\csname natexlab\endcsname\relax\def\natexlab#1{#1}\fi

\bibitem[{Aneiros et~al.(2019)Aneiros, Cao, Fraiman, Genest, and
  Vieu}]{Aneiros:2019aa}
Aneiros, G., Cao, R., Fraiman, R., Genest, C., and Vieu, P. (2019),
  \enquote{Recent advances in functional data analysis and high-dimensional
  statistics,} \textit{Journal of Multivariate Analysis}, 170, 3--9.

\bibitem[{Bai et~al.(2018)Bai, Choi, and Fujikoshi}]{Bai2018}
Bai, Z., Choi, K.~P., and Fujikoshi, Y. (2018), \enquote{Limiting behavior of
  eigenvalues in high-dimensional MANOVA via RMT,} \textit{The Annals of
  Statistics}, 46, 2985--3013.

\bibitem[{Bai and Saranadasa(1996)}]{Bai1996}
Bai, Z. and Saranadasa, H. (1996), \enquote{Effect of high dimension: By an
  example of a two sample problem,} \textit{Statistica Sinica}, 6, 311--329.

\bibitem[{Cai et~al.(2014)Cai, Liu, and Xia}]{Cai2014}
Cai, T.~T., Liu, W., and Xia, Y. (2014), \enquote{Two-sample test of high
  dimensional means under dependence,} \textit{Journal of Royal Statistical
  Society. Series B: Statistical Methodology}, 76, 349--372.

\bibitem[{Cai and Xia(2014)}]{Cai2014a}
Cai, T.~T. and Xia, Y. (2014), \enquote{High-dimensional sparse {MANOVA},}
  \textit{Journal of Multivariate Analysis}, 131, 174--196.

\bibitem[{Carey et~al.(2008)Carey, Harshman, Liedo, M\"{u}ller, Wang, and
  Zhen}]{Carey2008}
Carey, J., Harshman, L., Liedo, P., M\"{u}ller, H.-G., Wang, J.-L., and Zhen,
  Z. (2008), \enquote{Longevity-fertility trade-offs in the tephritid fruit
  fly, {A}nastrepha ludens, across dietary-restriction gradients.}
  \textit{Aging Cell}, 7, 470--477.

\bibitem[{Carey et~al.(2002)Carey, Liedo, Harshman, Zhang, M\"{u}ller,
  Partridge, and Wang}]{Carey2002}
Carey, J.~R., Liedo, P., Harshman, L., Zhang, Y., M\"{u}ller, H.-G., Partridge,
  L., and Wang, J.-L. (2002), \enquote{Life history response of {M}editerranean
  fruit flies to dietary restriction,} \textit{Aging Cell}, 1, 140--148.

\bibitem[{Carey et~al.(1998)Carey, Liedo, M\"{u}ller, Wang, and
  Vaupel}]{Carey1998}
Carey, J.~R., Liedo, P., M\"{u}ller, H.-G., Wang, J.-L., and Vaupel, J.~W.
  (1998), \enquote{Dual modes of aging in {M}editerranean fruit fly females,}
  \textit{Science}, 281, 996--998.

\bibitem[{Chang and McKeague(2020)}]{Chang2020}
Chang, H.-W. and McKeague, I.~W. (2020), \enquote{Nonparametric comparisons of
  activity profiles from wearable device data,} \textit{preprint}.

\bibitem[{Chang et~al.(2017)Chang, Zheng, Zhou, and Zhou}]{Chang2017}
Chang, J., Zheng, C., Zhou, W.-X., and Zhou, W. (2017),
  \enquote{Simulation-based hypothesis testing of high dimensional means under
  covariance heterogeneity,} \textit{Biometrics}, 73, 1300--1310.

\bibitem[{Chen and Qin(2010)}]{Chen2010}
Chen, S.~X. and Qin, Y.-L. (2010), \enquote{A two-sample test for
  high-dimensional data with applications to gene-set testing,} \textit{The
  Annals of Statistics}, 38, 808--835.

\bibitem[{Chernozhukov et~al.(2013)Chernozhukov, Chetverikov, and
  Kato}]{Chernozhukov2013}
Chernozhukov, V., Chetverikov, D., and Kato, K. (2013), \enquote{Gaussian
  approximations and multiplier bootstrap for maxima of sums of
  high-dimensional random vectors,} \textit{The Annals of Statistics}, 41,
  2786--2819.

\bibitem[{Chernozhukov et~al.(2017)Chernozhukov, Chetverikov, and
  Kato}]{Chernozhukov2017}
--- (2017), \enquote{Central limit theorems and bootstrap in high dimensions,}
  \textit{The Annals of Probability}, 45, 2309--2352.

\bibitem[{Cuesta-Albertos and Febrero-Bande(2010)}]{Cuesta-Albertos2010}
Cuesta-Albertos, J.~A. and Febrero-Bande, M. (2010), \enquote{A simple multiway
  {ANOVA} for functional data,} \textit{Test}, 19, 537--557.

\bibitem[{Dhillon et~al.(2003)Dhillon, Mallela, and Modha}]{Dhillon2003}
Dhillon, I., Mallela, S., and Modha, D. (2003), \enquote{Information-theoretic
  co-clustering,} in \textit{Proceedings of the 9th ACM SIGKDD International
  Conference on Knowledge Discovery and Data Mining}, pp. 89--98.

\bibitem[{Dvoretzky et~al.(1956)Dvoretzky, Kiefer, and
  Wolfowitz}]{Dvoretzky1956}
Dvoretzky, A., Kiefer, J., and Wolfowitz, J. (1956), \enquote{Asymptotic
  minimax character of the sample distribution Function and of the Classical
  Multinomial Estimator,} \textit{Annals of Mathematical Statistics}, 27,
  642--669.

\bibitem[{Fan and Lin(1998)}]{Fan1998}
Fan, J. and Lin, S.-K. (1998), \enquote{Test of significance when data are
  curves,} \textit{Journal of the American Statistical Association}, 93,
  1007--1021.

\bibitem[{Faraway(1997)}]{Faraway1997}
Faraway, J.~J. (1997), \enquote{Regression Analysis for a Functional Response,}
  \textit{Technometrics}, 39, 254--261.

\bibitem[{Feng and Sun(2015)}]{Feng2015}
Feng, L. and Sun, F. (2015), \enquote{A note on high-dimensional two-sample
  test,} \textit{Statistics and Probability Letters}, 105, 29--36.

\bibitem[{Feng et~al.(2015)Feng, Zou, Wang, and Zhu}]{Feng2015a}
Feng, L., Zou, C., Wang, Z., and Zhu, L. (2015), \enquote{Two-sample
  {B}ehrens-{F}isher problem for high-dimensional data,} \textit{Statistica
  Sinica}, 25, 1297--1312.

\bibitem[{Ferraty and Vieu(2006)}]{Ferraty2006}
Ferraty, F. and Vieu, P. (2006), \textit{Nonparametric Functional Data
  Analysis: Theory and Practice}, New York: Springer-Verlag.

\bibitem[{Fujikoshi et~al.(2004)Fujikoshi, Himeno, and Wakaki}]{Fujikoshi2004}
Fujikoshi, Y., Himeno, T., and Wakaki, H. (2004), \enquote{Asymptotic Results
  of a High Dimensional {MANOVA} Test and Power Comparison When the Dimension
  is Large Compared to the Sample Size,} \textit{Journal of the Japan
  Statistical Society}, 34, 19--26.

\bibitem[{G\'orecki and Smaga(2015)}]{Gorecki2015}
G\'orecki, T. and Smaga, L. (2015), \enquote{A comparison of tests for the
  one-way {ANOVA} problem for functional data,} \textit{Computational
  Statistics}, 30, 987--1010.

\bibitem[{G\'orecki and Smaga(2019)}]{Gorecki2019}
--- (2019), \enquote{{fdANOVA}: an {R} software package for analysis of
  variance for univariate and multivariate functional data,}
  \textit{Computational Statistics}, 34, 571--597.

\bibitem[{Gregory et~al.(2015)Gregory, Carroll, Baladandayuthapani, and
  Lahiri}]{Gregory2015}
Gregory, K.~B., Carroll, R.~J., Baladandayuthapani, V., and Lahiri, S.~N.
  (2015), \enquote{A Two-Sample Test for Equality of Means in High Dimension,}
  \textit{Journal of the American Statistical Association}, 110, 837--849.

\bibitem[{Horv{\'a}th and Kokoszka(2012)}]{Horvath2012}
Horv{\'a}th, L. and Kokoszka, P. (2012), \textit{Inference for {F}unctional
  {D}ata with {A}pplications}, Springer Series in Statistics, Springer.

\bibitem[{Hotelling(1947)}]{Hotelling1947}
Hotelling, H. (1947), \enquote{Multivariate Quality Control Illustrated by Air
  Testing of Sample Bombsights,} in \textit{Techniques of Statistical
  Analysis}, McGraw Hill, New York, pp. 111--184.

\bibitem[{Hsing and Eubank(2015)}]{Hsing2015}
Hsing, T. and Eubank, R. (2015), \textit{Theoretical Foundations of Functional
  Data Analysis, with an Introduction to Linear Operators}, Wiley.

\bibitem[{Hu et~al.(2017)Hu, Bai, Wang, and Wang}]{Hu2017}
Hu, J., Bai, Z., Wang, C., and Wang, W. (2017), \enquote{On testing the
  equality of high dimensional mean vectors with unequal covariance matrices,}
  \textit{Annals of the Institute of Statistical Mathematics}, 69, 365--387.

\bibitem[{Inouye et~al.(2017)Inouye, Yang, Allen, and Ravikumar}]{Inouye2017}
Inouye, D.~I., Yang, E., Allen, G.~I., and Ravikumar, P. (2017), \enquote{A
  review of multivariate distributions for count data derived from the Poisson
  distribution,} \textit{WIREs Computational Statistics}, 9, e1398.

\bibitem[{Kokoszka and Reimherr(2017)}]{Kokoszka2017}
Kokoszka, P. and Reimherr, M. (2017), \textit{Introduction to Functional Data
  Analysis}, Chapman and Hall/CRC.

\bibitem[{Krovetz(1993)}]{Krovetz1993}
Krovetz, R. (1993), \enquote{Viewing morphology as an inference process,} in
  \textit{Proceedings of the 16th Annual International ACM SIGIR Conference on
  Research and Development in Information Retrieval}, ACM Press, pp. 191--202.

\bibitem[{Lawley(1938)}]{Lawley1938}
Lawley, D.~N. (1938), \enquote{A generalization of Fisher's z test,}
  \textit{Biometrika}, 30, 180--187.

\bibitem[{Li et~al.(2020)Li, Aue, and Paul}]{Li2020}
Li, H., Aue, A., and Paul, D. (2020), \enquote{High-dimensional general linear
  hypothesis tests via non-linear spectral shrinkage,} \textit{Bernoulli}, 26,
  2541--2571.

\bibitem[{Li et~al.(2017)Li, Hu, Bai, Yin, and Zou}]{Li2017}
Li, H., Hu, J., Bai, Z., Yin, Y., and Zou, K. (2017), \enquote{Test on the
  linear combinations of mean vectors in high-dimensional data,} \textit{Test},
  26, 188--208.

\bibitem[{Lopes et~al.(2011)Lopes, Jacob, and Wainwright}]{Lopes2011}
Lopes, M.~E., Jacob, L., and Wainwright, M.~J. (2011), \enquote{A more powerful
  two-sample test in high dimensions using random projection,} in
  \textit{Advances in Neural Information Processing Systems}, pp. 1206--1214.

\bibitem[{Lopes et~al.(2020)Lopes, Lin, and M\"uller}]{Lopes2020}
Lopes, M.~E., Lin, Z., and M\"uller, H.-G. (2020), \enquote{Bootstrapping max
  statistics in high dimensions: Near-parametric rates under weak variance
  decay and application to functional data analysis,} \textit{The Annals of
  Statistics}, 48, 1214--1229.

\bibitem[{Massart(1990)}]{Massart1990}
Massart, P. (1990), \enquote{The tight constant in the
  {D}voretzky--{K}iefer--{W}olfowitz inequality,} \textit{Annals of
  Probability}, 18, 1269--1283.

\bibitem[{Mrkvi\v{c}ka et~al.(2020)Mrkvi\v{c}ka, Myllym\"aki, J\'ilek, and
  Hahn}]{Mrkvicka2020}
Mrkvi\v{c}ka, T., Myllym\"aki, M., J\'ilek, M., and Hahn, U. (2020), \enquote{A
  one-way {ANOVA} test for functional data with graphical interpretation,}
  \textit{Kybernetika}, 56, 432--458.

\bibitem[{Paparoditis and Sapatinas(2016)}]{Paparoditis2016}
Paparoditis, E. and Sapatinas, T. (2016), \enquote{Bootstrap-based testing of
  equality of mean functions or equality of covariance operators for functional
  data,} \textit{Biometrika}, 103, 727--733.

\bibitem[{Ramsay and Silverman(2005)}]{Ramsay2005}
Ramsay, J.~O. and Silverman, B.~W. (2005), \textit{Functional Data Analysis},
  Springer Series in Statistics, New York: Springer, 2nd ed.

\bibitem[{Schott(2007)}]{Schott2007}
Schott, J.~R. (2007), \enquote{Some high-dimensional tests for a one-way
  {MANOVA},} \textit{Journal of Multivariate Analysis}, 98, 1825--1839.

\bibitem[{Shen and Faraway(2004)}]{Shen2004}
Shen, Q. and Faraway, J. (2004), \enquote{An {F} test for linear models with
  functional responses,} \textit{Statistica Sinica}, 14, 1239--1257.

\bibitem[{Srivastava and Fujikoshi(2006)}]{SRIVASTAVA2006}
Srivastava, M.~S. and Fujikoshi, Y. (2006), \enquote{Multivariate analysis of
  variance with fewer observations than the dimension,} \textit{Journal of
  Multivariate Analysis}, 97, 1927 -- 1940.

\bibitem[{Srivastava and Kubokawa(2013)}]{SRIVASTAVA2013}
Srivastava, M.~S. and Kubokawa, T. (2013), \enquote{Tests for multivariate
  analysis of variance in high dimension under non-normality,} \textit{Journal
  of Multivariate Analysis}, 115, 204--216.

\bibitem[{St\"adler and Mukherjee(2016)}]{Staedler2016}
St\"adler, N. and Mukherjee, S. (2016), \enquote{Two-sample testing in high
  dimensions,} \textit{Journal of Roayl Statistical Society. Series B:
  Statistical Methodology}, 79, 225--246.

\bibitem[{Thulin(2014)}]{Thulin2014}
Thulin, M. (2014), \enquote{A high-dimensional two-sample test for the mean
  using random subspaces,} \textit{Computational Statistics and Data Analysis},
  74, 26--38.

\bibitem[{Wang et~al.(2016)Wang, Chiou, and M\"{u}ller}]{WangCM16}
Wang, J.-L., Chiou, J.-M., and M\"{u}ller, H.-G. (2016), \enquote{Functional
  data analysis,} \textit{Annual Review of Statistics and Its Application}, 3,
  257--295.

\bibitem[{Xu et~al.(2016)Xu, Lin, Wei, and Pan}]{Xu2016}
Xu, G., Lin, L., Wei, P., and Pan, W. (2016), \enquote{An adaptive two-sample
  test for highdimensional means,} \textit{Biometrika}, 103, 609--624.

\bibitem[{Xue and Yao(2020)}]{Xue2020}
Xue, K. and Yao, F. (2020), \enquote{Distribution and correlation free
  two-sample test of high-dimensional means,} \textit{The Annals of
  Statistics}.

\bibitem[{Yamada and Himeno(2015)}]{YAMADA2015}
Yamada, T. and Himeno, T. (2015), \enquote{Testing homogeneity of mean vectors
  under heteroscedasticity in high-dimension,} \textit{Journal of Multivariate
  Analysis}, 139, 7 -- 27.

\bibitem[{Yamada and Srivastava(2012)}]{Yamada2012}
Yamada, T. and Srivastava, M.~S. (2012), \enquote{A Test for Multivariate
  Analysis of Variance in High Dimension,} \textit{Communications in Statistics
  - Theory and Methods}, 41, 2602--2615.

\bibitem[{Zhang and Pan(2016)}]{ZHANG2016a}
Zhang, J. and Pan, M. (2016), \enquote{A high-dimension two-sample test for the
  mean using cluster subspaces,} \textit{Computational Statistics \& Data
  Analysis}, 97, 87 -- 97.

\bibitem[{Zhang(2011)}]{Zhang2011}
Zhang, J.-T. (2011), \enquote{Statistical inferences for linear models with
  functional responses,} \textit{Statistica Sinica}, 21, 1431--1451.

\bibitem[{Zhang(2013)}]{Zhang2013}
--- (2013), \textit{Analysis of {V}ariance for {F}unctional {D}ata}, London:
  Chapman \& Hall.

\bibitem[{Zhang and Chen(2007)}]{Zhang2007}
Zhang, J.-T. and Chen, J. (2007), \enquote{Statistical inferences for
  functional data,} \textit{The Annals of Statistics}, 35, 1052--1079.

\bibitem[{Zhang et~al.(2019{\natexlab{a}})Zhang, Cheng, Wu, and
  Zhou}]{Zhang2019}
Zhang, J.-T., Cheng, M.-Y., Wu, H.-T., and Zhou, B. (2019{\natexlab{a}}),
  \enquote{A new test for functional one-way {ANOVA} with applications to
  ischemic heart screening,} \textit{Computational Statistics \& Data
  Analysis}, 132, 3--17.

\bibitem[{Zhang et~al.(2017)Zhang, Guo, and Zhou}]{ZHANG2017}
Zhang, J.-T., Guo, J., and Zhou, B. (2017), \enquote{Linear hypothesis testing
  in high-dimensional one-way MANOVA,} \textit{Journal of Multivariate
  Analysis}, 155, 200 -- 216.

\bibitem[{Zhang et~al.(2019{\natexlab{b}})Zhang, Guo, Zhou, and
  Cheng}]{Zhang2019a}
Zhang, J.-T., Guo, J., Zhou, B., and Cheng, M.-Y. (2019{\natexlab{b}}),
  \enquote{A Simple Two-Sample Test in High Dimensions Based on $L^2$-Norm,}
  \textit{Journal of the American Statistical Association}.

\bibitem[{Zhang and Liang(2014)}]{Zhang2014}
Zhang, J.-T. and Liang, X. (2014), \enquote{One-Way {ANOVA} for Functional Data
  via Globalizing the Pointwise {F}-test,} \textit{Scandinavian Journal of
  Statistics}, 41, 51--71.

\bibitem[{Zhang and Xu(2009)}]{Zhang2009}
Zhang, J.-T. and Xu, J. (2009), \enquote{On the $k$-sample {B}ehrens-{F}isher
  problem for high-dimensional data,} \textit{Science in China, Series A:
  Mathematics}, 52, 1285--1304.

\bibitem[{Zhang et~al.(2018)Zhang, Zhou, He, , and Liu}]{Zhang2018b}
Zhang, M., Zhou, C., He, Y., , and Liu, B. (2018), \enquote{Data-adaptive test
  for high-dimensional multivariate analysis of variance problem,}
  \textit{Australian \& New Zealand Journal of Statistics}, 60, 447--470.

\bibitem[{Zhou et~al.(2017)Zhou, Guo, and Zhang}]{Zhou2017}
Zhou, B., Guo, J., and Zhang, J.-T. (2017), \enquote{High-dimensional general
  linear hypothesis testing under heteroscedasticity,} \textit{Journal of
  Statistical Planning and Inference}, 188, 36--54.

\bibitem[{Zipf(1949)}]{Zipf1949}
Zipf, G.~K. (1949), \textit{Human Behavior and the Principle of Least Effort:
  An Introduction to Human Ecology}, Addison-Wesley Press.

\end{thebibliography}



\section*{Supplementary material}

# Supplementary Materials to "High-dimensional MANOVA via Bootstrapping and its Application to Functional and Sparse Count Data"

Zhenhua Lin[*]

Department of Statistics and Applied Probability, National University of Singapore

Miles E. Lopes[†]

Department of Statistics, University of California, Davis

Hans-Georg Müller[‡]

Department of Statistics, University of California, Davis

*General remarks and notation.* Throughout we refer to the notations introduced in the main text and define $m_\circ = \min\{m_1, \ldots, m_K\}$, $m_{\max} = \max\{m_1, \ldots, m_K\}$, $\ell_\circ = \min\{\ell_1, \ldots, \ell_K\}$, and $\ell_{\max} = \max\{\ell_1, \ldots, \ell_K\}$. Let $\mathcal{J}_{k,l}(m_k, m_l) = \mathcal{J}_k(m_k) \cup \mathcal{J}_l(m_l)$ and $m_\circ \leq m_{k,l} := |\mathcal{J}_{k,l}(m_k, m_l)| \leq m_k + m_l \leq 2m_{\max}$. Define $\ell_{k,l}$ analogously. Define $\lambda_{k,l}^2 = n_l/(n_k + n_l)$ and

$$M_m(k,l) = \max_{j \in \mathcal{J}_{k,l}(m_k, m_l)} \left( \lambda_{k,l} S_{k,j}/\sigma_{k,l,j}^\tau - \lambda_{l,k} S_{l,j}/\sigma_{k,l,j}^\tau \right),$$

and define $\tilde{M}_m(k,l)$ and $M_m^\star(k,l)$ analogously. Let $N = |\mathcal{P}|$ and suppose we enumerate the pairs in $\mathcal{P}$ by $(k_1, l_1), \ldots, (k_N, l_N)$. Let $\mathbf{m} = (m_{k_1,l_1}, \ldots, m_{k_N,l_N})$. Define

$$M_{\mathbf{m}} = \max_{(k,l) \in \mathcal{P}} M_m(k,l),$$

and $\tilde{M}_{\mathbf{m}}$ and $M_{\mathbf{m}}^\star$ analogously. In addition, define

$$\kappa = \alpha(1 - \tau).$$

---

1

Lastly, the constant $c > 0$ used in the proofs below may vary from place to place; however, it does not depend on $K, N, p$ or $n_1, \ldots, n_K$.

*Remark.* Under Assumption 3, all $\ell_k$ and thus all $\ell_{k,l}$ are of the same order as $\ell_\circ$ and $\ell_{\max}$, and similarly, all $m_k$ and $m_{k,l}$ are of the same order as $m_\circ$ and $m_{\max}$.

*Remark.* It is sufficient to show that the results in the theorems hold for all large values of $n$. The proofs below implicitly assume $p > \min_{1 \le q \le N} m_{k_q, l_q}$ (unless otherwise stated). If $p \le \min_{1 \le q \le N} m_{k_q, l_q}$, then $p = m_{k_q, l_q}$ for all $1 \le q \le N$, and consequently, the quantities I and III in the proof of Theorem 3.1, $\mathrm{I}'(X)$ and $\mathrm{III}'(X)$ in the proof of Theorem 3.2, and $\mathrm{I}''$ and $\mathrm{III}''$ in the proof of Theorem 3.4 become exactly zero. In this case, the related proofs are simplified to bounding II, $\mathrm{II}'(X)$ and $\mathrm{II}''$.

## A Proof of Theorem 3.1

*Proof.* Consider the inequality
$$d_\mathrm{K}\left(\mathcal{L}(M), \mathcal{L}(\tilde{M})\right) \le \mathrm{I} + \mathrm{II} + \mathrm{III},$$
where we define
$$\mathrm{I} = d_\mathrm{K}\left(\mathcal{L}(M), \mathcal{L}(M_\mathbf{m})\right),$$
$$\mathrm{II} = d_\mathrm{K}\left(\mathcal{L}(M_\mathbf{m}), \mathcal{L}(\tilde{M}_\mathbf{m})\right),$$
$$\mathrm{III} = d_\mathrm{K}\left(\mathcal{L}(\tilde{M}_\mathbf{m}), \mathcal{L}(\tilde{M})\right).$$

Then the conclusion of the theorem follows from Propositions A.1 and A.2 below. □

**Proposition A.1.** *Under the conditions of Theorem 3.1, we have* $\mathrm{II} \lesssim n^{-\frac{1}{2}+\delta}$.

*Proof.* Let $\Pi$ denote the projection onto the coordinates indexed by $\mathcal{J} = \bigcup_{(k,l) \in \mathcal{P}} \mathcal{J}_{k,l}(m_k, m_l)$. Let $J = |\mathcal{J}|$. Define the $J \times J$ diagonal matrix $D_{k,l} = \mathrm{diag}(\sigma_{k,l,j} : j \in \mathcal{J})$. It follows that
$$M_m(k,l) = \max_{j \in \mathcal{I}(k,l)} e_j^\top D_{k,l}^{-\tau} \Pi(\lambda_{k,l} S_k - \lambda_{l,k} S_l),$$
where $e_j \in \mathbb{R}^J$ is the $j$th standard basis vector, and $\mathcal{I}(k,l)$ denotes the row indices involving $\mathcal{J}_{k,l}(m_k, m_l)$ in the projection $\Pi$. Let $\mathfrak{C}_{k,l}^\top = \lambda_{k,l} D_{k,l}^{-\tau} \Pi \Sigma_k^{1/2}$, which is of size $J \times p$.

Consider the QR decomposition $\Sigma_k^{1/2} \Pi^\top = Q_k V_k$ so that
$$\mathfrak{C}_{k,l} = Q_k V_k (\lambda_{k,l} D_{k,l}^{-\tau}) \equiv Q_k R_{k,l},$$
where the columns of $Q_k \in \mathbb{R}^{p \times J}$ are an orthonormal basis for the image of $\mathfrak{C}_{k,l}$ and $R_{k,l} \in \mathbb{R}^{J \times J}$. Define the

2

random vectors

$$\check{Z}_k = n_k^{-1/2} \sum_{i=1}^{n_k} Q_k^\top Z_{k,i}.$$

Then

$$D_{k,l}^{-\tau}\Pi(\lambda_{k,l}S_k - \lambda_{l,k}S_l) = R_{k,l}^\top \check{Z}_k - R_{l,k}^\top \check{Z}_l.$$

Let $R^\top$ be a $JN \times JK$ block matrix with $N \times K$ blocks of size $J \times J$ such that, for $q = 1, \ldots, N$ and $k = 1, \ldots, K$, the $(q,k)$-block is $R_{k_q,l_q}^\top$ if $k = k_q$, is $-R_{l_q,k_q}^\top$ if $k = l_q$, and is $\mathbf{0}$ otherwise. Then

$$\begin{pmatrix} D_{k_1,l_1}^{-\tau}\Pi(\lambda_{k_1,l_1}S_{k_1} - \lambda_{l_1,k_1}S_{l_1}) \\ D_{k_2,l_2}^{-\tau}\Pi(\lambda_{k_2,l_2}S_{k_2} - \lambda_{l_2,k_2}S_{l_2}) \\ \vdots \\ D_{k_N,l_N}^{-\tau}\Pi(\lambda_{k_N,l_N}S_{k_N} - \lambda_{l_N,k_N}S_{l_N}) \end{pmatrix} = R^\top \check{Z},$$

where $\check{Z}$ is the $(JK) \times 1$ column vector obtained by stacking the vectors $\check{Z}_1, \ldots, \check{Z}_K$.

It can be checked that for any fixed $t \in \mathbb{R}$, there exists a Borel convex set $\mathcal{A}_t \subset \mathbb{R}^r$, with $r = JK$, such that $\mathbb{P}(M_\mathbf{m} \leq t) = \mathbb{P}(\check{Z} \in \mathcal{A}_t)$. By the same reasoning, we also have $\mathbb{P}(\tilde{M}_\mathbf{m} \leq t) = \gamma_r(\mathcal{A}_t)$, where $\gamma_r$ is the standard Gaussian distribution on $\mathbb{R}^r$. Thus,

$$\text{II} \leq \sup_{\mathcal{A} \in \mathscr{A}} |\mathbb{P}(\check{Z} \in \mathcal{A}) - \gamma_r(\mathcal{A})|,$$

where $\mathscr{A}$ denotes the collection of all Borel convex subsets of $\mathbb{R}^r$.

Now we apply Theorem 1.2 of (Bentkus, 2005), as follows. Let $n_{1:k} = \sum_{j=1}^k n_j$. Define $Y_i \in \mathbb{R}^r$ in the following way: For $k = 1, \ldots, K$ and $i' = 1, \ldots, n_k$, set $i = n_{1:k} - n_k + i'$ and set all coordinates of $Y_i$ to zero except that $Y_{i,(Jk-J+1):(Jk)} = n_k^{-1/2} Q_k^\top Z_{k,i'}$, i.e., the subvector of $Y_i$ at coordinates $Jk - J + 1, \ldots, Jk$ is equal to the vector $n_k^{-1/2} Q_k^\top Z_{k,i'}$.

Then $\check{Z} = \sum_{i=1}^\mathbf{n} Y_i$, i.e., $\check{Z}$ is a sum of $\mathbf{n} = \sum_{k=1}^K n_k$ independent random vectors. We also observe that $\text{cov}(\check{Z}) = I_r$. For $n_{1:k} - n_k + 1 \leq i \leq n_{1:k}$, $\beta_i := \mathbb{E}\|\{\text{cov}(\check{Z})\}^{-1}Y_i\|^3 = \mathbb{E}\|Y_i\|^3 = n_k^{-3/2}\mathbb{E}\|Q_k^\top Z_{k,1}\|^3 \leq n_k^{-3/2}[\mathbb{E}(Z_{k,1}^\top Q_k Q_k^\top Z_{k,1})^2]^{3/4}$, where the inequality is due to Lyapunov's inequality. Let $v_j$ be the $j$th column of $Q_1$. If we put $\zeta_j = Z_{1,1}^\top v_j$, then

$$\mathbb{E}(Z_{1,1}^\top Q_1 Q_1^\top Z_{1,1})^2 = \left\|\sum_{j=1}^J \zeta_j^2\right\|_2^2 \leq \left(\sum_{j=1}^J \|\zeta_j^2\|_2\right)^2 \lesssim J^2,$$

where we used the fact that $\|Z_{1,1}^\top v_j\|_4^2 \leq c$ based on Assumption 1, where $c > 0$ is a constant depending only on $c_0$ of Assumption 1. The same argument applies to the quantity $\mathbb{E}(Z_{k,1}^\top Q_k Q_k^\top Z_{k,1})^2$ for a generic $k$ with the same constant $c$. This implies that $\beta_i \leq c n_k^{-3/2} J^{3/2}$ for all $n_{1:k} - n_k + 1 \leq i \leq n_{1:k}$, and some constant $c > 0$

not depending on $n$. Therefore,

$$\text{II} \lesssim J^{1/4} \sum_{i=1}^{n_1+\cdots+n_K} \beta_i \lesssim J^{7/4} \sum_{k=1}^{K} n_k^{-1/2} \lesssim N^{7/4} m_{\max}^{7/4} K n^{-1/2} \lesssim n^{-1/2+\delta},$$

where the third inequality is due to $J \le 2Nm_{\max}$, and the last one follows from $\max\{K, N\} \lesssim e^{\sqrt{\log n}} \lesssim n^\delta$ and $m_{\max} \lesssim n_{\max}^\delta \asymp n^\delta$ for any fixed $\delta > 0$. $\square$

**Proposition A.2.** *Under the conditions of Theorem 3.1, we have* $\text{I} \lesssim n^{-1/2+\delta}$ *and* $\text{III} \lesssim n^{-1/2+\delta}$.

*Proof.* We only establish the bound for I, since the same argument applies to III. For any fixed $t \in \mathbb{R}$,

$$|\mathbb{P}(M \le t) - \mathbb{P}(M_\mathbf{m} \le t)| = \mathbb{P}(A(t) \cap B(t)),$$

where

$$A(t) = \left\{ \max_{(k,l) \in \mathcal{P}} \max_{j \in \mathcal{J}_{k,l}(m_k, m_l)} (\lambda_{k,l} S_{k,j}/\sigma_{k,l,j}^\tau - \lambda_{l,k} S_{l,j}/\sigma_{k,l,j}^\tau) \le t \right\},$$

$$B(t) = \left\{ \max_{(k,l) \in \mathcal{P}} \max_{j \in \mathcal{J}_{k,l}^c(m_k, m_l)} (\lambda_{k,l} S_{k,j}/\sigma_{k,l,j}^\tau - \lambda_{l,k} S_{l,j}/\sigma_{k,l,j}^\tau) > t \right\},$$

and $\mathcal{J}_{k,l}^c(m_k, m_l)$ denotes the complement of $\mathcal{J}_{k,l}(m_k, m_l)$ in $\{1, \ldots, p\}$. Also, if $t_1 \le t_2$, it is seen that

$$A(t) \cap B(t) \subset A(t_2) \cup B(t_1)$$

for all $t \in \mathbb{R}$. By a union bound, we have

$$\text{I} \le \mathbb{P}(A(t_2)) + \mathbb{P}(B(t_1)).$$

Take

$$t_1 = cm_\circ^{-\kappa} \log n$$
$$t_2 = c_2 c_\circ \ell_{\max}^{-\kappa} \sqrt{\log \ell_{\max}}$$

for a certain constant $c > 0$, where we recall that $c_2 \in (0,1)$ is defined in Assumption 3. Then, $\mathbb{P}(A(t_2))$ and $\mathbb{P}(B(t_1))$ are at most of order $n^{-1/2+\delta}$, according to Lemma A.3 below. Moreover, the inequality $t_1 \le t_2$ holds for all large $n$, due to the definitions of $\ell_{\max}$, $m_\circ$, and $\kappa$, as well as the condition $(1-\tau)\sqrt{\log n} \gtrsim 1$. $\square$

**Lemma A.3.** *Under the conditions of Theorem 3.1, there is a positive constant $c$, not depending on $n$, that can be selected in the definition of $t_1$ and $t_2$, so that*

$$\mathbb{P}(A(t_2)) \lesssim n^{-\frac{1}{2}+\delta}, \tag{S1}$$

*and*
$$\mathbb{P}(B(t_1)) \lesssim n^{-1}. \tag{S2}$$

*Proof of* (S1). Let $\mathcal{I}_{k,l}$ be a subset of $\mathcal{J}_{k,l}(\ell_k, \ell_l)$ constructed in the following way: if $\mathcal{J}_k(\ell_k) \cap \mathcal{J}_l(\ell_l)$ contains at least $\ell_\circ/2$ elements, then $\mathcal{I}_{k,l} = \mathcal{J}_k(\ell_k) \cap \mathcal{J}_l(\ell_l)$, and otherwise, $\mathcal{I}_{k,l} = \mathcal{J}_k(\ell_k) \cap \mathcal{J}_l^c(\ell_l)$ when $\sigma_{k,(\ell_k)} \geq \sigma_{l,(\ell_l)}$ and $\mathcal{I}_{k,l} = \mathcal{J}_k^c(\ell_k) \cap \mathcal{J}_l(\ell_l)$ when $\sigma_{k,(\ell_k)} < \sigma_{l,(\ell_l)}$. According to Proposition A.1 and the fact that $\mathcal{I}_{k,l} \subset \mathcal{J}_{k,l}(m_k, m_l)$, we have

$$\begin{aligned}
\mathbb{P}(A(t_2)) &\leq \mathbb{P}\left(\max_{(k,l)\in\mathcal{P}} \max_{j\in\mathcal{J}_{k,l}(m_k,m_l)} (\lambda_{k,l}\tilde{S}_{k,j}/\sigma_{k,l,j}^\tau - \lambda_{l,k}\tilde{S}_{l,j}/\sigma_{k,l,j}^\tau) \leq t_2\right) + \text{II} \\
&\leq \mathbb{P}\left(\max_{(k,l)\in\mathcal{P}} \max_{j\in\mathcal{I}_{k,l}} (\lambda_{k,l}\tilde{S}_{k,j}/\sigma_{k,l,j}^\tau - \lambda_{l,k}\tilde{S}_{l,j}/\sigma_{k,l,j}^\tau) \leq t_2\right) + cn^{-\frac{1}{2}+\delta}.
\end{aligned}$$

As $\sigma_{k,l,j} = \sqrt{\lambda_{k,l}^2\sigma_{k,j}^2 + \lambda_{l,k}^2\sigma_{l,j}^2} \geq \lambda_{k,l}\sigma_{k,j}$ and $\sigma_{k,(j)} \geq c_\circ j^{-\alpha}$ for $j \in \{1,\ldots,m_k\}$, and due to Assumption 2 with $c_\circ \in (0,1)$ and Assumption 3 with $c_2 \in (0,1)$, we have $\sigma_{k,l,j}^{\tau-1} \leq \lambda_{k,l}^{\tau-1}\sigma_{k,j}^{\tau-1} \leq c_2^{\tau-1}\ell_k^{\alpha(1-\tau)}c_\circ^{\tau-1} \leq \ell_{\max}^\kappa/(c_2 c_\circ)$ for $j \in \mathcal{I}_{k,l}$. With an argument similar to that of Lemma B.1 of Lopes et al. (2020), we can show that

$$\begin{aligned}
&\mathbb{P}\left(\max_{(k,l)\in\mathcal{P}} \max_{j\in\mathcal{I}_{k,l}} (\lambda_{k,l}\tilde{S}_{k,j}/\sigma_{k,l,j}^\tau - \lambda_{l,k}\tilde{S}_{l,j}/\sigma_{k,l,j}^\tau) \leq t_2\right) \\
&\leq \mathbb{P}\left(\max_{(k,l)\in\mathcal{P}} \max_{j\in\mathcal{I}_{k,l}} (\lambda_{k,l}\tilde{S}_{k,j}/\sigma_{k,l,j} - \lambda_{l,k}\tilde{S}_{l,j}/\sigma_{k,l,j}) \leq \sqrt{\log \ell_{\max}}\right) \\
&\leq \sum_{(k,l)\in\mathcal{P}} \mathbb{P}\left(\max_{j\in\mathcal{I}_{k,l}} (\lambda_{k,l}\tilde{S}_{k,j}/\sigma_{k,l,j} - \lambda_{l,k}\tilde{S}_{l,j}/\sigma_{k,l,j}) \leq \sqrt{\log \ell_{\max}}\right).
\end{aligned}$$

Note that the cardinality of $\mathcal{I}_{k,l}$ is at least $\ell_\circ/2$. Based on Assumption 3, for all sufficiently large $n$, for all $1 \leq k < j \leq K$, we have $\log(\ell_{\max}) \leq 1.01 \log \ell_\circ \leq 1.01^2 \log(2|\mathcal{I}_{k,l}|) \leq 1.1^2 \log|\mathcal{I}_{k,l}|$. Then,

$$\begin{aligned}
&\mathbb{P}\left(\max_{j\in\mathcal{I}_{k,l}} (\lambda_{k,l}\tilde{S}_{k,j}/\sigma_{k,l,j} - \lambda_{l,k}\tilde{S}_{l,j}/\sigma_{k,l,j}) \leq \sqrt{\log \ell_{\max}}\right) \\
&\leq \mathbb{P}\left(\max_{j\in\mathcal{I}_{k,l}} (\lambda_{k,l}\tilde{S}_{k,j}/\sigma_{k,l,j} - \lambda_{l,k}\tilde{S}_{l,j}/\sigma_{k,l,j}) \leq 1.1\sqrt{\log |\mathcal{I}_{k,l}|}\right). \tag{S3}
\end{aligned}$$

To apply Lemma B.2 of Lopes et al. (2020), let $Q$ denote the correlation matrix of the random variables $\{\lambda_{k,l}\tilde{S}_{k,j}/\sigma_{k,l,j} - \lambda_{l,k}\tilde{S}_{l,j}/\sigma_{k,l,j} : 1 \leq j \leq p\}$. When $\mathcal{I}_{k,l} = \mathcal{J}_k(\ell_k) \cap \mathcal{J}_l(\ell_l)$, for $j, r \in \mathcal{I}_{k,l}$, one has

$$\begin{aligned}
Q_{j,r} &= \frac{\lambda_{k,l}^2 R_{k,j,r}(p)\sigma_{k,j}\sigma_{k,r} + \lambda_{l,k}^2 R_{l,j,r}(p)\sigma_{l,j}\sigma_{l,r}}{\sqrt{\lambda_{k,l}^2\sigma_{k,j}^2 + \lambda_{l,k}^2\sigma_{l,j}^2}\sqrt{\lambda_{k,l}^2\sigma_{k,r}^2 + \lambda_{l,k}^2\sigma_{l,r}^2}} \\
&\leq (1-\epsilon_0)\frac{\lambda_{k,l}^2 \sigma_{k,j}\sigma_{k,r} + \lambda_{l,k}^2\sigma_{l,j}\sigma_{l,r}}{\sqrt{\lambda_{k,l}^2\sigma_{k,j}^2 + \lambda_{l,k}^2\sigma_{l,j}^2}\sqrt{\lambda_{k,l}^2\sigma_{k,r}^2 + \lambda_{l,k}^2\sigma_{l,k}^2}} \\
&\leq 1-\epsilon_0,
\end{aligned}$$

since the construction of $\mathcal{I}_{k,l}$ implies that $\max\{R_{k,j,r}, R_{l,j,r}\} \leq 1-\epsilon_0$. When $\mathcal{I}_{k,l} = \mathcal{J}_k(\ell_k) \cap \mathcal{J}_l^c(\ell_l)$ (so that

$\sigma_{k,(\ell_k)} \geq \sigma_{l,(\ell_l)})$, we have

$$Q_{j,k} \leq 1 - \frac{\epsilon_0 \lambda_{k,l}^2 \sigma_{k,j} \sigma_{k,r}}{\sqrt{\lambda_{k,l}^2 \sigma_{k,j}^2 + \lambda_{l,k}^2 \sigma_{l,j}^2} \sqrt{\lambda_{k,l}^2 \sigma_{k,r}^2 + \lambda_{l,k}^2 \sigma_{l,r}^2}}$$

$$\leq 1 - \epsilon_0 \frac{\lambda_{k,l}^2}{\lambda_{k,l}^2 + \lambda_{l,k}^2}$$

$$\leq 1 - \epsilon_0,$$

where the first inequality is obtained by using $R_{k,j,r} \leq 1 - \epsilon_0$ for $j, r \in \mathcal{I}_{k,l}$ and the inequality $\lambda_{k,l}^2 \sigma_{k,j} \sigma_{k,r} + \lambda_{l,k}^2 \sigma_{l,j} \sigma_{l,r} \leq \sqrt{\lambda_{k,l}^2 \sigma_{k,j}^2 + \lambda_{l,k}^2 \sigma_{l,j}^2} \sqrt{\lambda_{k,l}^2 \sigma_{k,r}^2 + \lambda_{l,k}^2 \sigma_{l,r}^2}$ and the second is due to $\sigma_{k,j} \geq \sigma_{k,(\ell_k)} \geq \sigma_{l,(\ell_l)} \geq \sigma_{l,j}$ as $j \in \mathcal{J}_k(\ell_k)$ and $j \in \mathcal{J}_l^c(\ell_l)$. A similar argument shows that the inequality $Q_{j,k} \leq 1 - \epsilon_0$ also holds when $\sigma_{k,(\ell_k)} < \sigma_{l,(\ell_l)}$, in which $\mathcal{I}_{k,l} = \mathcal{J}_k^c(\ell_k) \cap \mathcal{J}_l(\ell_l)$.

To apply Lemma B.2 of Lopes et al. (2020), we note that the bound $\sqrt{\log |\mathcal{I}_{k,l}|}$ is required instead of $1.1\sqrt{\log |\mathcal{I}_{k,l}|}$. However, by carefully examining the proof of Lemma B.2 of Lopes et al. (2020), we find that the lemma is still valid for $1.1\sqrt{\log |\mathcal{I}_{k,l}|}$, potentially with constants different from $C$ and $\frac{1}{2}$ in (B.19) of Lopes et al. (2020). This shows that (S3) is bounded by $cn^{-1}$ for some constant $c$ not depending on $n$. Then $N \lesssim n^\delta$ for any $\delta > 0$ implies (S1). □

*Proof of* (S2). The following argument is similar to the proof for part (b) of Lemma B.1 in Lopes et al. (2020). Define the random variable

$$V = \max_{(k,l) \in \mathcal{P}} \max_{j \in \mathcal{J}_{k,l}^c(m_k, m_l)} (\lambda_{k,l} S_{k,j}/\sigma_{k,l,j}^\tau - \lambda_{l,k} S_{l,j}/\sigma_{k,l,j}^\tau)$$

and let $q = \max\{2\kappa^{-1}, 3, \log n\}$. To bound $\|V\|_q$, we observe that

$$\|V\|_q^q = \mathbb{E}\left[ \Big| \max_{(k,l) \in \mathcal{P}} \max_{j \in \mathcal{J}_{k,l}^c(m_k, m_l)} \lambda_{k,l} S_{k,j}/\sigma_{k,l,j}^\tau - \lambda_{l,k} S_{l,j}/\sigma_{k,l,j}^\tau \Big|^q \right]$$

$$\leq \sum_{(k,l) \in \mathcal{P}} \sum_{j \in \mathcal{J}_{k,l}^c(m_k, m_l)} \sigma_{k,l,j}^{q(1-\tau)} \mathbb{E} |\lambda_{k,l} S_{k,j}/\sigma_{k,l,j} - \lambda_{l,k} S_{l,j}/\sigma_{k,l,j}|^q.$$

Further, we have

$$\sum_{(k,l) \in \mathcal{P}} \sum_{j \in \mathcal{J}_{k,l}^c(m_k, m_l)} \sigma_{k,l,j}^{q(1-\tau)} \leq \sum_{(k,l) \in \mathcal{P}} \sum_{j \in \mathcal{J}_{k,l}^c(m_k, m_l)} \max\{\sigma_{k,j}, \sigma_{l,j}\}^{q(1-\tau)}$$

$$\leq \sum_{(k,l) \in \mathcal{P}} \sum_{j \in \mathcal{J}_{k,l}^c(m_k, m_l)} (\sigma_{k,j}^{q(1-\tau)} + \sigma_{l,j}^{q(1-\tau)})$$

$$\leq c_1^{q(1-\tau)} \sum_{(k,l) \in \mathcal{P}} \left( \sum_{j=m_k+1}^{p} j^{-\alpha q(1-\tau)} + \sum_{j=m_l+1}^{p} j^{-\alpha q(1-\tau)} \right)$$

$$\leq c_1^{q(1-\tau)} \sum_{(k,l) \in \mathcal{P}} \left( 2 \int_{m_\circ}^{p} x^{-q\kappa} dx \right)$$

6

$$\leq 2c_1^{q(1-\tau)} N \frac{m_\circ^{-q\kappa+1}}{q\kappa - 1}, \tag{S4}$$

where we recall $\kappa = \alpha(1-\tau)$, and note that $q\kappa \geq 2$. Then, with $\|\lambda_{k,l}S_{k,j}/\sigma_{k,l,j} - \lambda_{l,k}S_{l,j}/\sigma_{k,l,j}\|_q \leq cq$ according to Lemma E.3, we deduce that

$$\|V\|_q^q \leq 2c_1^{q(1-\tau)}(cq)^q N \frac{m_\circ^{-q\kappa+1}}{q\kappa - 1},$$

and with $C = \frac{c}{(q\kappa-1)^{1/q}} m_\circ^{1/q}(2N)^{1/q} \lesssim 1$ that

$$\|V\|_q \leq Cqm_\circ^{-\kappa}.$$

Also, the assumption that $(1-\tau)\sqrt{\log n} \gtrsim 1$ implies that $q \lesssim \log n$. Therefore, with $t = e\|V\|_q$ so that $t \leq cm_\circ^{-\kappa} \log n$ for some constant $c > 0$ not depending on $n$, by Chebyshev's inequality $\mathbb{P}(V \geq t) \leq t^{-q}\|V\|_q^q$, we obtain that

$$\mathbb{P}\left(V \geq cm_\circ^{-\kappa} \log n\right) \leq \mathbb{P}(V \geq t) \leq e^{-q} \leq n^{-1},$$

completing the proof. □

## B  Proof of Theorem 3.2

*Proof.* Consider the inequality

$$d_K\left(\mathcal{L}(\tilde{M}), \mathcal{L}(M^\star|X)\right) \leq \mathrm{I}' + \mathrm{II}'(X) + \mathrm{III}'(X),$$

where we define

$$\mathrm{I}' = d_K\left(\mathcal{L}(\tilde{M}), \mathcal{L}(\tilde{M}_\mathbf{m})\right),$$
$$\mathrm{II}'(X) = d_K\left(\mathcal{L}(\tilde{M}_\mathbf{m}), \mathcal{L}(M_\mathbf{m}^\star|X)\right),$$
$$\mathrm{III}'(X) = d_K\left(\mathcal{L}(M_\mathbf{m}^\star|X), \mathcal{L}(M^\star|X)\right).$$

The first term is equal to III in the proof of Theorem 3.1 and requires no further treatment. The second term is addressed in Proposition B.2.

To derive the bound for $\mathrm{III}'(X)$, we partially reuse the proof of Proposition A.2. For any real numbers $t_1' \leq t_2'$, the following bound holds

$$\mathrm{III}'(X) \leq \mathbb{P}(A'(t_2')|X) + \mathbb{P}(B'(t_1')|X),$$

7

where we define the following events for any $t \in \mathbb{R}$,

$$A'(t) = \left\{ \max_{(k,l) \in \mathcal{P}} \max_{j \in \mathcal{J}_{k,l}(m_k, m_l)} (\lambda_{k,l} S^\star_{k,j}/\hat{\sigma}^\tau_{k,l,j} - \lambda_{l,k} S^\star_{l,j}/\hat{\sigma}^\tau_{k,l,j}) \le t \right\},$$

$$B'(t) = \left\{ \max_{(k,l) \in \mathcal{P}} \max_{j \in \mathcal{J}^c_{k,l}(m_k, m_l)} (\lambda_{k,l} S^\star_{k,j}/\hat{\sigma}^\tau_{k,l,j} - \lambda_{l,k} S^\star_{l,j}/\hat{\sigma}^\tau_{k,l,j}) > t \right\}.$$

Lemma B.1 ensures that $t'_1$ and $t'_2$ can be chosen so that the random variables $\mathbb{P}(A'(t'_2)|X)$ and $\mathbb{P}(B'(t'_1)|X)$ are at most $cn^{-\frac{1}{2}+\delta}$ with probability at least $1 - cn^{-1}$. Under Assumption 2, it can be checked that the choices of $t'_1$ and $t'_2$ given in Lemma B.1 satisfy $t'_1 \le t'_2$ when $n$ (and hence all $n_k$) is sufficiently large. $\square$

**Lemma B.1.** *Under the conditions of Theorem 3.2, there are positive constants $c'_1$, $c'_2$, and $c$, not depending on $n$, for which the following statement is true: If $t'_1$ and $t'_2$ are chosen as*

$$t'_1 = c'_1 m_\circ^{-\kappa} \log^{3/2} n$$

$$t'_2 = c'_2 \ell_{\max}^{-\kappa} \sqrt{\log \ell_{\max}},$$

*then the events*

$$\mathbb{P}(A'(t'_2)|X) \le cn^{-\frac{1}{2}+\delta} \tag{S5}$$

*and*

$$\mathbb{P}(B'(t'_1)|X) \le n^{-1} \tag{S6}$$

*each hold with probability at least $1 - cn^{-1}$.*

*Proof.* By the triangle inequality and the definition of Kolmogorov distance,

$$\mathbb{P}(A'(t'_2)|X) \le \mathbb{P}\left( \max_{(k,l) \in \mathcal{P}} \max_{j \in \mathcal{J}_{k,l}(m_k, m_l)} (\lambda_{k,l} \tilde{S}_{k,j} - \lambda_{l,k} \tilde{S}_{l,j})/\sigma^\tau_{k,l,j} \le t'_2 \right) + \mathrm{II}'(X).$$

Taking $t'_2 = t_2$ as in the proof of Proposition A.2, the proof of Lemma A.3 shows that the first term is of order $n^{-1/2+\delta}$. Proposition B.2 shows that the second term is bounded by $cn^{-\frac{1}{2}+\delta}$ with probability at least $1 - cn^{-1}$ for some constant $c > 0$ not depending on $n$. This establishes (S5).

To deal with (S6), we define the random variable

$$V^\star = \max_{(k,l) \in \mathcal{P}} \max_{j \in \mathcal{J}^c_{k,l}(m_k, m_l)} (\lambda_{k,l} S^\star_{k,j}/\hat{\sigma}^\tau_{k,l,j} - \lambda_{l,k} S^\star_{l,j}/\hat{\sigma}^\tau_{k,l,j}),$$

and let $q = \max\{2\kappa^{-1}, 3, \log n\}$. We shall construct a function $b(\cdot)$ such that the following bound holds for every realization of $X$,

$$(\mathbb{E}[|V^\star|^q \mid X])^{1/q} \le b(X),$$

8

and then Chebyshev's inequality gives the following inequality for any number $b_n$ satisfying $b(X) \le b_n$,

$$\mathbb{P}(V^\star \ge eb_n \mid X) \le e^{-q} \le n^{-1}.$$

We will then find $b_n$ so that the event $\{b(X) \le b_n\}$ holds with high probability. Finally, we will see that $t_1' \asymp b_n$.

To construct $b$, we adopt the same argument of the proof of Lemma B.1(b) of Lopes et al. (2020) and show that for any realization of $X$,

$$\mathbb{E}(|V^\star|^q \mid X) \le \sum_{(k,l)\in\mathcal{P}} \sum_{j\in\mathcal{J}_{k,l}^c(m_k,m_l)} \hat{\sigma}_{k,l,j}^{q(1-\tau)} \mathbb{E}(|\lambda_{k,l}S_{k,j}^\star/\hat{\sigma}_{k,l,j}^\tau - \lambda_{l,k}S_{l,j}^\star/\hat{\sigma}_{k,l,j}^\tau|^q \mid X).$$

By Lemma E.3, for every $j \in \{1, \ldots, p\}$, the event

$$\mathbb{E}(|\lambda_{k,l}S_{k,j}^\star/\hat{\sigma}_{k,l,j}^\tau - \lambda_{l,k}S_{l,j}^\star/\hat{\sigma}_{k,l,j}^\tau|^q \mid X) \le (cq)^q$$

holds with probability 1. Consequently, if we set $s = q(1-\tau)$ and consider the random variable

$$\hat{\mathfrak{s}} = \left( \sum_{(k,l)\in\mathcal{P}} \sum_{j\in\mathcal{J}_{k,l}^c(m_k,m_l)} \hat{\sigma}_{k,l,j}^s \right)^{1/s},$$

as well as

$$b(X) = cq\hat{\mathfrak{s}}^{(1-\tau)},$$

we obtain the bound

$$[\mathbb{E}(|V^\star|^q \mid X)]^{1/q} \le b(X),$$

with probability 1. Now, Lemma E.2 implies that

$$\mathbb{P}\left(b(X) \ge q \frac{(c\sqrt{q})^{1-\tau}}{(q\kappa - 1)^{1/q}} m_\circ^{-\kappa+1/q}(2N)^{1/q}\right) \le e^{-q} \le n^{-1}$$

for some constant $c > 0$ not depending on $n$. By weakening this tail bound slightly, it can be simplified to

$$\mathbb{P}\left(b(X) \ge C'q^{3/2}m_\circ^{-\kappa}\right) \le n^{-1},$$

where $C' = cm_\circ^{1/q}(q\kappa - 1)^{-1/q}(2N)^{1/q}$. Since $C' \lesssim 1$ and $(1-\tau)\sqrt{\log n} \gtrsim 1$ gives $q \asymp \log n$, it follows that there is a constant $c_1'$ not depending on $n$, such that if $b_n = c_1' m_\circ^{-\kappa} \log^{3/2} n$, then $\mathbb{P}(b(X) \ge b_n) \le n^{-1}$, which completes the proof. □

**Proposition B.2.** *Under the conditions of Theorem 3.1, there is a constant $c > 0$, not depending on $n$, such*

9

*that the event*

$$\Pi'(X) \le cn^{-\frac{1}{2}+\delta}$$

*holds with probability at least* $1 - cn^{-1}$.

*Proof.* Define the random variable

$$\breve{M}_{\mathbf{m}}^\star = \max_{(k,l)\in\mathcal{P}} \max_{j\in\mathcal{J}_{k,l}(m_k,m_l)} (\lambda_{k,l}S_{k,j}^\star - \lambda_{l,k}S_{l,j}^\star)/\sigma_{k,l,j}^\tau \tag{S7}$$

and consider the triangle inequality

$$\Pi'(X) \le d_{\mathrm{K}}\left(\mathcal{L}(\tilde{M}_{\mathbf{m}}), \mathcal{L}(\breve{M}_{\mathbf{m}}^\star|X)\right) + d_{\mathrm{K}}\left(\mathcal{L}(\breve{M}_{\mathbf{m}}^\star|X), \mathcal{L}(M_{\mathbf{m}}^\star|X)\right). \tag{S8}$$

*Addressing the first term of (S8).* Let $S$ be the vector obtained by stacking column vectors $\lambda_{k,l}S_{k,j}^\star - \lambda_{l,k}S_{l,j}^\star$ for $(k,l) = (k_1,l_1),\ldots,(k_N,l_N)$. As in the proof of Proposition A.1, $\breve{M}_{\mathbf{m}}^\star$ can be expressed as coordinate-wise maximum of $\Pi_{\mathbf{m}} R^\top \zeta$ with $\zeta \sim N(0, \breve{\mathfrak{S}})$, where $\Pi_{\mathbf{m}}$ denotes the projection matrix onto the superindices $\mathcal{I} = \{(k,l,j) : k,l \in \mathcal{P}, j \in \mathcal{J}_{k,l}(m_k,m_l)\}$, $R$ is a matrix, and

$$\breve{\mathfrak{S}} = \begin{pmatrix} \Pi\hat{\Sigma}_1\Pi^\top & & & \\ & \Pi\hat{\Sigma}_2\Pi^\top & & \\ & & \ddots & \\ & & & \Pi\hat{\Sigma}_K\Pi^\top \end{pmatrix}$$

with $\Pi$ being defined in the proof of Proposition A.1. Similarly, $\tilde{M}_{\mathbf{m}}$ can be expressed as coordinate-wise maximum of $\Pi_{\mathbf{m}} R^\top \xi$, where $\xi \sim N(0,\mathfrak{S})$ with

$$\mathfrak{S} = \begin{pmatrix} \Pi\Sigma_1\Pi^\top & & & \\ & \Pi\Sigma_2\Pi^\top & & \\ & & \ddots & \\ & & & \Pi\Sigma_K\Pi^\top \end{pmatrix}.$$

For $\mathfrak{C}_k^\top = \Pi\Sigma_k^{1/2}$ consider the singular value decomposition

$$\mathfrak{C}_k = U_k \Lambda_k V_k^\top,$$

where $r_k \lesssim J \equiv |\mathcal{I}|$ denotes the rank of $\mathfrak{C}_k$. We may assume that $U_k \in \mathbb{R}^{p\times r_k}$ has orthonormal columns, $\Lambda_k \in \mathbb{R}^{r_k \times r_k}$ to be invertible, and $V_k^\top$ to have orthonormal rows. Define

$$W_k = n_k^{-1} \sum_{i=1}^{n_k} (Z_{k,i} - \bar{Z}_k)(Z_{k,i} - \bar{Z}_k)^\top,$$

10

where $\bar{Z}_k = n_k^{-1} \sum_{i=1}^{n_k} Z_{k,i}$, and

$$W = \begin{pmatrix} W_1 & & & \\ & W_2 & & \\ & & \ddots & \\ & & & W_K \end{pmatrix}.$$

Then $\mathfrak{S} = \mathfrak{C}^\top \mathfrak{C}$ and $\breve{\mathfrak{S}} = \mathfrak{C}^\top W \mathfrak{C}$ with

$$\mathfrak{C} = \begin{pmatrix} \mathfrak{C}_1 & & & \\ & \mathfrak{C}_2 & & \\ & & \ddots & \\ & & & \mathfrak{C}_K \end{pmatrix}.$$

Define $r_k$-dimensional vectors $\tilde{\xi}_k = V_k^\top \xi_k$ and $\tilde{\zeta}_k = V_k^\top \zeta_k$, where $\xi_k$ and $\zeta_k$ are respectively the subvectors of $\xi$ and $\zeta$ corresponding to the $k$th sample. It can be shown that the columns of $\Pi\hat{\Sigma}_k\Pi^\top$ and $\Pi\Sigma_k\Pi^\top$ span the same subspace of $\mathbb{R}^J$ with probability at least $1 - cn_k^{-2}$ (due to Lemma D.5 of Lopes et al. (2020) and noting that the probability bound there can be strengthened to $1 - cn^{-2}$). Therefore, the event $E = \{$the columns of $\mathfrak{S}$ and $\hat{\mathfrak{S}}$ span the same subspace$\}$ holds with probability at least $1 - c\sum_{k=1}^K n_k^{-2} \geq 1 - cn^{-1}$, and furthermore, conditionally on $E$, the random vector $\xi$ lies in the column-span of $V$, where

$$V = \begin{pmatrix} V_1 & & & \\ & V_2 & & \\ & & \ddots & \\ & & & V_K \end{pmatrix},$$

since $\breve{\mathfrak{S}} = V\Lambda(U^\top W U)\Lambda V^\top$ with

$$U = \begin{pmatrix} U_1 & & & \\ & U_2 & & \\ & & \ddots & \\ & & & U_K \end{pmatrix} \quad \text{and} \quad \Lambda = \begin{pmatrix} \Lambda_1 & & & \\ & \Lambda_2 & & \\ & & \ddots & \\ & & & \Lambda_K \end{pmatrix}.$$

The argument below is conditional on the event $E$.

Given $E$, the random vector $\xi$ lies in the column-span of $V$ almost surely, which means $V\tilde{\xi} = \xi$ almost surely. The same argument applies to $\zeta$ and $\tilde{\zeta}$. It follows that for any $t \in \mathbb{R}$, the events $\{\tilde{M}_{\mathbf{m}} \leq t\}$ and $\{\breve{M}_{\mathbf{m}}^\star \leq t\}$ can be expressed as $\{\tilde{\xi} \in \mathcal{A}_t\}$ and $\{\tilde{\zeta} \in \mathcal{A}_t\}$, respectively, for a convex set $\mathcal{A}_t$. Hence $d_K\left(\mathcal{L}(\tilde{M}_{\mathbf{m}}), \mathcal{L}(\breve{M}_{\mathbf{m}}^\star|X)\right)$ is upper-bounded by the total variation distance between $\mathcal{L}(\tilde{\xi})$ and $\mathcal{L}(\tilde{\zeta})$, and in turn, Pinsker's inequality implies that this is upper-bounded by $c\sqrt{d_{\mathrm{KL}}(\mathcal{L}(\tilde{\zeta}), \mathcal{L}(\tilde{\xi}))}$, where $c > 0$ is an absolute constant, and $d_{\mathrm{KL}}$ denotes the KL divergence. Since the random vectors $\tilde{\xi} \sim N(0, V^\top \mathfrak{S} V)$ and $\tilde{\zeta} \sim N(0, V^\top \breve{\mathfrak{S}} V)$ are Gaussian (conditional on $X$), the following exact formula is available if we let $H = (V^\top \mathfrak{S} V)^{1/2}$ (so that $H^\top H = V^\top \mathfrak{S} V$)

and $\tilde{C} = H^{-\top}(V^\top \check{\mathfrak{S}} V)H^{-1} - I_r$,

$$d_{\mathrm{KL}}(\mathcal{L}(\tilde{\zeta}), \mathcal{L}(\tilde{\xi})) = \frac{1}{2}\{\mathrm{tr}(\tilde{C}) - \log\det(\tilde{C} + I_r)\}$$
$$= \frac{1}{2}\sum_{j=1}^{r}\{\theta_j(\tilde{C}) - \log(\theta_j(\tilde{C}) + 1)\},$$

where $r = \sum_{k=1}^{K} r_k \le KJ$ and $\theta_j(\tilde{C})$ denotes the eigenvalues of $\tilde{C}$. Note that $\|\tilde{C}\|_{\mathrm{op}} \le cKn^{-1/2}J\log n_{\max}$ by utilizing Lemma D.5 of Lopes et al. (2020) and the diagonal block structure of $\tilde{C}$. Using the inequality $|x - \log(x+1)| \le x^2/(1+x)$ that holds for any $x \in (-1, \infty)$, as well as the condition $|\theta_j(\tilde{C})| \le \|\tilde{C}\|_{\mathrm{op}} \le cKn^{-1/2}J\log n_{\max} \le 1/2$ for sufficiently large $n$, we have

$$d_{\mathrm{KL}}(\mathcal{L}(\tilde{\zeta}), \mathcal{L}(\tilde{\xi})) \le cr\|\tilde{C}\|_{\mathrm{op}}^2 \le cKJ\left(Kn^{-1/2}J\log n_{\max}\right)^2,$$

for some absolute constant $c > 0$. Thus,

$$d_{\mathrm{K}}\left(\mathcal{L}(\tilde{M}_{\mathbf{m}}), \mathcal{L}(\check{M}_{\mathbf{m}}^\star | X)\right) \le cJ^{3/2}K^{3/2}n^{-1/2}\log n_{\max}$$

with probability at least $1 - cn^{-1}$. With $J \le Km_{\max}$ and observing

$$cJ^{3/2}K^{3/2}n^{-1/2}\log n_{\max} \lesssim K^3 m_{\max}^{3/2} n^{-1/2} \log n_{\max} \lesssim n^{-\frac{1}{2}+\delta},$$

the first term of (S8) is bounded by $cn^{-\frac{1}{2}+\delta}$ with probability at least $1 - cn^{-1}$.

*Addressing the second term of (S8).* We proceed by considering the general inequality

$$d_{\mathrm{K}}(\mathcal{L}(\xi), \mathcal{L}(\zeta)) \le \sup_{t\in\mathbb{R}}\mathbb{P}(|\zeta - t| \le \varepsilon) + \mathbb{P}(|\xi - \zeta| > \varepsilon),$$

which holds for any random variables $\xi$ and $\zeta$, and any real number $\varepsilon > 0$. We will let $\mathcal{L}(\check{M}_{\mathbf{m}}^\star | X)$ play the role of $\mathcal{L}(\xi)$, and $\mathcal{L}(M_{\mathbf{m}}^\star | X)$ play the role of $\mathcal{L}(\zeta)$. Thus we need to establish an anti-concentration inequality for $\mathcal{L}(M_{\mathbf{m}}^\star | X)$, as well as a coupling inequality for $M_{\mathbf{m}}^\star$ and $\check{M}_{\mathbf{m}}^\star$, conditionally on $X$.

For the coupling inequality, we put

$$\varepsilon = cn^{-1/2}\log^{5/2} n_{\max}$$

for a suitable constant $c > 0$ not depending on $n$. Then Lemma E.6 shows that the event

$$\mathbb{P}\left(|\check{M}_{\mathbf{m}}^\star - M_{\mathbf{m}}^\star| > \varepsilon \mid X\right) \le cn^{-1}$$

holds with probability at least $1 - cn^{-1}$.

12

For the anti-concentration inequality, we use Nazarov's inequality (Lemma G.2, Lopes et al., 2020). Let

$$\hat{\underline{\sigma}}_{\mathbf{m}} = \min_{(k,l)\in\mathcal{P}} \min_{j\in\mathcal{J}_{k,l}(m_k,m_l)} \hat{\sigma}_{k,l,j}.$$

Then Nazarov's inequality implies that the event

$$\sup_{t\in\mathbb{R}} \mathbb{P}(|M_{\mathbf{m}}^\star - t| \leq \varepsilon | X) \leq c\varepsilon \hat{\underline{\sigma}}_{\mathbf{m}}^{\tau-1} \sqrt{\log m} \leq c\varepsilon \hat{\underline{\sigma}}_{\mathbf{m}}^{\tau-1} \sqrt{\log(2Nm_{\max})}$$

holds with probability 1, where $m = \sum_{(k,l)\in\mathcal{P}} m_{k,l} \leq 2Nm_{\max}$. Meanwhile, we observe that

$$\begin{aligned}
\sigma_{k,l,j} &= \sqrt{\lambda_{k,l}^2 \sigma_{k,j}^2 + \lambda_{l,k}^2 \sigma_{l,j}^2} \geq \max\{\lambda_{k,l}\sigma_{k,j}, \lambda_{l,k}\sigma_{l,j}\} \\
&\geq c_2 \max\{\sigma_{k,j}, \sigma_{l,j}\} \geq c_2 c_\circ \max\{m_k^{-\alpha}, m_l^{-\alpha}\} \geq cm_{\max}^{-\alpha}
\end{aligned} \tag{S9}$$

for all $(k,l) \in \mathcal{P}$ and $j \in \mathcal{J}_{k,l}(m_k, m_l)$. Then, Lemma E.4 and Assumption 2 imply that the event

$$\hat{\underline{\sigma}}_{\mathbf{m}}^{\tau-1} \leq cm_{\max}^{\kappa}$$

holds with probability at least $1 - Nn^{-2} \geq 1 - cn^{-1}$. Given the above, we conclude that

$$\sup_{t\in\mathbb{R}} \mathbb{P}(|M_{\mathbf{m}}^\star - t| \leq \varepsilon \mid X) \leq cm_{\max}^\kappa \sqrt{\log(2Nm_{\max})} n^{-1/2} \log^{5/2} n_{\max} \leq cn^{-1/2+\delta}$$

holds with probability at least $1 - cn^{-1}$, which completes the proof. □

## C  Proof of Theorem 3.4

Define

$$\hat{M}_{\mathbf{m}} = \max_{(k,l)\in\mathcal{P}} \hat{M}_{m_{k,l}}(k,l), \tag{S10}$$

*Proof.* We first observe that

$$d_{\mathrm{K}}(\mathcal{L}(\hat{M}), \mathcal{L}(M)) \leq \mathrm{I}'' + \mathrm{II}'' + \mathrm{III}'',$$

where

$$\mathrm{I}'' = d_{\mathrm{K}}\left(\mathcal{L}(\hat{M}), \mathcal{L}(\hat{M}_{\mathbf{m}})\right),$$
$$\mathrm{II}'' = d_{\mathrm{K}}\left(\mathcal{L}(\hat{M}_{\mathbf{m}}), \mathcal{L}(M_{\mathbf{m}})\right),$$
$$\mathrm{III}'' = d_{\mathrm{K}}\left(\mathcal{L}(M_{\mathbf{m}}), \mathcal{L}(M)\right).$$

The last term III″ requires no further consideration, as it is equal to I in the proof of Theorem 3.1. The

13

second term is handled in Proposition C.1, while the first term is handled in Proposition C.2. □

**Proposition C.1.** *Let $\delta$ be as in Theorem 3.4. Under Assumptions 1–3, one has $\mathrm{II}'' \lesssim n^{-\frac{1}{2}+\delta}$.*

*Proof.* We again proceed by considering the general inequality

$$d_{\mathrm{K}}(\mathcal{L}(\xi), \mathcal{L}(\zeta)) \leq \sup_{t \in \mathbb{R}} \mathbb{P}(|\zeta - t| \leq \varepsilon) + \mathbb{P}(|\xi - \zeta| > \varepsilon),$$

which holds for any random variables $\xi$ and $\zeta$, and any real number $\varepsilon > 0$. We will let $\mathcal{L}(\hat{M}_{\mathbf{m}})$ play the role of $\mathcal{L}(\xi)$, and let $\mathcal{L}(M_{\mathbf{m}})$ play the role of $\mathcal{L}(\zeta)$. As before, we then need to establish an anti-concerntration inequality for $\mathcal{L}(M_{\mathbf{m}})$, as well as a coupling inequality for $\hat{M}_{\mathbf{m}}$ and $M_{\mathbf{m}}$.

For the coupling inequality, we put

$$\varepsilon = cn^{-1/2} \log^{5/2} n_{\max}$$

for a suitable constant $c$ not depending on $n$. Then Lemma E.7 shows that

$$\mathbb{P}\left(|\hat{M}_{\mathbf{m}} - M_{\mathbf{m}}| > \varepsilon\right) \lesssim n^{-1}.$$

For the anti-concentration inequality, we utilize $d_{\mathrm{K}}(\mathcal{L}(M_{\mathbf{m}}), \mathcal{L}(\tilde{M}_{\mathbf{m}})) \lesssim n^{-1/2+\delta}$, which was established in the proof of Theorem 3.1, whence

$$\sup_{t \in \mathbb{R}} \mathbb{P}(|M_{\mathbf{m}} - t| \leq \varepsilon) = \sup_{t \in \mathbb{R}} \{\mathbb{P}(M_{\mathbf{m}} \leq t + \varepsilon) - \mathbb{P}(M_{\mathbf{m}} \leq t - \varepsilon)\}$$

$$= \sup_{t \in \mathbb{R}} \{\mathbb{P}(\tilde{M}_{\mathbf{m}} \leq t + \varepsilon) - \mathbb{P}(\tilde{M}_{\mathbf{m}} \leq t - \varepsilon)\} + cn^{-1/2+\delta}.$$

Let

$$\underline{\sigma}_{\mathbf{m}} = \min_{(k,l) \in \mathcal{P}} \min_{j \in \mathcal{J}_{k,l}(m_k, m_l)} \sigma_{k,l,j}.$$

Then Nazarov's inequality implies that

$$\sup_{t \in \mathbb{R}} \mathbb{P}\left(|\tilde{M}_{\mathbf{m}} - t| \leq \varepsilon\right) \lesssim \varepsilon \underline{\sigma}_{\mathbf{m}}^{\tau-1} \sqrt{\log m} \lesssim \varepsilon \underline{\sigma}_{\mathbf{m}}^{\tau-1} \sqrt{\log(2Nm_{\max})} \lesssim \varepsilon m_{\max}^{\alpha(1-\tau)} \sqrt{\log(2Nm_{\max})},$$

where $m = \sum_{(k,l) \in \mathcal{P}} m_{k,l}$, and the last inequality is due to (S9). Given the above, we conclude that

$$\sup_{t \in \mathbb{R}} \mathbb{P}\left(|\tilde{M}_{\mathbf{m}} - t| \leq \varepsilon\right) \leq cn^{-1/2}(\log^{5/2} n_{\max}) m_{\max}^{\alpha(1-\tau)} \sqrt{\log(2Nm_{\max})} \leq cn^{-1/2+\delta}.$$

This completes the proof. □

**Proposition C.2.** *Under the conditions of Theorem 3.4, one has $\mathrm{I}'' \lesssim n^{-\frac{1}{2}+\delta}$.*

14

*Proof.* Define

$$A''(t) = \left\{\max_{(k,l)\in\mathcal{P}} \max_{j\in\mathcal{J}_{k,l}(m_k,m_l)} (\lambda_{k,l}S_{k,j}/\hat{\sigma}^\tau_{k,l,j} - \lambda_{l,k}S_{l,j}/\hat{\sigma}^\tau_{k,l,j}) \le t\right\},$$

$$B''(t) = \left\{\max_{(k,l)\in\mathcal{P}} \max_{j\in\mathcal{J}^c_{k,l}(m_k,m_l)} (\lambda_{k,l}S_{k,j}/\hat{\sigma}^\tau_{k,l,j} - \lambda_{l,k}S_{l,j}/\hat{\sigma}^\tau_{k,l,j}) > t\right\},$$

where $\mathcal{J}^c_{k,l}(m_k,m_l)$ denotes the complement of $\mathcal{J}_{k,l}(m_k,m_l)$ in $\{1,\ldots,p\}$. Also, if $t''_1 \le t''_2$, it is seen that

$$A''(t) \cap B''(t) \subset A''(t''_2) \cup B''(t''_1)$$

for all $t \in \mathbb{R}$. By a union bound, we have

$$\text{I}'' \le \mathbb{P}(A''(t''_2)) + \mathbb{P}(B''(t''_1)).$$

Setting

$$t''_1 = cm_\circ^{-\kappa} \log n$$
$$t''_2 = c_\circ \ell_{\max}^{-\kappa} \sqrt{\log \ell_{\max}}$$

for a constant $c > 0$, we proceed to show that $\mathbb{P}(A''(t''_2))$ and $\mathbb{P}(B''(t''_1))$ are bounded by $cn^{-1/2+\delta}$. We note the inequality $t''_1 \le t''_2$ holds for all large $n$, due to the definitions of $\ell_{\max}$, $m_\circ$, and $\kappa$, as well as the condition $(1-\tau)\sqrt{\log n} \gtrsim 1$. Specifically we will establish that

$$\mathbb{P}(A''(t''_2)) \lesssim n^{-\frac{1}{2}+\delta}, \tag{S11}$$

and

$$\mathbb{P}(B''(t''_1)) \lesssim n^{-1}. \tag{S12}$$

According to Propositions C.1 and A.1, we have

$$\mathbb{P}(A''(t''_2)) \le \mathbb{P}\left(\max_{(k,l)\in\mathcal{P}} \max_{j\in\mathcal{J}_{k,l}(m_k,m_l)} (\lambda_{k,l}S_{k,j}/\sigma^\tau_{k,l,j} - \lambda_{l,k}S_{l,j}/\sigma^\tau_{k,l,j}) \le t''_2\right) + \text{II}''$$

$$\le \mathbb{P}\left(\max_{(k,l)\in\mathcal{P}} \max_{j\in\mathcal{J}_{k,l}(m_k,m_l)} (\lambda_{k,l}\tilde{S}_{k,j}/\sigma^\tau_{k,l,j} - \lambda_{l,k}\tilde{S}_{l,j}/\sigma^\tau_{k,l,j}) \le t''_2\right) + \text{II} + \text{II}''$$

$$\le \mathbb{P}\left(\max_{(k,l)\in\mathcal{P}} \max_{j\in\mathcal{J}_{k,l}(m_k,m_l)} (\lambda_{k,l}\tilde{S}_{k,j}/\sigma^\tau_{k,l,j} - \lambda_{l,k}\tilde{S}_{l,j}/\sigma^\tau_{k,l,j}) \le t''_2\right) + cn^{-\frac{1}{2}+\delta}.$$

Then (S11) follows from a similar argument as given in the proof of Lemma A.3.

To derive (S12), consider

$$U = \max_{(k,l)\in\mathcal{P}} \max_{j\in\mathcal{J}^c_{k,l}(m_k,m_l)} \frac{\lambda_{k,l}S_{k,j} - \lambda_{l,k}S_{l,j}}{\hat{\sigma}^\tau_{k,l,j}}.$$

15

For $q = \max\{2\kappa^{-1}, 3, \log n\}$, we first observe that

$$\|U\|_q^q \leq \sum_{(k,l)\in\mathcal{P}, j\in\mathcal{J}_{k,l}^c(m_k,m_l)} \mathbb{E}\left|\frac{\lambda_{k,l}S_{k,j} - \lambda_{l,k}S_{l,j}}{\hat{\sigma}_{k,l,j}^\tau}\right|^q \leq \sum_{(k,l)\in\mathcal{P}, j\in\mathcal{J}_{k,l}^c(m_k,m_l)} \|V_{k,l,j}\|_{2q}^q \|Y_{k,l,j}\|_{2q}^q$$

with

$$V_{k,l,j} = \left|\frac{\sigma_{k,l,j}^\tau}{\hat{\sigma}_{k,l,j}^\tau}\right|,$$

$$Y_{k,l,j} = \left|\frac{\lambda_{k,l}S_{k,j} - \lambda_{l,k}S_{l,j}}{\sigma_{k,l,j}^\tau}\right|.$$

By Lemma E.10, we further have

$$\|U\|_q^q \leq c^q \sum_{(k,l)\in\mathcal{P}, j\in\mathcal{J}_{k,l}^c(m_k,m_l)} \|Y_{k,l,j}\|_{2q}^q$$

$$\leq c^q \sum_{(k,l)\in\mathcal{P}, j\in\mathcal{J}_{k,l}^c(m_k,m_l)} \left(\sigma_{k,l,j}^{2q(1-\tau)} \mathbb{E}\left|\frac{\lambda_{k,l}S_{k,j} - \lambda_{l,k}S_{l,j}}{\sigma_{k,l,j}}\right|^{2q}\right)^{1/2}$$

$$\leq c^q \sum_{(k,l)\in\mathcal{P}, j\in\mathcal{J}_{k,l}^c(m_k,m_l)} \sigma_{k,l,j}^{q(1-\tau)} \left(\mathbb{E}\left|\frac{\lambda_{k,l}S_{k,j} - \lambda_{l,k}S_{l,j}}{\sigma_{k,l,j}}\right|^{2q}\right)^{1/2}$$

$$\leq (cq)^q \sum_{(k,l)\in\mathcal{P}, j\in\mathcal{J}_{k,l}^c(m_k,m_l)} \sigma_{k,l,j}^{q(1-\tau)}$$

$$\leq (cq)^q c_1^{q(1-\tau)} N \frac{m_\circ^{-q\kappa+1}}{q\kappa - 1},$$

where the last inequality is due to (S4). If we put $C = \frac{c}{(q\kappa-1)^{1/q}} m_\circ^{1/q} N^{1/q} \lesssim 1$, then

$$\|U\|_q \leq Cqm_\circ^{-\kappa}.$$

Since $q \asymp \log n$, we have

$$\mathbb{P}\left(U \geq ecm_\circ^{-\kappa} \log n\right) \leq e^{-q} \leq \frac{1}{n},$$

as needed. $\square$

*Remark.* If Assumption 4 is replaced with the condition $n^{-1/2}\log^3 p \ll 1$, then (S12) can be established in the following way. With the same notations in the proof of Proposition C.2, we first observe that

$$U \leq \left(\max_{(k,l)\in\mathcal{P}} \max_{j\in\mathcal{J}_{k,l}^c(m_k,m_l)} \left|\frac{\sigma_{k,l,j}^\tau}{\hat{\sigma}_{k,l,j}^\tau}\right|\right) V$$

with

$$V = \max_{(k,l)\in\mathcal{P}} \max_{j\in\mathcal{J}_{k,l}^c(m_k,m_l)} \left|\frac{\lambda_{k,l}S_{k,j} - \lambda_{l,k}S_{l,j}}{\sigma_{k,l,j}^\tau}\right|.$$

16

Under the condition $n^{-1/2} \log^3 p \ll 1$,

$$\max_{(k,l)\in\mathcal{P}} \max_{j\in\mathcal{J}_{k,l}^c(m_k,m_l)} \left|\frac{\sigma_{k,l,j}^{\tau}}{\hat{\sigma}_{k,l,j}^{\tau}}\right| \asymp 1$$

with probability at least $1 - cNn^{-2}$, according to Lemma E.8. With the aid of Lemma E.3, the term $V$ then can be handled by an argument similar to the proof of Lemma A.3.

*Remark.* The above proofs relied on the condition $p > m_\circ$ and this implies $m_\circ \geq n^{\log^{-a} n}$ and $\ell_\circ \geq \log^3 n$. These conditions are used in the analysis of I and III, as well as I′, III′(X), I″ and III″. If $p \leq m_\circ$, then the definition of $m_\circ$ implies that $p = m_1 = \cdots = m_K$, and the quantities I, III, I′, III′(X), I″ and III″ become exactly 0. In this case, the proofs of Theorems 3.1, 3.2 and 3.4 reduce to bounding II, II′(X) and II″, and these arguments can be repeated as before.

## D  Proof of Theorem 3.3

*Proof.* Part (ii) is handled in Proposition D.1. Below we establish part (i).

Let $q_{M^*}$ be the quantile of the conditional probability distribution $\mathcal{L}(M^*|X)$. By Theorem 3.1 and 3.2, the event $E = \{d_K(\mathcal{L}(M), \mathcal{L}(M^*|X)) \leq ca_n\}$ holds with probability at least $1 - cn^{-1}$, where $a_n = n^{-1/2+\delta}$. Below we condition on the event $E$ and observe that $q_M(\varrho - ca_n) \leq q_{M^*}(\varrho) \leq q_M(\varrho + ca_n)$ conditional on $E$.

In the derivation of Proposition A.2, with the notation there, we have

$$\mathbb{P}(M \leq t) = \mathbb{P}\left(A(t) \cap B^c(t)\right) = P(A(t)) - \mathbb{P}\left(A(t) \cap B(t)\right)$$
$$\geq \mathbb{P}\left(A(t)\right) - \mathbb{P}\left(A(t_2)\right) - \mathbb{P}\left(B(t_1)\right)$$
$$\geq \mathbb{P}\left(A(t)\right) - cn^{-1/2+\delta}.$$

By an argument similar to Lemma A.3, one can show that if

$$V = \max_{(k,l)\in\mathcal{P}} \max_{j\in\mathcal{J}_{k,l}(m_k,m_l)} \frac{\lambda_{k,l} S_{k,j} - \lambda_{l,k} S_{l,j}}{\sigma_{k,l,j}^{\tau}},$$

then $\|V\|_Q^Q \leq (cQ)^Q N$ if we define $Q = \max\{2\kappa^{-1}, 3, \sqrt{\log n}\}$. Thus,

$$\mathbb{P}\left(A(t)\right) = 1 - \mathbb{P}(V > t) \geq 1 - \frac{\|V\|_Q^Q}{t^Q} \geq 1 - e^{-Q} \to 1$$

if $t \geq e\|V\|_Q$. Therefore, $q_M(\varrho + ca_n) \lesssim \|V\|_Q \lesssim \sqrt{\log n}$, otherwise $\mathbb{P}(M \leq t) \to 1 > \varrho$. Similar arguments show that $q_M(\varrho - ca_n) \lesssim \sqrt{\log n}$.

The above shows that $|q_{M^*}(\varrho)| \leq c\sqrt{\log n}$ with probability at least $1 - cn^{-1}$. Since conditional on the data $X$ the random variable $M^*$ is the maximum of a multivariate Gaussian distribution, according to

Nadarajah et al. (2019), the conditional distribution of $M^*$ has a probability density and thus its cumulative distribution function is continuous. This enables us to apply Theorem 2 of Xia (2019), with an appropriately chosen $t \asymp \sqrt{\log n}$ in that theorem, to conclude that $|\hat{q}_M(\varrho) - q_{M^*}(\varrho)| \leq c\sqrt{\log n}$ with probability at least $1 - e^{-2h^2 B} \geq 1 - cn^{-1}$ for all sufficiently large $n$ when the event $E$ holds, where $h > 0$ is a constant not depending on $n$. To complete the proof, we should verify that the quantities $|\varphi_F^+(t)|$ and $|\varphi_F^-(t)|$ defined in that theorem, with the chosen $t \asymp \sqrt{\log n}$, are bounded from below by $h > 0$ with probability at least $1 - cn^{-1}$ for all sufficiently large $n$, where $F$ is the cumulative distribution function of $M^*$ conditional on the data $X$. For this, set $h = \min\{(1-\varrho)/4, \varrho/4\} > 0$. Then replacing $\varrho$ with $\varrho + 2h$ in the inequality $|q_{M^*}(\varrho)| \leq c\sqrt{\log n}$, we observe that $|q_{M^*}(\varrho + 2h)| \leq c\sqrt{\log n}$ with probability at least $1 - cn^{-1}$. This means that the event $|q_{M^*}(\varrho + 2h) - q_{M^*}(\varrho)| \leq c\sqrt{\log n}$ occurs with probability at least $1 - cn^{-1}$, and when this event holds we are able to find an appropriate $t \asymp \sqrt{\log n}$ such that $\varphi_F^+(t) = \lfloor(1-\varrho)n\rfloor/n + F(q_{M^*}(\varrho) + t) - 1 = \lfloor(1-\varrho)n\rfloor/n + \varrho + 2h - 1 \geq h$ for all sufficiently large $n$. The claim for $|\varphi_F^-(t)|$ can be established in a similar fashion. $\square$

**Proposition D.1.** *Under Assumptions 1–3, for some constant $c > 0$ not depending on $n$, one has*

$$\mathbb{P}\left(\max_{(k,l)\in\mathcal{P}} \max_{1\leq j\leq p} \hat{\sigma}_{k,l,j}^2 < 2\sigma_{\max}^2\right) \geq 1 - cn^{-1},$$

*where $\sigma_{\max} = \max\{\sigma_{k,j} : 1 \leq j \leq p, 1 \leq k \leq K\}$.*

*Proof.* Define

$$A^\circ(t) = \left\{\max_{(k,l)\in\mathcal{P}} \max_{j\in\mathcal{J}_{k,l}(n,n)} \hat{\sigma}_{k,l,j}^2 > t\right\},$$
$$B^\circ(t) = \left\{\max_{(k,l)\in\mathcal{P}} \max_{j\in\mathcal{J}_{k,l}^c(n,n)} \hat{\sigma}_{k,l,j}^2 > t\right\},$$

where as before $\mathcal{J}_{k,l}^c(n,n)$ denotes the complement of $\mathcal{J}_{k,l}(n,n)$ in $\{1,\ldots,p\}$. With $t^\circ = 2\sigma_{\max}^2$ we will establish that

$$\mathbb{P}(A^\circ(t^\circ)) \lesssim n^{-1}, \tag{S13}$$

and when $\mathcal{J}_{k,l}^c(n,n) \neq \varnothing$ for some $(k,l)$ that

$$\mathbb{P}(B^\circ(t^\circ)) \lesssim n^{-1}. \tag{S14}$$

For (S13), we first observe that

$$\mathbb{P}(\hat{\sigma}_{k,l,j}^2 > t^\circ) \leq \mathbb{P}(|\hat{\sigma}_{k,l,j}^2 - \sigma_{k,l,j}^2| > t^\circ - \sigma_{\max}^2).$$

18

With the above inequality, by using Lemma E.5 and a union bound, we conclude that

$$\mathbb{P}\left(A^\circ(t^\circ)\right) \leq \sum_{(k,l)\in\mathcal{P}} \sum_{j\in\mathcal{J}_{k,l}(n,n)} \mathbb{P}(|\hat{\sigma}^2_{k,l,j} - \sigma^2_{k,l,j}| > t^\circ - \sigma^2_{\max})$$

$$\leq cNn \cdot n^{-3} \lesssim n^{-1}.$$

To derive (S14), consider

$$U = \max_{(k,l)\in\mathcal{P}} \max_{j\in\mathcal{J}^c_{k,l}(m_k,m_l)} \hat{\sigma}^2_{k,l,j}.$$

For $q = \max\{\alpha^{-1}, 3, \log n\}$, we first observe that

$$\|U\|^q_q \leq \sum_{(k,l)\in\mathcal{P},\, j\in\mathcal{J}^c_{k,l}(n,n)} \mathbb{E}\left|\hat{\sigma}^2_{k,l,j}\right|^q$$

By Lemma E.1, we further have

$$\|U\|^q_q \leq \sum_{(k,l)\in\mathcal{P},\, j\in\mathcal{J}^c_{k,l}(n,n)} \|\hat{\sigma}_{k,l,j}\|^{2q}_{2q}$$

$$\leq \sum_{(k,l)\in\mathcal{P},\, j\in\mathcal{J}^c_{k,l}(n,n)} (c\sigma_{k,l,j}\sqrt{2q})^{2q}$$

$$\leq c(cq)^q \sum_{(k,l)\in\mathcal{P},\, j\in\mathcal{J}^c_{k,l}(n,n)} \sigma^{2q}_{k,l,j}$$

$$\leq c(cq)^q N n^{-2q\alpha+1},$$

where the last inequality is derived in analogy to (S4), and this implies

$$\|U\|_q \lesssim q n^{-2\alpha+1/q} N^{1/q} \ll \sigma^2_{\max}.$$

Since $q \asymp \log n$, we have

$$\mathbb{P}(U > 2\sigma^2_{\max}) \leq \mathbb{P}\left(U \geq e\|U\|_q\right) \leq e^{-q} \leq \frac{1}{n}$$

for all sufficiently large $n$. $\square$

# E  Technical Lemmas

**Lemma E.1.** *Suppose the conditions of Theorem 3.1 hold. For any fixed $b > 0$, if $3 \leq q \leq \max\{2\kappa^{-1}, 3, \log^b n\}$, there exists a constant $c > 0$ not depending on $q$, $K$, $N$, $p$ or $n_1, \ldots, n_K$, such that for any $k, l \in \{1, \ldots, K\}$ and $j \in \{1, \ldots, p\}$, we have $\|\hat{\sigma}_{k,j}\|_q \leq c\sigma_{k,j}\sqrt{q}$ and $\|\hat{\sigma}_{k,l,j}\|_q \leq c\sigma_{k,l,j}\sqrt{q}$.*

*Proof.* According to Lemma D.1 of Lopes et al. (2020) (which still holds when $q = \log^b n \geq 3$), we have $\|\hat{\sigma}_{k,j}\|_q \leq c\sigma_{k,j}\sqrt{q}$. Therefore, due to $\hat{\sigma}_{k,l,j} = \sqrt{\lambda^2_{k,l}\hat{\sigma}^2_{k,j} + \lambda^2_{l,k}\hat{\sigma}^2_{l,j}} \leq \lambda_{k,l}\hat{\sigma}_{k,j} + \lambda_{l,k}\hat{\sigma}_{l,j}$, and using the fact that

19

$\|Y\|_q^2 = \|Y^2\|_{q/2}$ for any random variable $Y$, we deduce that

$$\|\hat{\sigma}_{k,l,j}\|_q^2 = \|\hat{\sigma}_{k,l,j}^2\|_{q/2} = \|\lambda_{k,l}^2\hat{\sigma}_{k,j}^2 + \lambda_{l,k}^2\hat{\sigma}_{l,j}^2\|_{q/2}$$

$$\leq \lambda_{k,l}^2\|\hat{\sigma}_{k,j}^2\|_{q/2} + \lambda_{l,k}^2\|\hat{\sigma}_{l,j}^2\|_{q/2} = \lambda_{k,l}^2\|\hat{\sigma}_{k,j}\|_q^2 + \lambda_{l,k}^2\|\hat{\sigma}_{k,j}\|_q^2$$

$$\leq c^2 q(\lambda_{k,l}^2\sigma_{k,j}^2 + \lambda_{l,k}^2\sigma_{l,j}^2) = c^2 q \sigma_{k,l,j}^2.$$

□

**Lemma E.2.** *Let $q = \max\{2\kappa^{-1}, 3, \log n\}$ and $s = q(1-\tau)$. Consider the random variables $\hat{\mathfrak{s}}$ and $\hat{\mathfrak{t}}$ defined by*

$$\hat{\mathfrak{s}} = \left(\sum_{(k,l)\in\mathcal{P}} \sum_{j\in\mathcal{J}_{k,l}^c(m_k,m_l)} \hat{\sigma}_{k,l,j}^s\right)^{1/s}$$

*and*

$$\hat{\mathfrak{t}} = \left(\sum_{(k,l)\in\mathcal{P}} \sum_{j\in\mathcal{J}_{k,l}(m_k,m_l)} \hat{\sigma}_{k,l,j}^s\right)^{1/s}.$$

*Under the conditions of Theorem 3.1, there is a constant $c > 0$, not depending on $q$, $K$, $N$, $p$ or $n_1, \ldots, n_K$, such that*

$$\mathbb{P}\left(\hat{\mathfrak{s}} \geq \frac{c\sqrt{q}}{(q\kappa - 1)^{1/s}} m_\circ^{-\alpha + 1/s}(2N)^{1/s}\right) \leq e^{-q} \tag{S15}$$

*and*

$$\mathbb{P}\left(\hat{\mathfrak{t}} \geq \frac{c\sqrt{q}}{(q\kappa - 1)^{1/s}}(2N)^{1/s}\right) \leq e^{-q}. \tag{S16}$$

*Proof.* Using Lemma E.1, this lemma follows from similar arguments as in the proof of Lemma D.2 in Lopes et al. (2020). For further details, consider

$$\|\hat{\mathfrak{s}}\|_q = \left\|\sum_{(k,l)\in\mathcal{P}} \sum_{j\in\mathcal{J}^c(m_k,m_l)} \hat{\sigma}_{k,l,j}^s\right\|_{q/s}^{1/s} \leq \left(\sum_{(k,l)\in\mathcal{P}} \sum_{j\in\mathcal{J}^c(m_k,m_l)} \|\hat{\sigma}_{k,l,j}^s\|_{q/s}\right)^{1/s}$$

$$= \left(\sum_{(k,l)\in\mathcal{P}} \sum_{j\in\mathcal{J}^c(m_k,m_l)} \|\hat{\sigma}_{k,l,j}\|_q^s\right)^{1/s} \leq c\sqrt{q}\left(\sum_{(k,l)\in\mathcal{P}} \sum_{j\in\mathcal{J}^c(m_k,m_l)} \sigma_{k,l,j}^s\right)^{1/s}$$

$$\leq c\sqrt{q}\left(\sum_{(k,l)\in\mathcal{P}} \sum_{j\in\mathcal{J}^c(m_k,m_l)} \max\{\sigma_{k,j}, \sigma_{l,j}\}^s\right)^{1/s} \leq c\sqrt{q}\left(\sum_{(k,l)\in\mathcal{P}} \sum_{j\in\mathcal{J}^c(m_k,m_l)} (\sigma_{k,j}^s + \sigma_{l,j}^s)\right)^{1/s}$$

$$\leq c\sqrt{q}\left(\sum_{(k,l)\in\mathcal{P}} \left\{\int_{m_k}^p x^{-s\alpha}dx + \int_{m_l}^p x^{-s\alpha}dx\right\}\right)^{1/s} \leq c\sqrt{q}\left(2N \int_{m_\circ}^p x^{-s\alpha}dx\right)^{1/s}$$

$$\leq c\sqrt{q}(2N)^{1/s}\frac{m_\circ^{-\alpha+1/s}}{(s\alpha - 1)^{1/s}},$$

where for the last step, we use $s\alpha = q\kappa > 1$. The proof for $\hat{\mathfrak{t}}$ can be obtained by the same argument, except that the bound becomes $\sum_{j\in\mathcal{J}_{k,l}(m_k,m_l)} \sigma_{k,j}^s \lesssim 1$. □

**Lemma E.3.** *Suppose the conditions of Theorem 3.1 hold, and for any fixed $b > 0$, let $q = \max\{2\kappa^{-1}, \log^b n, 3\}$. Then for a constant $c > 0$, not depending on $q$, $K$, $N$, $p$ or $n_1, \ldots, n_K$, such that for any $(k,l) \in \mathcal{P}$ and $j \in \{1, \ldots, p\}$, it holds that*

$$\left\| \frac{\lambda_{k,l} S_{k,j}}{\sigma_{k,l,j}} - \frac{\lambda_{l,k} S_{l,j}}{\sigma_{k,l,j}} \right\|_q \leq cq, \tag{S17}$$

*and the following event holds with probability 1,*

$$\left( \mathbb{E}\left[ \left| \frac{\lambda_{k,l} S_{k,j}^\star}{\hat{\sigma}_{k,l,j}} - \frac{\lambda_{l,k} S_{l,j}^\star}{\hat{\sigma}_{k,l,j}} \right|^q \mid X \right] \right)^{1/q} \leq cq. \tag{S18}$$

*Proof.* Without loss of generality, let $(k,l) = (1,2)$, and set $\lambda_1 = \lambda_{k,l}$, $\lambda_2 = \lambda_{l,k}$, and $\sigma_j = \sigma_{k,l,j}$. We reuse the notation $k$ for some index from $\{1,2\}$, i.e., $k \in \{1,2\}$ in what follows.

Since $q > 2$, by Minkowski's inequality and Lemma G.4 of Lopes et al. (2020), we have

$$\|\lambda_1 S_{1,j}/\sigma_j - \lambda_2 S_{2,j}/\sigma_j\|_q \leq \|\lambda_1 S_{1,j}/\sigma_j\|_q + \|\lambda_2 S_{2,j}/\sigma_j\|_q$$

$$\leq q \max\{\|\lambda_1 S_{1,j}/\sigma_j\|_2, \lambda_1 n_1^{-1/2+1/q} \|(X_{1,1,j} - \mu_{1,j})/\sigma_j\|_q\}$$

$$+ q \max\{\|\lambda_2 S_{2,j}/\sigma_j\|_2, \lambda_2 n_2^{-1/2+1/q} \|(X_{2,1,j} - \mu_{1,j})/\sigma_j\|_q\},$$

and furthermore

$$\|S_{k,j}\|_2^2 = \operatorname{var}(S_{k,j}) = \sigma_{k,j}^2.$$

Thus $\|\lambda_k S_{k,j}/\sigma_j\|_2 = \lambda_k \sigma_{k,j} \sigma_j^{-1} \leq 1$, where we note that $\lambda_k^2 \sigma_{k,j}^2 \sigma_j^{-2} = \lambda_k^2 \sigma_{k,j}^2 / (\lambda_1^2 \sigma_{1,j}^2 + \lambda_2^2 \sigma_{2,j}^2) \leq 1$. Also, if we define the vector $u_k = \sigma_{k,j}^{-1} \Sigma_k^{1/2} e_j$ in $\mathbb{R}^p$ for standard basis $e_1, \ldots, e_p$ in $\mathbb{R}^p$, which satisfies $\|u\|_2 = 1$, then

$$\lambda_k \|(X_{k,1,j} - \mu_{k,j})/\sigma_j\|_q = \lambda_k \sigma_{k,j} \sigma_j^{-1} \|(X_{k,1,j} - \mu_{k,j})/\sigma_{k,j}\|_q \leq \|Z_{k,1}^\top u\|_q \lesssim q,$$

proving (S17). Inequality (S18) follows from the same argument, conditioning on $X$. □

Define the correlation

$$\rho_{k,l,j,j'} = \frac{\Sigma_{k,l}(j,j')}{\sigma_{k,l,j} \sigma_{k,l,j'}},$$

and its sample version

$$\hat{\rho}_{k,l,j,j'} = \frac{\widehat{\Sigma}_{k,l}(j,j')}{\hat{\sigma}_{k,l,j} \hat{\sigma}_{k,l,j'}},$$

for any $j, j' \in \{1, \ldots, p\}$.

**Lemma E.4.** *Under Assumption 1 and 3, there is a constant $c > 0$, not depending on $n$, such that the following events*

$$\max_{j \in \mathcal{J}_{k,l}(m_k, m_l)} \left| \frac{\hat{\sigma}_{k,l,j}}{\sigma_{k,l,j}} - 1 \right| \leq ca_n,$$

21

$$\min_{j\in\mathcal{J}_{k,l}(m_k,m_l)} \hat\sigma_{k,l,j}^{1-\tau} \geq \left(\min_{j\in\mathcal{J}_{k,l}(m_k,m_l)} \sigma_{k,l,j}^{1-\tau}\right)(1-ca_n),$$

*and*

$$\max_{j,j'\in j\in\mathcal{J}_{k,l}(m_k,m_l)} |\hat\rho_{j,j'} - \rho_{j,j'}| \leq ca_n$$

*each hold with probability at least $1 - cn^{-2}$, where $a_n = n^{-1/2}\log n_{\max}$.*

*Proof.* These conclusions are direct consequences of Lemma E.5. $\square$

**Lemma E.5.** *Suppose Assumptions 1 and 3 hold, and fix any $1 \leq k < l \leq K$ and any two (possibly equal) indices $j, j' \in \{1,\ldots,p\}$. Then, for any number $\vartheta \geq 1$, there are positive constants $c$ and $c_1(\vartheta)$, not depending on $n$, such that the event*

$$\left|\frac{\widehat\Sigma_{k,l}(j,j')}{\sigma_{k,l,j}\sigma_{k,l,j'}} - \rho_{k,l,j,j'}\right| \leq c_1(\vartheta)n^{-1/2}\log n_{\max}$$

*holds with probability at least $1 - cn^{-\vartheta}$.*

*Proof.* It is equivalent to showing that

$$\left|\widehat\Sigma_{k,l}(j,j') - \Sigma_{k,l}(j,j')\right| \leq c_1(\vartheta)n^{-1/2}(\log n_{\max})\sigma_{k,l,j}\sigma_{k,l,j'}.$$

Furthermore

$$\begin{aligned}
\left|\widehat\Sigma_{k,l}(j,j') - \Sigma_{k,l}(j,j')\right| &= \left|\lambda_{k,l}^2\widehat\Sigma_k(j,j') - \lambda_{k,l}^2\Sigma_k(j,j') + \lambda_l^2\widehat\Sigma_l(j,j') - \lambda_{l,k}^2\Sigma_l(j,j')\right| \\
&\leq \lambda_{k,l}^2\left|\widehat\Sigma_k(j,j') - \Sigma_k(j,j')\right| + \lambda_{l,k}^2\left|\widehat\Sigma_l(j,j') - \Sigma_l(j,j')\right| \\
&\leq c_1(\vartheta)(n_k^{-1/2}\lambda_{k,l}^2\sigma_{k,j}\sigma_{k,j'}\log n_k + n_l^{-1/2}\lambda_{l,k}^2\sigma_{l,j}\sigma_{l,j'}\log n_l) \\
&\leq c_1(\vartheta)(\log n_{\max})n^{-1/2}(\lambda_{k,l}^2\sigma_{k,j}\sigma_{k,j'} + \lambda_{l,k}^2\sigma_{l,j}\sigma_{l,j'}),
\end{aligned}$$

with probability at least $1 - cn_k^{-\vartheta} - cn_l^{-\vartheta} \geq 1 - 2cn^{-\vartheta}$, where the second inequality is due to Lemma D.7 of Lopes et al. (2020). Now, by the Cauchy–Schwarz inequality,

$$2\sigma_{k,j}\sigma_{k,j'}\sigma_{l,j}\sigma_{l,j'} \leq \sigma_{k,j}^2\sigma_{l,j'}^2 + \sigma_{l,j}^2\sigma_{k,j'}^2,$$

and further

$$\begin{aligned}
\lambda_{k,l}^2\sigma_{k,j}\sigma_{k,j'} + \lambda_{l,k}^2\sigma_{l,j}\sigma_{l,j'} &= \sqrt{(\lambda_{k,l}^2\sigma_{k,j}\sigma_{k,j'} + \lambda_{l,k}^2\sigma_{l,j}\sigma_{l,j'})^2} \\
&\leq \sqrt{(\lambda_{k,l}^2\sigma_{k,j}^2 + \lambda_{l,k}^2\sigma_{l,j}^2)}\sqrt{(\lambda_{k,l}^2\sigma_{k,j'}^2 + \lambda_{l,k}^2\sigma_{l,j'}^2)} \\
&= \sigma_{k,l,j}\sigma_{k,l,j'},
\end{aligned}$$

which completes the proof. $\square$

*Remark.* In the above proof, we note that Lemma D.7 of Lopes et al. (2020) does not depend on Assumption 2 of Lopes et al. (2020).

**Lemma E.6.** *Under the conditions of Theorem 3.1, there is a constant $c > 0$, not depending on $n$, such that*

$$\mathbb{P}\left(|\breve{M}_{\mathbf{m}}^{\star} - M_{\mathbf{m}}^{\star}| > r_n | X\right) \le cn^{-1}$$

*holds with probability at least $1 - cn^{-1}$, where $\breve{M}_{\mathbf{m}}^{\star}$ is defined in (S7) and $r_n = cn^{-1/2} \log^{5/2} n_{\max}$.*

*Proof.* Using a similar argument as in the proof of Lemma D.8 of Lopes et al. (2020), we find that

$$|\breve{M}_{\mathbf{m}}^{\star} - M_{\mathbf{m}}^{\star}| \le \max_{(k,l)\in\mathcal{P}} \max_{j\in\mathcal{J}_{k,l}(m_k,m_l)} \left|\left(\frac{\hat{\sigma}_{k,l,j}}{\sigma_{k,l,j}}\right)^{\tau} - 1\right| \cdot \max_{(k,l)\in\mathcal{P}} \max_{j\in\mathcal{J}_{k,l}(m_k,m_l)} \left|\frac{S_{k,l,j}^{\star}}{\hat{\sigma}_{k,l,j}^{\tau}}\right|.$$

It follows from Lemma E.4 that the event

$$\max_{(k,l)\in\mathcal{P}} \max_{j\in\mathcal{J}_{k,l}(m_k,m_l)} \left|\left(\frac{\hat{\sigma}_{k,l,j}}{\sigma_{k,l,j}}\right)^{\tau} - 1\right| \le cn^{-1/2} \log n_{\max}$$

holds with probability at least $1 - cNn^{-2} \ge 1 - cn^{-1}$. Now consider

$$U^{\star} = \max_{(k,l)\in\mathcal{P}} \max_{j\in\mathcal{J}_{k,l}(m_k,m_l)} \left|\frac{S_{k,l,j}^{\star}}{\hat{\sigma}_{k,l,j}^{\tau}}\right|.$$

Showing that

$$\mathbb{P}\left(U^{\star} \ge c \log^{3/2} n_{\max} \mid X\right) \le cn^{-1}$$

holds with probability at least $1 - cn^{-1}$ will complete the proof.

Using Chebyshev's inequality with $q = \{2\kappa^{-1}, 3, \log n\}$ gives

$$\mathbb{P}\left(U^{\star} \ge e[\mathbb{E}(|U^{\star}|^q \mid X)]^{1/q} \mid X\right) \le e^{-q}.$$

Now it suffices to show that the event

$$[\mathbb{E}(|U^{\star}|^q \mid X)]^{1/q} \le c \log^{3/2} n_{\max}$$

holds with probability at least $1 - cn^{-1}$. This is done by repeating the argument in Lemma B.1 with the aid of (S16) from Lemma E.2. □

**Lemma E.7.** *Under the conditions of Theorem 3.4, for some constant $c > 0$, not depending on $n$, we have*

$$\mathbb{P}\left(|\hat{M}_{\mathbf{m}} - M_{\mathbf{m}}| > r_n\right) \le cn^{-1},$$

*where $\hat{M}_{\mathbf{m}}$ is defined in (S10) and $r_n = cn^{-1/2} \log^{5/2} n_{\max}$.*

*Proof.* A similar argument as in the proof of Lemma D.8 of Lopes et al. (2020) leads to

$$|\hat{M}_{\mathbf{m}} - M_{\mathbf{m}}| \leq \max_{(k,l)\in\mathcal{P}} \max_{j\in\mathcal{J}_{k,l}(m_k,m_l)} \left|\left(\frac{\sigma_{k,l,j}}{\hat{\sigma}_{k,l,j}}\right)^{\tau} - 1\right| \cdot \max_{(k,l)\in\mathcal{P}} \max_{j\in\mathcal{J}_{k,l}(m_k,m_l)} \left|\frac{\lambda_{k,l}S_{k,j} - \lambda_{l,k}S_{l,j}}{\sigma_{k,l,j}^{\tau}}\right|.$$

It follows from Lemma E.4 that the event

$$\max_{(k,l)\in\mathcal{P}} \max_{j\in\mathcal{J}_{k,l}(m_k,m_l)} \left|\left(\frac{\sigma_{k,l,j}}{\hat{\sigma}_{k,l,j}}\right)^{\tau} - 1\right| \leq cn^{-1/2} \log n_{\max}$$

holds with probability at least $1 - cNn^{-2} \geq 1 - cn^{-1}$. Now consider

$$U = \max_{(k,l)\in\mathcal{P}} \max_{j\in\mathcal{J}_{k,l}(m_k,m_l)} \left|\frac{\lambda_{k,l}S_{k,j} - \lambda_{l,k}S_{l,j}}{\sigma_{k,l,j}^{\tau}}\right|.$$

Then

$$\mathbb{P}\left(U \geq c \log^{3/2} n_{\max}\right) \leq cn^{-1}$$

will complete the proof.

Using Chebyshev's inequality with $q = \max\{2\kappa^{-1}, 3, \log n\}$ gives

$$\mathbb{P}\left(U \geq e(\mathbb{E}|U|^q)^{1/q}\right) \leq e^{-q}.$$

Now it suffices to show that

$$\|U\|_q = (\mathbb{E}|U|^q)^{1/q} \lesssim \log^{3/2} n_{\max}.$$

Observe that

$$\|U\|_q^q \leq \sum_{(k,l)\in\mathcal{P}, j\in\mathcal{J}_{k,l}(m_k,m_l)} \sigma_{k,l,j}^{q(1-\tau)} \mathbb{E}|\sigma_{k,l,j}^{-1}(\lambda_{k,l}S_{k,j} - \lambda_{l,k}S_{l,j})|^q.$$

By Lemma E.3, and noting that $q\alpha(1-\tau) = q\kappa \geq 2$, we further have

$$\|U\|_q^q \leq (cq)^q \sum_{(k,l)\in\mathcal{P}, j\in\mathcal{J}_{k,l}(m_k,m_l)} \sigma_{k,l,j}^{q(1-\tau)} \lesssim N(cq)^q,$$

or equivalently,

$$\|U\|_q \lesssim qN^{1/q} \lesssim \log^{3/2} n_{\max},$$

where we use the fact that $N^{1/q} \lesssim 1$ given the choice of $q$. □

Define the correlation

$$\rho_{k,j,j'} = \frac{\Sigma_k(j,j')}{\sigma_{k,j}\sigma_{k,j'}},$$

and its sample version

$$\hat{\rho}_{k,j,j'} = \frac{\widehat{\Sigma}_k(j,j')}{\hat{\sigma}_{k,j}\hat{\sigma}_{k,j'}},$$

for any $j, j' \in \{1, \ldots, p\}$.

**Lemma E.8.** *Under Assumption 1 and 3, if $n^{-1/2}\log^3 p \ll 1$, then for any number $\theta \geq 2$, there are positive constants $c$ and $c_\theta$, not depending on $n$, such that the event*

$$\sup_{1 \leq k \leq K} \sup_{1 \leq j,j' \leq p} \left| \frac{\widehat{\Sigma}_k(j,j')}{\sigma_{k,j}\sigma_{k,j'}} - \rho_{k,j,j'} \right| \leq c_\theta (\log n_{\max} + \log^3 p) n^{-1/2}$$

*holds with probability at least $1 - cKn^{-\theta}$.*

*Proof.* It suffices to show that

$$\sup_{1 \leq j,j' \leq p} \left| \frac{\widehat{\Sigma}_k(j,j')}{\sigma_{k,j}\sigma_{k,j'}} - \rho_{k,j,j'} \right| \leq \frac{c_\theta (\log n_k + \log^3 p)}{\sqrt{n_k}}$$

with probability at least $1 - cn_k^\theta$. Consider $\ell_2$-unit vectors $u = \Sigma_k^{1/2} e_j \sigma_{k,j}^{-1}$ and $v = \Sigma_k^{1/2} e_{j'} \sigma_{k,j'}^{-1}$ in $\mathbb{R}^p$. Define

$$W_k = n_k^{-1} \sum_{i=1}^{n_k} (Z_{k,i} - \bar{Z}_k)(Z_{k,i} - \bar{Z}_k)^\top,$$

where $\bar{Z}_k = \sum_{i=1}^{n_k} Z_{k,i}$. Observe that

$$\frac{\widehat{\Sigma}_k(j,j')}{\sigma_{k,j}\sigma_{k,j'}} - \rho_{k,j,j'} = u^\top (W_k - I_p) v. \tag{S19}$$

For each $1 \leq i \leq n_k$, define the random variable $\zeta_{i,u} = Z_{k,i}^\top u$ and $\zeta_{i,v} = Z_{k,i}^\top v$. In this notation, the relation (S19) becomes

$$\frac{\widehat{\Sigma}_k(j,j')}{\sigma_{k,j}\sigma_{k,j'}} - \rho_{k,j,j'} = \Delta(u,v) + \Delta'(u,v)$$

where

$$\Delta(u,v) = \frac{1}{n_k} \sum_{i=1}^{n_k} \zeta_{i,u}\zeta_{i,v} - u^\top v,$$

$$\Delta'(u,v) = \left( \frac{1}{n_k} \sum_{i=1}^{n_k} \zeta_{i,u} \right) \left( \frac{1}{n_k} \sum_{i=1}^{n_k} \zeta_{i,v} \right).$$

Note that $\mathbb{E}(\zeta_{i,u}\zeta_{i,v}) = u^\top v$. Also, if we let $q = \max\{\theta(\log n_k + \log^3 p), 3\}$, then

$$\|\zeta_{i,u}\zeta_{i,v} - u^\top v\|_q \leq 1 + \|\zeta_{i,u}\zeta_{i,v}\|_q \leq 1 + \|\zeta_{i,u}\|_{2q}\|\zeta_{i,v}\|_{2q} \leq cq^2,$$

where the second inequality is due to the Cauchy–Schwarz inequality, and the third to Assumption 1. The constant $c$, although it varies from place to place, does not depend on $n_k$ or $p$. Then, Lemma G.4 of Lopes

25

et al. (2020) gives the following bound for $q > 2$,

$$\|\Delta(u,v)\|_q \leq cq \max \left\{ \|\Delta(u,v)\|_2, n_k^{-1} \left( \sum_{i=1}^{n_k} \|\zeta_{i,u}\zeta_{i,v} - u^\top v\|_q^q \right)^{1/q} \right\}$$

$$\leq cq \max\{n_k^{-1/2}, n_k^{-1+1/q} q^2\}$$

$$\leq c(\log n_k + \log^3 p) n_k^{-1/2}.$$

By the Chebyshev inequality

$$\mathbb{P}\left(|\Delta(u,v)| \geq e\|\Delta(u,v)\|_q\right) \leq e^{-q},$$

whence

$$\mathbb{P}\left(|\Delta(u,v)| \geq \frac{c\theta(\log n_k + \log^3 p)}{\sqrt{n_k}}\right) \leq \frac{1}{n^\theta p^\theta}.$$

Similar arguments apply to $\Delta'(u,v)$. Thus,

$$\mathbb{P}\left(\left|\frac{\widehat{\Sigma}_k(j,j')}{\sigma_{k,j}\sigma_{k,j'}} - \rho_{k,j,j'}\right| \geq \frac{c_\theta(\log n_k + \log^3 p)}{\sqrt{n_k}}\right) \leq \frac{1}{n^\theta p^\theta}$$

and furthermore by a union bound

$$\mathbb{P}\left(\sup_{1 \leq j,j' \leq p} \left|\frac{\widehat{\Sigma}_k(j,j')}{\sigma_{k,j}\sigma_{k,j'}} - \rho_{k,j,j'}\right| \geq \frac{c_\theta(\log n_k + \log^3 p)}{\sqrt{n_k}}\right)$$

$$\leq \sum_{1 \leq j,j' \leq p} \frac{1}{n_k^\theta p^\theta} = \frac{1}{n_k^\theta} \frac{p^2}{p^\theta} \leq \frac{1}{n_k^\theta}.$$

$\square$

Observing that $\sigma_{k,l,j} = \sqrt{\lambda_{k,l}^2 \sigma_{k,j}^2 + \lambda_{l,k}^2 \sigma_{l,j}^2}$, one obtains the following corollary.

**Corollary E.9.** *Under Assumption 1 and 3, if $n^{-1/2}\log^3 p \ll 1$, for any number $\theta \geq 2$, there are positive constants $c$ and $c_\theta$, not depending on $n$, such that the event*

$$\sup_{(k,l) \in \mathcal{P}} \sup_{1 \leq j \leq p} \left|\frac{\hat{\sigma}_{k,l,j}}{\sigma_{k,l,j}} - 1\right| \leq c_\theta (\log n_{\max} + \log^3 p) n^{-1/2}$$

*holds with probability at least $1 - cNn^{-\theta}$.*

**Lemma E.10.** *Suppose Assumptions 1–4 hold. Then, for any fixed $\theta \in (0, \infty)$ and $Q \asymp \log n$, for some constant $c$, not depending on $n$, one has*

$$\sup_{1 \leq k \leq K, 1 \leq j \leq p} \left\|\frac{\sigma_{k,j}^\theta}{\hat{\sigma}_{k,j}^\theta}\right\|_Q \leq c.$$

*Proof.* Below we suppress the subscripts from $\hat{\sigma}_{k,j}$, $\mu_k$ and $n_k$. Also, the constant $c$ might change its value

26

from place to place and depend on $\theta$. In addition, observing that

$$\frac{\sigma^2}{\hat{\sigma}^2} = \frac{1}{n^{-1}\sum_{i=1}^n[\{(X_i-\mu)-(\overline{X}-\mu)\}/\sigma]^2} = \frac{1}{n^{-1}\sum_{i=1}^n(Y_i-\overline{Y})^2}$$

with $Y_i = (X_i - \mu)/\sigma$ and $\overline{Y} = \sum_{i=1}^n Y_i$, without loss of generality, we assume $\mathbb{E}X = 0$ and $\mathbb{E}X^2 = 1$.

Let $\omega = Q\theta \asymp \log n$ and $c_1 = 1/2$. We first observe that

$$\mathbb{E}\hat{\sigma}^{-\omega} = \int_0^\infty \mathbb{P}(\hat{\sigma}^{-\omega} > t)\mathrm{d}t = \int_0^\infty \mathbb{P}(\hat{\sigma}^2 < t^{-2/\omega})\mathrm{d}t$$
$$= \int_0^{c_1^{-\omega/2}} \mathbb{P}(\hat{\sigma}^2 < t^{-2/\omega})\mathrm{d}t + \int_{c_1^{-\omega/2}}^{n^\omega} \mathbb{P}(\hat{\sigma}^2 < t^{-2/\omega})\mathrm{d}t + \int_{n^\omega}^\infty \mathbb{P}(\hat{\sigma}^2 < t^{-2/\omega})\mathrm{d}t.$$

For the last term, we have

$$\begin{aligned}
\mathbb{P}(\hat{\sigma}^2 < t^{-2/\omega}) &= \mathbb{P}\left(n^{-1}\sum_{i=1}^n (X_i - \overline{X})^2 \le t^{-2/\omega}\right) \\
&\le \mathbb{P}\left(\forall\, 1 \le i \le n : (X_i - \overline{X})^2 \le nt^{-2/\omega}\right) \\
&\le \mathbb{P}\left(\forall\, 1 \le i \le n : |X_i - \overline{X}| \le \sqrt{n}t^{-1/\omega}\right) \\
&\le \mathbb{P}\left(\forall\, 1 \le i \le n-1 : |X_i - X_n| \le 2\sqrt{n}t^{-1/\omega}\right) \\
&= \mathbb{E}\mathbb{P}\left(\forall\, 1 \le i \le n-1 : |X_i - X_n| \le 2\sqrt{n}t^{-1/\omega} \mid X_n\right) \\
&= \mathbb{E}\{\mathbb{P}\left(|X_1 - X_n| \le 2\sqrt{n}t^{-1/\omega} \mid X_n\right)\}^{n-1} \\
&\le (c\sqrt{n}t^{-1/\omega})^{(n-1)\nu}
\end{aligned}$$

for some universal constant $c > 0$ and for all sufficiently large $n$, where the last inequality is due to Assumption 4, and the last equality is due to the conditional independence of the random variables $|X_1 - X_n|, \ldots, |X_{n-1} - X_n|$ given $X_n$ and that these variables have identical conditional distributions. Therefore,

$$\int_{n^\omega}^\infty \mathbb{P}(\hat{\sigma}^2 < t^{-2/\omega})\mathrm{d}t \le (-(n-1)\nu/\omega + 1)^{-1} c^{\nu(n-1)} n^{(n-1)\nu/2} t^{-(n-1)\nu/\omega+1}\big|_{n^\omega}^\infty \asymp \nu c^{\nu(n-1)} n^{\omega - \frac{(n-1)\nu}{2}} \ll 1.$$

When $t \ge c_1^{-\omega/2}$ or equivalently $t^{-2/\omega} \le 1/2$, noting that $\sigma^2 = 1$ as we have assumed standardized $X$, one has

$$\begin{aligned}
\mathbb{P}(\hat{\sigma}^2 - 1 < t^{-2/\omega} - 1) &\le \mathbb{P}(\hat{\sigma}^2 - 1 < -1/2) \\
&\le \mathbb{P}(|\hat{\sigma}^2 - 1| \ge 1/2) \\
&\le \mathbb{P}(|\hat{\sigma}^2 - 1| \ge 2n^{-1/2}\omega \log n) \\
&\le cn^{-2\omega},
\end{aligned}$$

where the last inequality is obtained by an argument identical to that in the proof of Lemma D.7 of Lopes et al. (2020), except that the number $q = \max\{\kappa \log(n), 3\}$ there is replaced by $q = \max\{2\omega \log n, 3\}$. This

implies that
$$\int_{c_1^{-\omega/2}}^{n^\omega} \mathbb{P}(\hat{\sigma}^2 < t^{-2/\omega})\mathrm{d}t \leq n^\omega \cdot cn^{-2\omega} = cn^{-\omega} \ll 1.$$

Note that when $t \leq c_1^{-\omega/2}$, we have the trivial bound $\mathbb{P}(\hat{\sigma}^2 < t) \leq 1$. Therefore,
$$\mathbb{E}\hat{\sigma}^{-\omega} \leq c_1^{-\omega/2} + cn^{-\omega} + \nu c^{\nu(n-1)} n^{\omega - \frac{(n-1)\nu}{2}} \leq cc_1^{-\omega/2} = c2^{\omega/2},$$

or $\|\hat{\sigma}^{-\theta}\|_Q \leq c$. □

## F  Additional Simulation Studies on Functional ANOVA

As observed by Zhang et al. (2019), the level of within-function correlation impacts the power of the test. Following a suggestion of a reviewer, we assessed the effect of the within-function correlation on the proposed method by using the simulation setup of Zhang et al. (2019). Specifically, we set $\mu_k(t) = (1 + 2.3t + 3.4t^2 + 1.5t^3) + \theta(k-1)(1 + 2t + 3t^2 + 4t^3)/\sqrt{30}$ for $k = 1, 2, 3$ and $t \in [0, 1]$, and $X_k(t) = \mu_k(t) + \sum_{j=1}^{11} \sqrt{1.5}\rho^{j/2}\xi_{kj}\phi_j(t)$, where $\phi_1, \phi_2, \ldots$ are Fourier basis functions defined in Section 4. Two cases are considered for the random variables $\xi_{kj}$, namely, the Gaussian case $\xi_{kj} \overset{iid}{\sim} N(0,1)$ and the non-Gaussian case $\xi_{kj} \overset{iid}{\sim} t_4/\sqrt{2}$, where $t_4$ denotes Student's $t$ distribution with 4 degrees of freedom. As the number $m$ of design points has little impact on the results (Zhang et al., 2019), we fix $m = 100$ as in our previous simulation setting. As in Zhang et al. (2019), we consider $\rho = 0.1, 0.3, 0.5, 0.7, 0.9$, where small values of $\rho$ correspond to strong within-function correlation and large values signify weak correlation.

The results in Table S1 and Figure S1 show that our approach outperforms the other methods substantially. This may be partially explained by the fact that the basis functions $\phi_1, \phi_2, \ldots$ used to generate data in Zhang et al. (2019) coincide with the basis functions we use for projection in the proposed method. In light of this, we have provided an additional comparison in a more challenging setting, where the data are generated using a modified version of these basis functions (while our method still uses the original basis functions). Specifically, the modified basis functions, denoted by $\tilde{\phi}_1, \tilde{\phi}_2, \ldots$, are constructed in the following way: We first define $\tilde{\phi}_j(t) = \phi_j(0.8t + 0.1)$ for $t \in [0, 1]$ and $j = 1, 2, \ldots, 11$, and then apply the Gram–Schmidt procedure to orthonormalize $\tilde{\phi}_1, \tilde{\phi}_2, \ldots$ within $L^2([0,1])$. The results for simulation studies with these modified basis functions are shown in Table S2 and Figure S2, where the average value of the selected $\tau$ across all settings is $0.843 \pm 0.153$ and $0.824 \pm 0.143$ for Gaussian and non-Gaussian cases, respectively. We observe that the empirical sizes of most methods are close to the nominal level and the RP method tends to be conservative when $\rho$ is small. When the within-function correlation is strong, e.g., when $\rho = 0.1, 0.3$, the power of the proposed method is considerably larger than that of MPF and GET, which in turn is much larger than the power of the other methods except RP; the power of RP is close to that of the proposed method when $\rho \geq 0.3$ but substantially less when $\rho = 0.1$. When the within-function correlation becomes weaker, e.g., when $\rho = 0.7, 0.9$, all tests tend to have similar power. This shows that the proposed method is

28

Table S1: Empirical size of functional ANOVA in the simulation setting of Zhang et al. (2019)

|  | $\rho$ | $n$ | proposed | L2 | F | GPF | MPF | GET | RP |
|---|---|---|---|---|---|---|---|---|---|
| Gaussian | 0.1 | 50,50,50 | .044 | .044 | .044 | .044 | .038 | .039 | .011 |
| | | 30,50,70 | .054 | .066 | .060 | .062 | .066 | .040 | .017 |
| | 0.3 | 50,50,50 | .042 | .060 | .050 | .058 | .060 | .042 | .021 |
| | | 30,50,70 | .046 | .044 | .040 | .044 | .048 | .042 | .024 |
| | 0.5 | 50,50,50 | .058 | .060 | .056 | .062 | .052 | .038 | .021 |
| | | 30,50,70 | .052 | .062 | .062 | .064 | .058 | .039 | .037 |
| | 0.7 | 50,50,50 | .034 | .040 | .038 | .040 | .036 | .047 | .041 |
| | | 30,50,70 | .060 | .056 | .050 | .048 | .060 | .040 | .049 |
| | 0.9 | 50,50,50 | .042 | .050 | .048 | .050 | .054 | .025 | .044 |
| | | 30,50,70 | .044 | .058 | .054 | .060 | .032 | .039 | .051 |
| non-Gaussian | 0.1 | 50,50,50 | .048 | .060 | .054 | .056 | .050 | .030 | .004 |
| | | 30,50,70 | .052 | .054 | .050 | .052 | .064 | .041 | .009 |
| | 0.3 | 50,50,50 | .030 | .046 | .040 | .044 | .040 | .034 | .015 |
| | | 30,50,70 | .046 | .056 | .050 | .054 | .042 | .047 | .030 |
| | 0.5 | 50,50,50 | .046 | .076 | .072 | .080 | .060 | .032 | .027 |
| | | 30,50,70 | .036 | .036 | .036 | .036 | .024 | .037 | .030 |
| | 0.7 | 50,50,50 | .036 | .042 | .036 | .044 | .034 | .035 | .034 |
| | | 30,50,70 | .026 | .042 | .040 | .038 | .038 | .030 | .037 |
| | 0.9 | 50,50,50 | .034 | .064 | .062 | .058 | .054 | .036 | .037 |
| | | 30,50,70 | .044 | .060 | .060 | .054 | .056 | .030 | .047 |

preferred for ANOVA for functional data in which the within-function correlation is typically strong.

For simulation studies in the above and in Section 4, the tuning parameter $\tau$ is selected by the data-driven procedure described in Section 2. Below we investigate the effectiveness of this procedure by comparing it with using a fixed value of $\tau$. Specifically, in the simulation studies of Section 4, we also compute the empirical size and power for the proposed method by using each of the values $0, 0.1, \ldots, 0.9, 0.99$ for $\tau$. The results are presented in Table S3 and Figure S3, where the results for $\tau = 0, 0.2, 0.4, 0.6, 0.8, 0.99$ are provided. We observe that values close to 1 such as $\tau = 0.99$ yield inflated sizes for the resulting tests. For the family (M3), all values of $\tau$ lead to similar performance. Similar observations emerge for the family (M2) except $\tau = 0.99$ that yields much lower power. In families (M1) and (M4), the power increases as $\tau$. In these families, although $\tau = 0.99$ gives rise to the largest power, its empirical size also noticeably deviates from the nominal level. In all cases, the power of the data-driven method is close to that of $\tau = 0.8$. Overall, the data-driven selection procedure selects a value for $\tau$ that produces competitive power while keeping the size close to the nominal level.

Table S2: Empirical size of functional ANOVA in the modified simulation setting of Zhang et al. (2019)

|  | $\rho$ | $n$ | proposed | L2 | F | GPF | MPF | GET | RP |
|---|---|---|---|---|---|---|---|---|---|
| Gaussian | 0.1 | 50,50,50 | .053 | .061 | .055 | .059 | .049 | .043 | .017 |
|  |  | 30,50,70 | .048 | .052 | .050 | .050 | .038 | .042 | .010 |
|  | 0.3 | 50,50,50 | .055 | .062 | .057 | .062 | .063 | .043 | .020 |
|  |  | 30,50,70 | .061 | .056 | .054 | .055 | .060 | .034 | .021 |
|  | 0.5 | 50,50,50 | .055 | .061 | .047 | .064 | .039 | .043 | .037 |
|  |  | 30,50,70 | .056 | .047 | .044 | .048 | .049 | .040 | .033 |
|  | 0.7 | 50,50,50 | .053 | .050 | .047 | .044 | .047 | .037 | .035 |
|  |  | 30,50,70 | .052 | .059 | .058 | .056 | .050 | .026 | .051 |
|  | 0.9 | 50,50,50 | .057 | .054 | .053 | .050 | .053 | .043 | .050 |
|  |  | 30,50,70 | .051 | .053 | .052 | .045 | .044 | .038 | .057 |
| non-Gaussian | 0.1 | 50,50,50 | .043 | .048 | .045 | .046 | .051 | .036 | .009 |
|  |  | 30,50,70 | .050 | .049 | .047 | .047 | .048 | .057 | .016 |
|  | 0.3 | 50,50,50 | .052 | .050 | .048 | .047 | .043 | .041 | .017 |
|  |  | 30,50,70 | .050 | .067 | .064 | .063 | .056 | .042 | .016 |
|  | 0.5 | 50,50,50 | .048 | .048 | .043 | .042 | .033 | .031 | .018 |
|  |  | 30,50,70 | .043 | .045 | .043 | .055 | .049 | .038 | .029 |
|  | 0.7 | 50,50,50 | .042 | .053 | .054 | .046 | .053 | .032 | .033 |
|  |  | 30,50,70 | .048 | .053 | .051 | .054 | .037 | .048 | .044 |
|  | 0.9 | 50,50,50 | .053 | .054 | .050 | .048 | .047 | .030 | .040 |
|  |  | 30,50,70 | .055 | .036 | .037 | .039 | .042 | .031 | .050 |

# G   Additional Simulation Studies on MANOVA

We complement the numerical studies in Section 5 by assessing the performance of the proposed method in a more standard MANOVA setting, where data are sampled from continuous distributions. We consider three groups that are represented by random vectors $X_1, X_2, X_3 \in \mathbb{R}^p$, such that each $X_k$ follows an elliptical distribution (Fang et al., 1990) with mean $\mu_k$ and covariance $\Sigma_k$. Specifically, $X_k = \mu_k + U_k \Sigma_k^{1/2} Z_k$, where $U_k$ is a random variable following the exponential distribution with mean 1, $Z_k$ is a $p$-dimensional random vector following the standard multivariate normal distribution, and $U_k$ is independent of $Z_k$. Note that each $X_k$ is not Gaussian due to the random multiplicative factor $U_k$. We consider $p = 25$ and $p = 100$, and two scenarios of $\mu_k$, namely,

- the sparse case, where $\mu_k(j) = 1 + j \sin(2\pi j/p) \exp(j/p)/p + a_p \theta(k-1)((p-j+1)/p)^4$, so that when $k \neq l$, the difference $\mu_k(j) - \mu_l(j)$ between two mean vectors decays as $j$ increases, and

- the dense case, where $\mu_k(j) = 1 + j \sin(2\pi j/p) \exp(j/p)/p + a_p \theta(k-1)$, so that the difference $\mu_k(j) - \mu_l(j)$ remains constant across different coordinates $j$,

for $k = 1, 2, 3$, $j = 1, 2, \ldots, p$ and a constant $a_p$. In these settings, $\theta = 0$ corresponds to the null hypothesis,

Table S3: Empirical size of functional ANOVA with different values of $\tau$

| Covariance | M | $n$ | $\tau$ | | | | | | |
|---|---|---|---|---|---|---|---|---|---|
| | | | 0 | 0.2 | 0.4 | 0.6 | 0.8 | 0.99 | data-driven |
| common | M1 | 50,50,50 | .053 | .056 | .058 | .060 | .061 | .076 | .051 |
| | | 30,50,70 | .045 | .045 | .045 | .051 | .055 | .099 | .053 |
| | M2 | 50,50,50 | .048 | .047 | .049 | .050 | .056 | .082 | .042 |
| | | 30,50,70 | .043 | .044 | .045 | .046 | .050 | .094 | .057 |
| | M3 | 50,50,50 | .044 | .044 | .047 | .049 | .049 | .086 | .057 |
| | | 30,50,70 | .073 | .076 | .074 | .073 | .074 | .075 | .056 |
| | M4 | 50,50,50 | .045 | .047 | .046 | .046 | .052 | .073 | .046 |
| | | 30,50,70 | .046 | .045 | .050 | .050 | .058 | .084 | .053 |
| group-specific | M1 | 50,50,50 | .056 | .056 | .058 | .057 | .060 | .098 | .055 |
| | | 30,50,70 | .049 | .047 | .049 | .049 | .058 | .095 | .043 |
| | M2 | 50,50,50 | .051 | .050 | .050 | .049 | .052 | .095 | .056 |
| | | 30,50,70 | .061 | .061 | .060 | .062 | .068 | .108 | .052 |
| | M3 | 50,50,50 | .042 | .044 | .044 | .047 | .052 | .099 | .051 |
| | | 30,50,70 | .050 | .051 | .051 | .055 | .064 | .091 | .049 |
| | M4 | 50,50,50 | .060 | .058 | .059 | .056 | .054 | .095 | .052 |
| | | 30,50,70 | .055 | .055 | .056 | .054 | .062 | .114 | .050 |

under which the mean vectors of all groups are equal. We set $a_p = 0.025$ in the sparse scenario, and in the dense scenario we set $a_p = 0.014$ if $p = 25$ and $a_p = 0.008$ if $p = 100$; these values are chosen in the way that the power is approximately 1 when $\theta = 1$. We set $\Sigma_k(j_1, j_2) = j_1^{-1/4} j_2^{-1/4} \mathcal{C}\big((j_1-1)/(p-1), (j_2-1)/(p-1)\big)$, where $\mathcal{C}(s,t)$ is the Matérn correlation function defined in (7) but with different values of parameters; here we set $\sigma^2 = 1$, $\eta = 5$ and $\nu = 0.1$.

Similar to the study in Section 5, we compare the proposed method with the classic Lawley–Hotelling trace test (LH), the ridge-regularized Lawley–Hotelling trace test (RRLH) (Li et al., 2020), the procedure (S) of Schott (2007) and the data-adaptive $\ell_p$-norm-based test (DALp) (Zhang et al., 2018). As mentioned in Section 5, the method of Schott (2007) is favored for testing problems with a dense alternative, while the method of Zhang et al. (2018) has been reported to be powerful against different patterns of alternatives. The classic Lawley–Hotelling trace test is included as a baseline procedure. The empirical sizes in Table S4 show that those of the proposed method, Lawley–Hotelling trace test and Schott (2007) are close to the nominal level, while the size of Zhang et al. (2018) is inflated. This result is consistent with the observation made in Section 5, except that the inflation in the size of Zhang et al. (2018) seems more pronounced here, especially in the settings with a dense alternative. The empirical power functions shown in Figure S4 suggest that, in the sparse case, the proposed method consistently outperforms the others. In the dense setting, the proposed test has almost the same power of the test of Schott (2007), and substantially outperforms the Lawley–Hotelling trace test. With regard to the Zhang et al. (2018) test, it turns out that its type I

Table S4: Empirical size of ANOVA on multivariate Laplace data

|        | $p$ | $(n_1, n_2, n_3)$ | proposed | S | DALp | LH | RRLH |
|---|---|---|---|---|---|---|---|
| sparse | 25 | 50,50,50 | .056 | .048 | .067 | .049 | .032 |
|        |    | 30,50,70 | .058 | .049 | .063 | .046 | .038 |
|        | 100 | 50,50,50 | .053 | .052 | .080 | .032 | .054 |
|        |    | 30,50,70 | .056 | .046 | .063 | .029 | .062 |
| dense  | 25 | 50,50,50 | .055 | .059 | .070 | .036 | .053 |
|        |    | 30,50,70 | .057 | .054 | .070 | .040 | .054 |
|        | 100 | 50,50,50 | .052 | .045 | .067 | .026 | .046 |
|        |    | 30,50,70 | .054 | .045 | .080 | .039 | .044 |

Table S5: Empirical size of ANOVA on multivariate Laplace data with different values of $\tau$

|        | $p$ | $(n_1, n_2, n_3)$ | $\tau$ | | | | | | |
|---|---|---|---|---|---|---|---|---|---|
|        |    |          | 0 | 0.2 | 0.4 | 0.6 | 0.8 | 0.99 | data-driven |
| sparse | 25 | 50,50,50 | .055 | .051 | .047 | .043 | .048 | .051 | .056 |
|        |    | 30,50,70 | .032 | .033 | .042 | .046 | .056 | .063 | .058 |
|        | 100 | 50,50,50 | .038 | .037 | .039 | .045 | .051 | .067 | .053 |
|        |    | 30,50,70 | .048 | .046 | .049 | .055 | .061 | .075 | .056 |
| dense  | 25 | 50,50,50 | .039 | .040 | .048 | .048 | .055 | .061 | .055 |
|        |    | 30,50,70 | .048 | .044 | .042 | .046 | .049 | .057 | .057 |
|        | 100 | 50,50,50 | .047 | .045 | .045 | .052 | .058 | .076 | .052 |
|        |    | 30,50,70 | .042 | .049 | .048 | .049 | .056 | .062 | .054 |

error rate under the null is substantially higher than the nominal level, which makes it difficult to compare power. We also observe that the classic Lawley–Hotelling trace test, which is not specifically designed for the high-dimensional setting, deteriorates considerably as the dimension becomes larger, e.g., when $p = 100$ in both settings, while the regularized Lawley–Hotelling trace test (Li et al., 2020) improves upon the classic unregularized version when the dimension is relatively large in both settings.

As in the case of functional ANOVA, below we investigate the effectiveness of the data-driven selection procedure for $\tau$ by comparing it with using a fixed value of $\tau$. Specifically, in the above simulation studies, we also compute the empirical size and power for the proposed high-dimensional MANOVA by using each of the values $0, 0.1, \ldots, 0.9, 0.99$ for $\tau$. The results are presented in Table S5 and Figure S5, where we report the results for $\tau = 0, 0.2, 0.4, 0.6, 0.8, 0.99$. We observe that values rather close to 1 such as $\tau = 0.99$ result in slightly inflated size. In the sparse setting, all values of $\tau$ and the data-driven procedure lead to similar performance, while in the dense setting, large values of $\tau$ yield larger power, and the power of the data-driven method is close to that of $\tau = 0.8$. Similar to the observations in the functional ANOVA case, overall the data-driven selection procedure is effective in selecting a near optimal value for $\tau$.

# References

Bentkus, V. (2005). A Lyapunov-type bound in $R^d$. *SIAM Journal on Theory of Probability & Its Applications*, 49(2):311–323.

Fang, K.-T., Kotz, S., and Ng, K. W. (1990). *Symmetric multivariate and related distributions*. Chapman and Hall.

Li, H., Aue, A., and Paul, D. (2020). High-dimensional general linear hypothesis tests via non-linear spectral shrinkage. *Bernoulli*, 26(4):2541–2571.

Lopes, M. E., Lin, Z., and Müller, H.-G. (2020). Bootstrapping max statistics in high dimensions: Near-parametric rates under weak variance decay and application to functional data analysis. *The Annals of Statistics*, 48(2):1214–1229.

Nadarajah, S., Afuecheta, E., and Chan, S. (2019). On the distribution of maximum of multivariate normal random vectors. *Communications in Statistics - Theory and Methods*, 48(10):2425–2445.

Schott, J. R. (2007). Some high-dimensional tests for a one-way MANOVA. *Journal of Multivariate Analysis*, 98(9):1825–1839.

Xia, D. (2019). Non-asymptotic bounds for percentiles of independent non-identical random variables. *Statistics & Probability Letters*.

Zhang, J.-T., Cheng, M.-Y., Wu, H.-T., and Zhou, B. (2019). A new test for functional one-way ANOVA with applications to ischemic heart screening. *Computational Statistics & Data Analysis*, 132:3–17.

Zhang, M., Zhou, C., He, Y., , and Liu, B. (2018). Data-adaptive test for high-dimensional multivariate analysis of variance problem. *Australian & New Zealand Journal of Statistics*, 60(4):447–470.

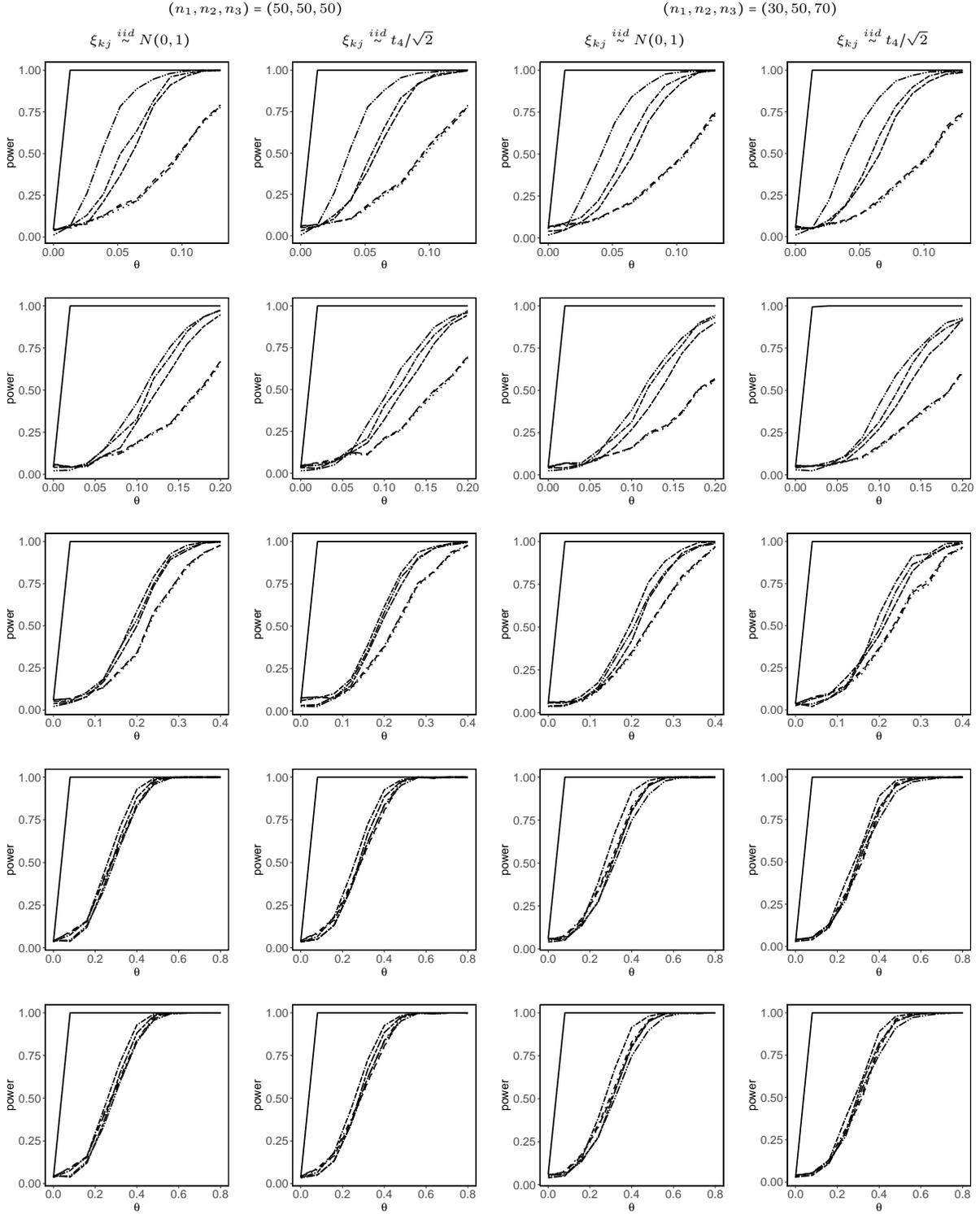

Figure S1: Empirical power of the proposed functional ANOVA (solid), L2 (dashed), F (dotted), GPF (dot-dashed), MPF (dot-dash-dashed), GET (short-long-dashed) and RP (dot-dot-dashed) in the simulation setting of Zhang et al. (2019). Rows 1 to 5 correspond to $\rho = 0.1, 0.3, 0.5, 0.7, 0.9$, respectively. The power functions of L2, F and GPF are nearly indistinguishable.

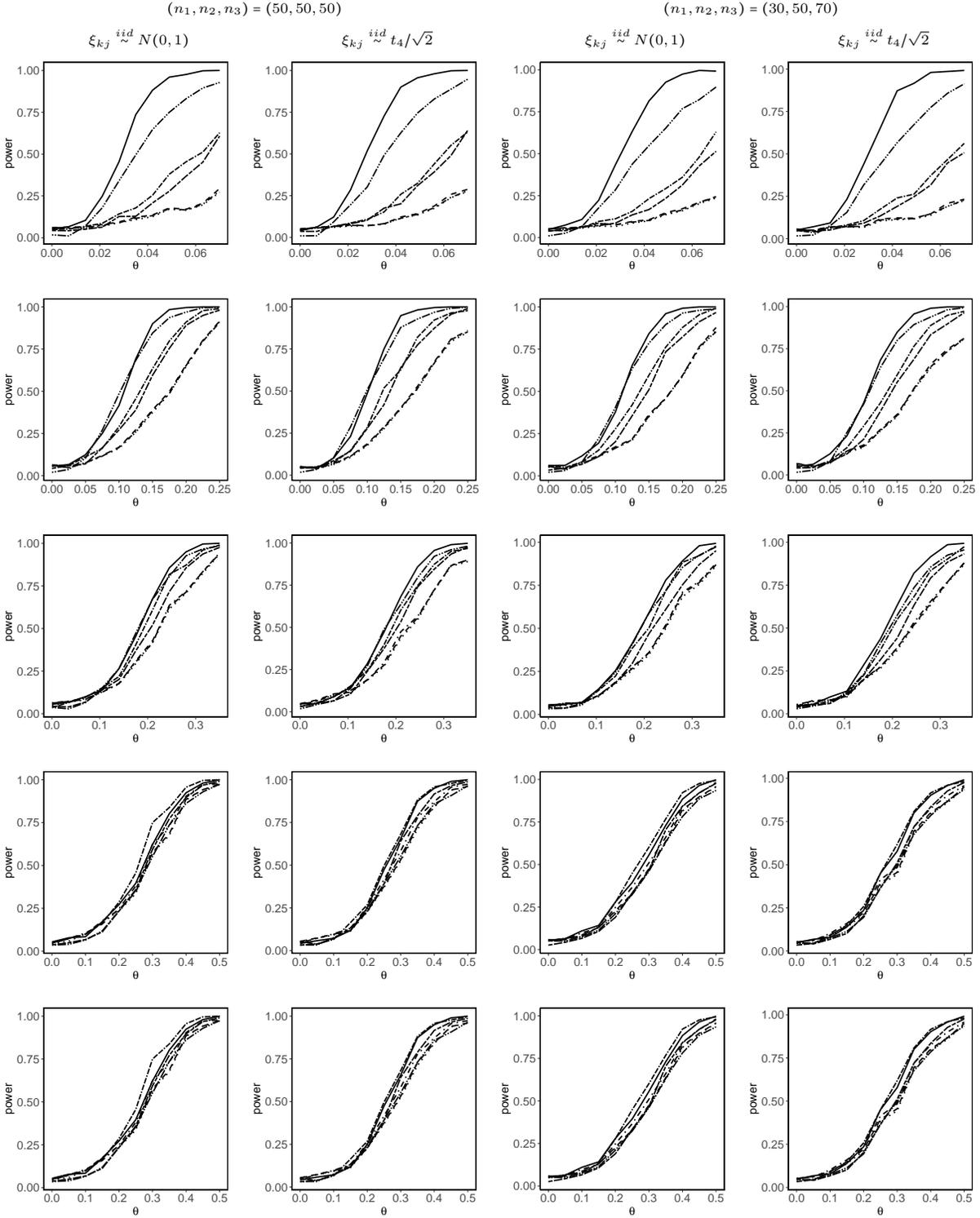

Figure S2: Empirical power of the proposed functional ANOVA (solid), L2 (dashed), F (dotted), GPF (dot-dashed), MPF (dot-dash-dashed), GET (short-long-dashed) and RP (dot-dot-dashed) in the modified simulation setting of Zhang et al. (2019). Rows 1 to 5 correspond to $\rho = 0.1, 0.3, 0.5, 0.7, 0.9$, respectively. The power functions of L2, F and GPF are nearly indistinguishable.

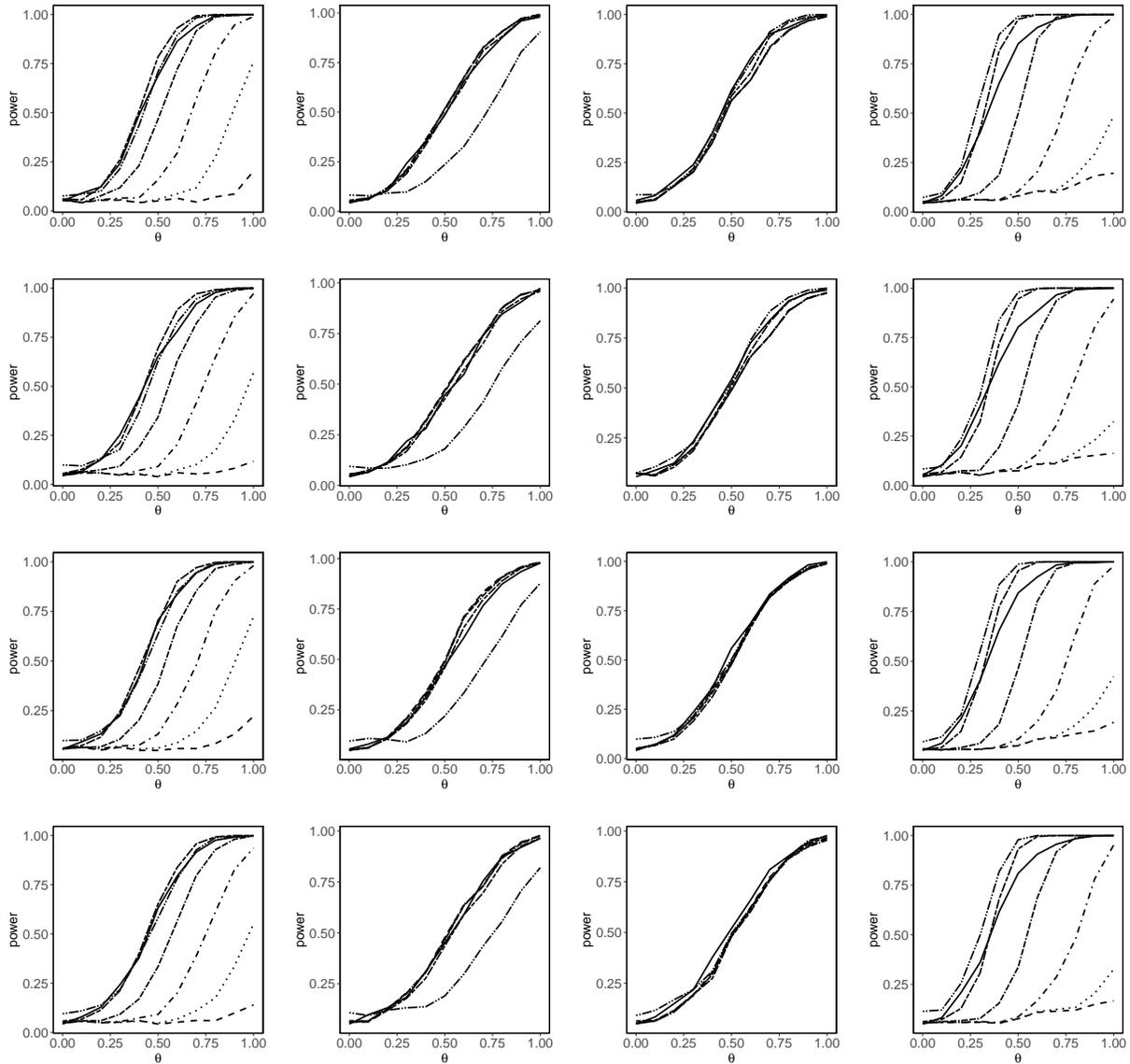

Figure S3: Empirical power of the proposed functional ANOVA with adaptively selected $\tau$ (solid), and fixed $\tau$ at 0 (dashed), 0.2 (dotted), 0.4 (dot-dashed), 0.6 (dot-dash-dashed), 0.8 (short-long-dashed) and 0.99 (dot-dot-dashed). From left to right the panels display the empirical power functions for families (M1), (M2), (M3) and (M4). The first two rows correspond to the "common covariance" setting and the last two correspond to the "group-specific covariance" setting. The sample sizes are $(n_1, n_2, n_3) = (50, 50, 50)$ for the first and third rows, and $(n_1, n_2, n_3) = (30, 50, 70)$ for the second and fourth rows.

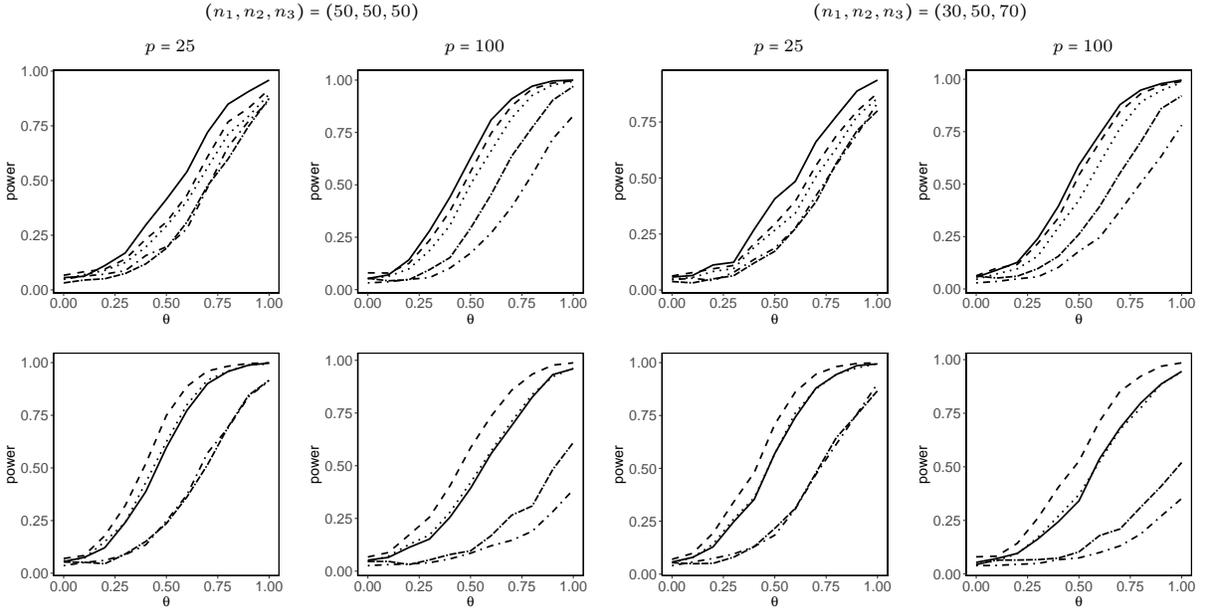

Figure S4: Empirical power of the proposed high-dimensional ANOVA (solid), DALp (dashed), S (dotted), LH (dot-dashed) and RRLH (dot-dash-dashed). Top, sparse setting; bottom, dense setting.

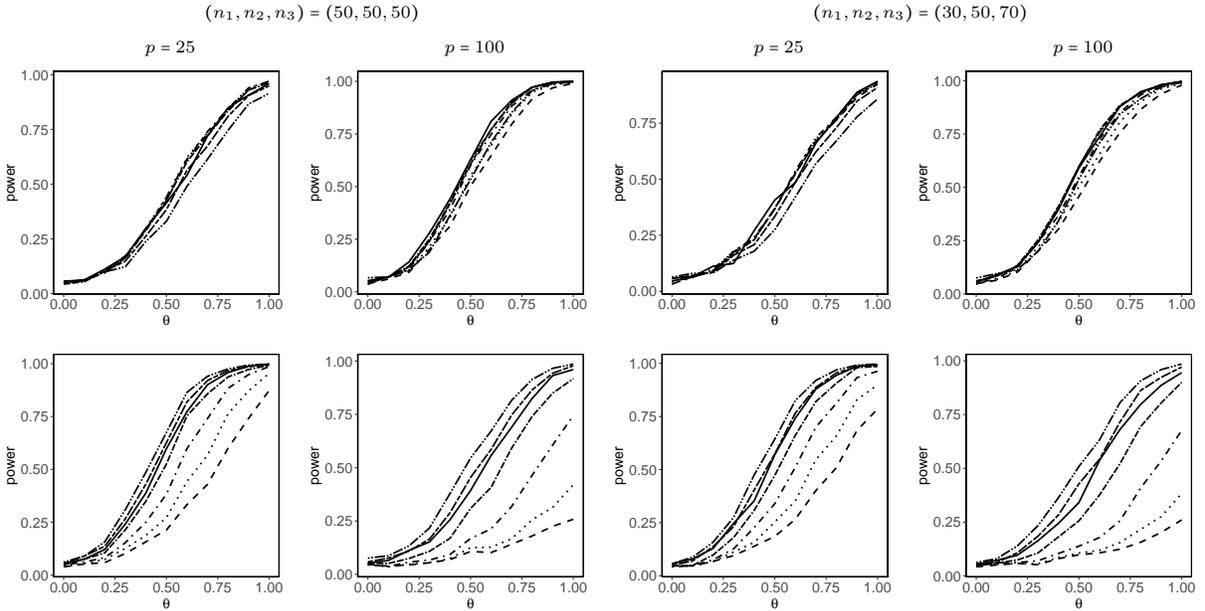

Figure S5: Empirical power of the proposed high-dimensional ANOVA with adaptively selected $\tau$ (solid), and fixed $\tau$ at 0 (dashed), 0.2 (dotted), 0.4 (dot-dashed), 0.6 (dot-dash-dashed), 0.8 (short-long-dashed) and 0.99 (dot-dot-dashed). Top, sparse setting; bottom, dense setting.



\end{document}